\documentclass[prd,preprint,eqsecnum,nofootinbib,amsmath,amssymb,preprintnumbers,tightenlines,longbibliography]{revtex4-1}

\usepackage{afterpage}
\usepackage{mathrsfs}
\usepackage{graphicx}
\usepackage{bm}
\usepackage{paralist}
\usepackage[colorlinks,linkcolor=blue,citecolor=blue]{hyperref}

\def\t{\tilde}

\def\F{{\mathcal F}}
\def\P{{\mathcal P}}

\def\L{{\mathcal L}}

\def\Im{{\rm Im}}

\def\tR{\tau_R}
\def\p{{\boldsymbol p}}

\def\Eq#1{Eq.~(\ref{#1})}
\def\Eqs#1{Eqs.~(\ref{#1})}

\def\Fig#1{Fig.~\ref{#1}}
\def\Sect#1{Section~\ref{#1}}

\def\Tab#1{Table~\ref{#1}}

\def\bra#1{\langle#1\vert}
\def\ket#1{\vert#1\rangle}

\newcommand{\be}{\begin{equation}}
\newcommand{\ee}{\end{equation}}

\newcommand{\beq}{\begin{eqnarray}}
\newcommand{\eeq}{\end{eqnarray}}
\newcommand{\nn}{\nonumber\\ }
\newcommand{\rme}{{\rm e}}
\newcommand{\rmd}{{\rm d}}
\newcommand{\del}{\partial}

\def\be{\begin{equation}}
\def\ee{\end{equation}}
\def\bea{\begin{eqnarray}}
\def\eea{\end{eqnarray}}

\begin{document}

\title{On attractor and fixed points in Bjorken flows}

\author{Jean-Paul Blaizot}
\affiliation{
	Institut de Physique Th{\'e}orique, Universit\'e Paris Saclay, 
        CEA, CNRS, 
	F-91191 Gif-sur-Yvette, France} 

\author{Li Yan} 
\affiliation{
Key Laboratory of Nuclear Physics and Ion-Beam Application (MOE) \& Institute of Modern Physics
Fudan University, 220 Handan Road, 200433, Yangpu District, Shanghai, China
}

\begin{abstract}

We consider a plasma of massless particles undergoing Bjorken expansion, mimicking the matter created in ultra-relativistic heavy ion collisions. We study the transition to hydrodynamics using kinetic theory in the relaxation time approximation. By  allowing  the relaxation time to depend on time, we can monitor the speed of the transition from the collisionless regime to hydrodynamics. By using a special set of moments of the momentum distribution, we reduce the kinetic equation to a coupled mode problem which encompasses all versions of  second order viscous hydrodynamics for Bjorken flows. This coupled mode problem is analysed first using techniques of linear algebra. Then we transform this two mode problem into a single non linear differential equation and proceed to a fixed point analysis. We identify an attractor solution as the particular solution of this non linear equation that joins two fixed points: one corresponding to the collisionless, early time regime, the other corresponding to late time hydrodynamics. We exploit the analytic solution of this equation in order to test several approximations and to identify generic features of the transition to hydrodynamics. We argue that extending the accuracy of hydrodynamics to early time, i.e. to the region of  large gradients, amounts essentially to improve the accuracy of the location of the collisionless fixed point. This is demonstrated by showing that a simple renormalisation of a second order transport coefficient puts the free streaming fixed point at the right location, and allows us to reproduce accurately the full solution of the kinetic equation within second order viscous hydrodynamics, even  in regimes far from local equilibrium. 
\end{abstract}

\maketitle


\section{Introduction}

A remarkable feature that emerges from  the study of heavy ion collisions at  RHIC and LHC is the success of relativistic viscous hydrodynamics in the description of the time evolution of the matter created there \cite{Heinz_2013} (for a recent review see \cite{shen2020recent} and references therein). It is even argued that viscous hydrodynamics  start  working in regimes where it is a priori not expected to work, such as for instance in small collisional systems, or when there subsists a seemingly large pressure anisotropy \cite{Weller:2017tsr}.  This ``unreasonable effectiveness'' of hydrodynamics, as it has been sometimes qualified \cite{Romatschke_2017}, has triggered recently a large amount of works questioning the very foundations of relativistic fluid dynamics, and the conditions under which it can be applied \cite{romatschke2017relativistic,Florkowski:2017olj}. One particular  issue that calls for a deeper understanding is indeed how hydrodynamics emerge as a universal description of many systems, with different underlying microphysics.

In the context of kinetic theory, hydrodynamic behavior typically emerges for small deviations away from local equilibrium: collisions drive the system toward local equilibrium, and the ensuing dynamics is mainly controlled by conservation laws and the associated long wavelength, low frequency excitations. By long wavelength, low frequency, we mean wavelengths large compared to the collision mean free path, and frequencies small compared to the collision frequency.  This is usually characterized in terms of a dimensionless number, referred to as a Knudsen number, the ratio between microscopic and macroscopic scales: hydrodynamics sets in typically when this number becomes small.  When local equilibrium is reached, the collisions do not play much of a role aside from maintaining the  local equilibrium as the system expands, conserving locally energy and  momentum. The evolution of the system is then amenable to a simple description in terms of a finite set of fields, energy density, pressure and fluid velocity, i.e., the hydrodynamic fields.

The modern view on hydrodynamics is that of an effective field theory for long wavelength, low frequency modes. This is not inconsistent with the kinetic description, but it does not require the existence of particles (or quasiparticles), and hence it makes no reference to collisions. It is applicable to systems that remain strongly coupled during their entire evolution, as described for instance by holographic techniques.  Indeed the description of a boost invariant plasma using such techniques was shown to lead  to hydrodynamics at late time\cite{Heller:2011ju}.  In such situations, hydrodynamics set in when short wavelength excitations responsible for various transient effects that are sensitive to the initial conditions have died out,  leaving only the long wavelength modes associated to conservation laws. In this context the quasi-normal modes play a role somewhat similar to the single particle excitations in the context of kinetic theory. These modes decay on short-time scales, leaving at late times the hydrodynamic fields as the only degrees of freedom.  In that point of view, the emphasis is put on the gradients of the fields, with the effective theory being built as a gradient expansion. The coefficients of the various gradient structures are naturally related to transport coefficients. 

In order to assess the validity of hydrodynamics, one may look for the consistency of the approach within the theory itself, comparing for instance successive orders in the gradient expansion. Evidently, for practical purpose, large orders in the gradient expansion are of limited use. However, conceptually, it is interesting  to have  an understanding of the mathematical properties of the whole series of gradients. For a 1+1 dimensional boost invariant plasma, the gradient expansion can be pushed to essentially arbitrary high order, and its nature, as a divergent, asymptotic series, revealed \cite{Heller:2013fn}. In the same context, using Borel summation techniques, it was shown that the full solution of  the Israel-Stewart version of second order viscous hydrodynamics takes the form of a  trans-series,  where each order of the hydrodynamic gradient expansion receives corrections that are exponentially suppressed at large time \cite{Heller:2015dha}. It was also found that all solutions evolve toward a particular solution, referred to as ``attractor'', which coincides with hydrodynamics at late time, irrespective of the initial conditions. At this point, we should recall an important feature of Bjorken flow (in the absence of transverse expansion): Because of the boost symmetry all gradients in the systems are proportional to $1/\tau$ where $\tau$ is the proper time.  Thus, when probing the validity of hydrodynamics in regimes of large gradients, one is implicitly  extending  hydrodynamics to early time. As we shall see, all versions of second order viscous hydrodynamics make, by construction,  an implicit assumption about the early time regime. 

The notion of attractor is a familiar one in the theory of dynamical systems. In fact, by expanding the distribution function in moments, one can transform the kinetic equations into a discrete (infinite) set of coupled differential equations that can be analyzed using concepts well developed for dynamical systems. Such an approach has been followed in \cite{Behtash_2019a}. Our analysis bears similarity with that strategy.  However we do not aim at the same level of generality, and we deliberately leave aside some aspects of the thermalisation process (for a recent review see \cite{berges2020thermalization}).  As we have mentioned above, provided they are sufficiently frequent, collisions drive the system to local equilibrium. In particular, they tend to wash out deviations of the momentum distribution from spherical  symmetry. This process of ``isotropisation'' is slowed down by the slow relaxation of conserved quantities. The longitudinal expansion acts in a similar way as the conservation laws and contribute to delay the isotropization. This competition between expansion and collisions is in fact a major feature of the problem that we want to address. To do so,  we define moments $\L_n$ in which the absolute value of the momentum is integrated out uniformly for all $n$, in such a way that all the $\L_n$'s  have the same dimension as the energy density \cite{Blaizot:2017lht,Blaizot:2017ucy}: the $\L_n$'s characterize the angular distorsion of the momentum distribution, with $\L_0$ a monopole, $\L_1$ a quadrupole, and so on. As was shown in \cite{Blaizot:2019scw}, the first two moments turn out to be sufficient to study the main features of isotropization, at a semi-quantitative level. These two moments correspond to the two independent components of the energy momentum tensor: $\L_0$  is the energy density,   $\L_1$ is the difference between the longitudinal and the transverse pressures (we consider only massless particles). These moments obey a simple set of coupled equations whose structure can be shown to be equivalent, in the context of Bjorken flow, to all versions of second order viscous hydrodynamics. These equations also describe, albeit approximately, the collisionless regime which dominates at early times. The kinetic equation reduces then to a simple coupled mode problem that can be analyzed, to a large extent, analytically. 


 By transforming the system of linear equations for $\L_0$ and $\L_1$ into a non linear differential equation for the pressure asymmetry (essentially $\L_1/\L_0$) one can perform a fixed point analysis. In the collisionless regime the two eigenvalues of the linear problem correspond to two fixed points, the lowest mode corresponding to a stable fixed point, the highest mode to an unstable one (as one moves forward in time). The stable fixed point evolves smoothly (adiabatically) under the effect of collisions towards another fixed  point characteristic of the hydrodynamic regime. This provides a simple and intuitive picture for the ``attractor'' as the particular solution of the non linear equation that joins the two fixed points.  The structure is robust, as the two fixed points represent the well identified collisionless and hydrodynamic regimes. It is therefore not surprising that it was found in all studies of Bjorken flows based on an underlying kinetic theory\footnote{The case of AdS/CFT is special. There is some argument \cite{Kovchegov_2007} that, in this context, the energy density goes to a constant as $\tau\to 0$, with $\P_L=-\varepsilon$ and $\P_T=\varepsilon$. However, it is unclear whether this short time behavior can be associated to  any fixed point.  Only the hydrodynamic fixed point can be  clearly identified, and numerical studies seem indeed to indicate that an attractor behavior is only manifest at late times~\cite{Kurkela:2019set}. See also \cite{Romatschke:2017vte}.}. 
 

In the present paper, we analyze the emergence of hydrodynamics in the framework of a simple kinetic theory, where collisions are treated in the relaxation time approximation. We  specialize to the paradigmatic case of Bjorken flow, ignoring the transverse expansion. An important aspect of the present work is that  we allow for the relaxation time $\tau_R$ to depend on time.  We  consider a simple ansatz, namely a power law  of the form
\be
\label{eq:delta}
\tau_R \sim \tau^{1-\Delta},
\ee
where $\Delta$ is a constant. This simple ansatz, which allows for an analytic solution \cite{Blaizot:2020gql}, captures  generic behaviors that have been observed in more sophisticated calculations, as we shall discuss shortly.  The ratio between the collision rate $\sim 1/\tau_R$ and the expansion rate $\sim 1/\tau$ (which can  be considered as the  inverse of a Knudsen number), will be denoted by 
\beq\label{wdef}
w\equiv \frac{\tau}{\tau_R(\tau)}=\left(\frac{\tau}{\tau_1} \right)^{\Delta}.
\eeq
The dimensionful parameter $\tau_1$ is the time at which the collision rate equals the expansion rate, i.e. $\tau_1=\tau_R(\tau_1)$. For a constant relaxation time ($\Delta=1$) we have of course, $\tau_1=\tau_R$. 
Note that as long as $\Delta>0$, we can use the variable $w$ as a measure of time: time evolution can then be seen as a flow in Knudsen number.   Indeed the mapping  between $\tau$ and $w$ is then monotonous positive, that is, $w$ increases as $\tau $ increases. The change between the collisionless regime and the hydrodynamic regime occurs when $w\sim 1$. Since $\delta w/w =\Delta\,\delta \tau/\tau$, the same relative change in $w$ corresponds to a relative change in $\tau$ that is multiplied by $1/\Delta$. Thus, the speed of the transition increases with increasing $\Delta$.  We shall indeed see that the regime of small $\Delta$ is well captured by an approximation akin to the adiabatic approximation, while the limit of a large $\Delta$ corresponds rather to a sudden transition. When $\Delta=0$, the collision rate equals the expansion rate throughout the evolution of the system which then reaches a (non hydrodynamic) stationary state. For $\Delta<0$ more peculiar behaviors can be observed.

 There are various physical reasons for which one may want to consider a time-dependent collision time $\tau_R$. For instance, in the bottom-up scenario of thermalization \cite{Baier:2000sb}, the initial regime is one in which the parton-parton scattering cross-section $\sigma$ grows linearly with time, so that the collision rate $\sim 1/(\sigma n)$ is constant. In the following stage however, one finds that $\tau_R\sim \tau^{1/2}$. When solving the Boltzmann equation in the small collision angle approximation, one finds that the collision rate is nearly equal to the expansion rate during a large part of the evolution \cite{Blaizot:2019dut}. A similar situation is met in \cite{Denicol:2019lio} when solving the Boltzmann equation for a gas of hard spheres (i.e. for constant cross section). It has also been argued that when the transverse expansion starts to become significant, the system rapidly approaches freeze-out, which can be mimicked by a rapidly increasing collision time \cite{Chattopadhyay:2019jqj} (corresponding in our setting to a negative $\Delta$). Thus, by varying $\Delta$, one may indeed explore many relevant physical situations.

\begin{figure}[h]
\begin{center}
\includegraphics[angle=0,scale=0.3]{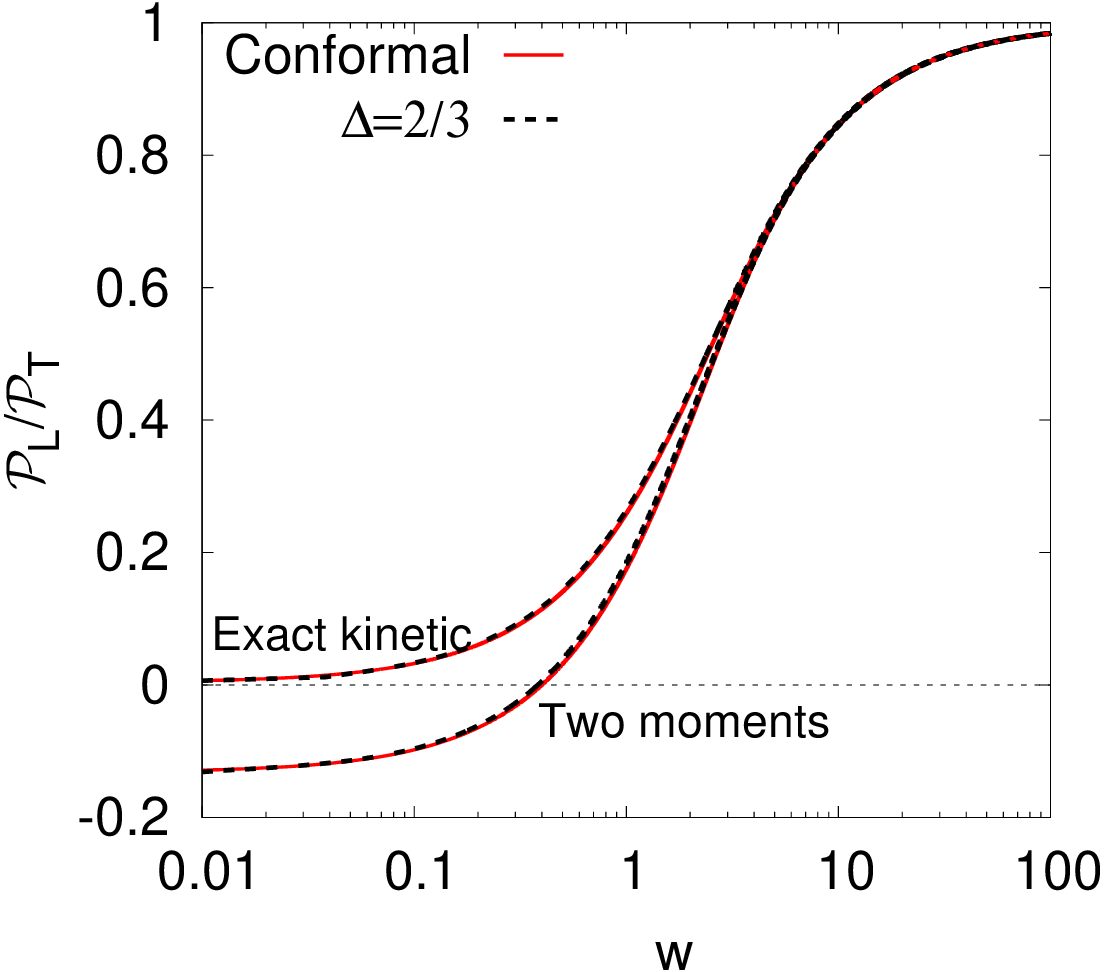} 
\end{center}
\caption{Evolution of the attractor solution for the ratio $\P_L/\P_T$ calculated from full kinetic theory and the two-moment truncation, with a conformal relaxation time $\tau_R\propto 1/T$ (full lines) and $\tau_R\propto \tau^{1-\Delta}$ with $\Delta=2/3$ (dashed lines) .  \label{fig:conformal}}
\end{figure}

 Physically, it would be more natural to relate the relaxation time to  the local properties of the matter, such as for instance the local density of particles, but doing so would  introduce non linearities  that hinder an analytical treatment. However, the example of conformal symmetry shows that our simple ansatz gives an accurate account of the evolution, at least for this case. Imposing conformal symmetry on the expanding system requires $\tau_R\sim T^{-1}$, where $T$ is the effective temperature, which, in the late time hydrodynamic regime, varies with time as  $T\sim\tau^{-1/3}$. This corresponds, in  this regime,  to $\Delta=2/3$. The plot in Fig.\ref{fig:conformal} reveals that the solution of the kinetic equation for $\Delta=2/3$ matches perfectly the exact solution (obtained for $\tau_R T={\rm cste}$). This plot contains also a number of interesting features that we shall comment shortly along with the outline of the paper to which we come now.

The paper contains three main sections. In the next section we first recall the general structure of the two moment  problem and show that all versions of second order hydrodynamics share this same mathematical structure. Then we analyse the solution of the coupled equation for the two moments using elementary techniques from linear algebra. We explore the behaviors of the solution in various regimes of early and late times, and as a function of the parameter $\Delta$. This allows us in particular to explore approximations such as time dependent perturbation theory or the adiabatic approximation which becomes exact in the limit $\Delta=0$. We also  comment on the role of higher moments, which are ignored in the two-moment truncation, but whose effects are amenable to a simple analysis when $\Delta=0$. 
In the following section we recast the linear system into a single non linear differential equation for the pressure asymmetry, defined here as $\L_1/\L_0$,  and proceed to its fixed point analysis. This provides a new vision of the attractor, as the solution that joins the two fixed points of the non linear equation. For the quantity $\P_L/\P_T$, the three parts of the attractor are well illustrated in Fig.~\ref{fig:conformal}: two nearly constant parts (in logarithmic scale) that  are characteristic of the evolution near the collisionless and hydrodynamic fixed points, and a transition region that takes place naturally in the region where $\tau\sim \tau_R$ ($w\sim 1$). The differential equation for the pressure asymmetry is amenable to an analytic solution which is used in particular to test various approximations. The representation of the solution as a trans-series is given, and approximations involving   effective viscosities are briefly discussed. The last major section is devoted to a more physical discussion where we study, by varying the parameter $\Delta$, various regimes that have been identified in numerical simulations. In particular we comment on the special regimes where the relaxation time nearly equals the expansion time, i.e. where collisions nearly compensate the effect of the expansion. In the latter case, a scaling regime is observed, somewhat reminiscent of the phenomenon of non-thermal fixed points \cite{Mazeliauskas_2019}. A particular feature of the two-moment truncation is that it allows for unphysical excursions into regions of negative pressures. This is visible in Fig.\ref{fig:conformal}. However, at the end of this paper, we show that a simple renormalization of a second order transport coefficient allows us to correct this unphysical feature, and bring the solution of the two-moment truncation in excellent agreement with the exact solution of the kinetic equation.  The last section summarizes the conclusions. Several appendices contain technical material.

\section{Hydrodynamics as a coupled mode problem}\label{sec:coupledmodes}

In this paper, we consider a system of massless particles, mimicking that produced in high energy heavy ion collisions. This system expands at the speed of light along the collision axis, the $z$-axis, in a boost invariant fashion \cite{Bjorken:1982qr}. It follows from the boost invariance that the full evolution of the system can be deduced from that of a slice of matter centered around the ``tranverse'' plane, $z=0$.  We describe the particles in this slice of matter by a phase space distribution function  $f(\p,\tau)$, independent of the transverse coordinates, with  $\tau=\sqrt{t^2-z^2}$ denoting the proper time.  The evolution of this distribution function is governed by a simple kinetic equation, where the collision term is treated in the relaxation time approximation~\cite{Baym:1984np}:
\beq\label{kineticeqn}
\left[ \del_\tau-\frac{p_z}{\tau}\del_{p_z} \right]f(\p,\tau)=-\frac{f(\p,\tau)-f_{\rm eq}(p/T)}{\tau_R}.
\eeq
Here,  $f_{\rm eq}(p/T)$ is the local equilibrium distribution function with $T$ the effective temperature determined at each time from the requirement that the energy density be the same whether evaluated with $f_{\rm eq}$ or the exact distribution $f(\p,\tau)$. This constraint on  $f(\p,\tau)$ is commonly referred to as the Landau matching condition. We  ignore particle number conservation, so that $f_{\rm eq}(p/T)$ depends solely on the temperature and not on a chemical potential. In Eq.~(\ref{kineticeqn}) $\tau_R$ is the relaxation  time, which may depend on $\tau$, as we have mentioned.

Although Eq.~(\ref{kineticeqn}) can be solved numerically to any desired accuracy, much insight can be gained by eliminating unnecessary details of the momentum distribution, and  focus on the most important degrees of freedom, those which play an essential role in driving the system to isotropy\footnote{We ignore in our discussion the plasma instabilities and their potential role in the isotropization \cite{Arnold_2005}.}and eventually to the hydrodynamic regime. To that aim, we define the following moments~\cite{Blaizot:2017lht}
\beq\label{Lndef}
\L_n(\tau)=\int\frac{\rmd^3\p}{(2\pi)^3 p^0} |\p|^2 P_{2n}(p_z/|\p|) f(\p,\tau).
\eeq
 The first two moments identify with the two independent components of the energy-momentum tensor, the energy density $\varepsilon$  and the  difference between the longitudinal and transverse pressures, respectively $\P_L$ and $\P_T$:
 \beq\label{L0L1andeP}
  \L_0=\varepsilon,\qquad \L_1=\P_L-\P_T.
  \eeq
  This identification is the main motivation for the choice of the moments (\ref{Lndef}). Other choices have been used in the literature (see e.g. Ref.~\cite{Denicol:2012cn}). As an interesting alternative to the definition (\ref{Lndef}) one may choose to expand the distribution function as follows \cite{behtash2020transasymptotics}, 
$f(\p,\tau)=f_{\rm eq}(p/T) \sum_{n=0}^\infty c_\ell(\tau) P_{2\ell}(p_z/|\p|)$. One would then obtain slightly different moments. However, for massless particles, and a suitable normalization, these first two moments are simply related to $\L_0$ and $\L_1$, to within trivial numerical factors\footnote{Thus $c_ {\ell=0}=\L_0$ and $c_{\ell=1}$ (denoted $c_{01}$ in \cite{Behtash:2019txb}) is $c_{\ell=1}=5 \L_1/\L_0$.}. 
Note that because of the special momentum weight  in Eq.~(\ref{Lndef}) ($|\p|^2/p^0=p$ for massless particles), all the moments $\L_n$ have the same dimension, that of the energy density. A useful quantity that is related to the first two moments is the pressure anisotropy, which we define here as the ratio
\beq\label{pressureasym}
\frac{\L_1}{\L_0}=\frac{\P_L-\P_T}{\varepsilon}.
\eeq
Since for massless particles, $\P_L+2\P_T=\varepsilon$, we have $-1/2\le \L_1/\L_0\le 1$, the lower bound corresponding to $\P_L=0$, while the upper bound corresponds to $\P_T=0$.

As was shown in \cite{Blaizot:2017ucy}, the $\L_n$'s satisfy an infinite set of coupled equations that can be easily deduced from the kinetic equation (\ref{kineticeqn}):
\begin{align}
\label{eq:eomL}
\frac{\partial \L_n}{\partial \tau} =& -\frac{1}{\tau}\left[a_n\L_n
+b_n\L_{n-1}+c_n\L_{n+1}\right]-\frac{(1-\delta_{n0})\L_n}{\tR}\,,
\end{align}
Since the details of the ``radial'' distribution are integrated out\footnote{The evolution of the ``radial'' momentum distribution plays an important role in the final thermalization of the system, but it is not essential to understand the main dynamics that drives the system towards isotropy, which is our main concern here. It is possible to improve the present description by introducing a more complete set of moments, as done for instance in \cite{Behtash:2019txb}, although there does not seem to be any strong argument that would justify a truncation in this case.  An alternative is to use an angular mode expansion, as  in \cite{Blaizot:2019dut}.  } (only the root mean squared radius of the distribution is taken into account), these moments do not allow us to reconstruct the full momentum distribution. There is therefore at this point a  loss of information which has, however, no impact on the energy-momentum tensor: 
if one solves the complete set of equations (\ref{eq:eomL}), even though one cannot reconstruct the distribution function entirely,  one obtains the exact evolution of the energy-momentum tensor~\cite{Blaizot:2017ucy,Blaizot:2019scw}.

Now, the main usefulness of a description in terms of moments is that, in some instances, the essential dynamics can be captured by a small set of moments. This happens to be the case here, even for the most drastic truncation, i.e., the two-moment truncation \cite{Blaizot:2017ucy,Blaizot:2019scw}. This may come as a surprise since the expansion drives the momentum distribution to a flat oblate distribution that requires many moments $\L_n$ to be accurately described (see later and in particular footnote~\ref{note:ellispoid}) . However, we are not interested here in an accurate description of the details of the shape of the momentum distribution, nor in a precise description of the collisionless regime,  but mainly in the energy momentum tensor whose non trivial components are only the first two moments. Indeed, in the case of Bjorken flow the two independent components in the energy-momentum tensor are related to $\L_0$ and $\L_1$, namely, 
\be\label{Txxyyzz}
T^{00} = T^{xx} + T^{yy} + T^{zz} = \L_0\,,\qquad
\frac{1}{2}(T^{xx}+T^{yy}) - T^{zz} = \L_1
\ee
The two-moment truncation provides then an effective theory for these  two degrees of freedom, whose dynamics is only moderately renormalized when higher moments are taken into account \cite{Blaizot:2019scw}.  The  dynamical reasons behind the the success of this truncation will be discussed  further as we proceed. 


The two-moment truncation provides a transparent description of the transition between the collisionless regime and the regime dominated by collisions, leading eventually to viscous hydrodynamics. In particular, the effect of the collisions  is neatly isolated in the last term of Eq.~(\ref{eq:eomL}).   The two-moment truncation gives an approximate description of the collisionless regime in terms of two coupled modes. As we shall recall in the next subsection, the same two coupled modes, once collisions start to dominate the dynamics, also account for all versions of second order viscous hydrodynamics applied to Bjorken flow. 

Before we move on, it is perhaps useful to mention an analogy with a somewhat similar situation encountered in quantum liquids \cite{Pines,baym1991landau}. We expect that, quite generally,  the effect of the collisions is to  damp efficiently all high order moments of the momentum distribution, but those related to conservation laws.  However, as the famous example of the zero sound in liquid helium shows, interactions between particles  may contribute to maintain non trivial distorsions of the momentum distribution, independently of those attached to the conservation laws. It is only when the temperature is high enough that these interaction effects are overwhelmed by those of the collisions, and the zero sound turns into first sound where only the first two moments of the momentum distribution play a significant role (those moments that are associated with conservation laws). Although the analogy is not perfect, one may think of the expansion as playing a role analogous to that of the interactions in the zero sound mode, and view the competition between expansion and collisional effects as the analog of that at work in the transition between zero and first sound. 
  
 \subsection{The two-moment truncation and second order viscous hydrodynamics}
 \label{sec:2ndhyd}

 The two-moment truncation derives from Eqs.~(\ref{eq:eomL}), where we set $\L_n=0$ for all $n\ge 2$. We then end up with the two coupled linear equations 
  \begin{subequations}
\label{eq:Lequ}
\begin{align}
\label{eq:Lequa}
\frac{\rmd \L_0}{\rmd \tau} =& - \frac{1}{\tau}(a_0 \L_0 + c_0 \L_1)\,,\\
\label{eq:Lequb}
\frac{\rmd \L_1}{\rmd \tau} =& - \frac{1}{\tau}(a_1 \L_1 + b_1 \L_0) - \frac{\L_1}{\tau_R}\,.
\end{align}
\end{subequations}
The coefficients 
\be 
a_0=4/3,\quad 
a_1=38/21,\quad
b_1=8/15,\quad
c_0=2/3\,,
\ee 
are pure numbers that are determined by the geometry of the expansion.   The coefficients $a_0$ and $c_0$  are actually fixed by energy-momentum conservation, $\partial_\mu T^{\mu\nu}=0$, which translates into the equation 
\begin{eqnarray} \label{eq:energy_expansion}
\rmd  (\tau \varepsilon )+ \P_L \rmd \tau=0.
\end{eqnarray}
This equation  is in fact Eq.~(\ref{eq:Lequa}). The values of $a_0$ and $c_0$ can then be read off, after noticing that $\P_L+2\P_T=\varepsilon$ (we are assuming massless particles), so that $\P_L =\varepsilon/3+(2/3)\L_1$ (where we have used (\ref{L0L1andeP})).

When the collision rate is small compared to the expansion rate, the system is effectively collisionless, and Eqs.~(\ref{eq:Lequ}) describe, approximately, the free streaming of the particles. We shall return to this regime later in this section. At the moment, we focus on the late time regime where, generically, the collision rate overcomes the expansion rate and the hydrodynamic regime is reached. As already emphasized,  all versions of second order viscous hydrodynamics for Bjorken flow  
share the same mathematical structure  as that encoded in the linear system (\ref{eq:Lequ}), modulo an adjustment of the parameters $a_1$ and $b_1$. This is what we review briefly now.

We just saw that Eq.~(\ref{eq:Lequa}), which translates energy conservation, has a universal character and  is common to all  formulations of hydrodynamics. It is more commonly written as follows (with $\P=\varepsilon/3$)
\beq
\frac{\rmd \varepsilon}{\rmd \tau}+\frac{\varepsilon+\P}{\tau} =\frac{\pi}{\tau},\qquad \pi=-c_0\L_1\eeq
where $\pi$ is the viscous tensor. 
In\emph{ ideal} hydrodynamics, the viscous tensor is neglected. The equation (\ref{eq:energy_expansion}) can  be then solved and yield $\varepsilon(\tau)\sim \tau^{-a_0}$. 
By taking  viscous effects into account via the leading order constitutive equation for $\pi$, namely $\pi=4\eta/(3\tau)$ with $\eta$ the shear viscosity, one obtains the  Navier-Stokes (NS) equation\footnote{In the kinetic framework that we are using here,  the shear viscosity is given by (with $s$ the entropy density)
\beq\label{viscosity}
\eta=\frac{b_1}{2}\varepsilon \tau_R\qquad  \frac{\eta}{s}=\frac{b_1}{2a_0} T\tau_R =\frac{1}{5} T\tau_R\qquad (Ts=a_0\varepsilon).
\eeq 
\label{note:eta}}:
\beq
\frac{\rmd \varepsilon}{\rmd \tau} =- \frac{a_0}{\tau}\left( \varepsilon -\frac{\eta}{\tau}\right).
\eeq
In fact, this equation can be also deduced from \Eqs{eq:Lequb} if one effectively takes 
$b_1^{\rm NS}=2\eta/(\varepsilon \tau_R)$ and lets $\tau_R \to 0$.

An equation similar to 
 Eq.~(\ref{eq:Lequb}) was introduced by Israel and Stewart (IS) \cite{Israel:1979wp} (see also \cite{muller1967paradoxon}) in order to cure  causality issues  of the relativistic Navier-Stokes equation. It takes into account a finite relaxation time $\tau_\pi$, 
over which the viscous pressure $\pi$  relax towards its Navier-Stokes value $4\eta/(3\tau)$. In the present context the corresponding equation is easily obtained by rewriting Eq.~(\ref{eq:Lequb}) as follows
 \beq
\del_\tau \pi+\frac{a_1^{\rm IS}}{\tau}\pi=-\frac{1}{\tau_\pi} \left( \pi-\frac{4\eta}{3\tau}  \right)
\eeq
where again the exression (\ref{viscosity}) of the viscosity has been used and related to $b_1^{\rm IS}=2\eta/(\varepsilon \tau_R)$, and we have the identification $\tau_R=\tau_\pi$. 
Note that $a_1^{\rm IS}$ is evaluated differently in variants of the IS hydrodynamics. For instance, in the simple IS formulation, it is often taken as $a_1^{\rm IS}=a_0=4/3$, while in Refs.~\cite{Denicol:2017lxn,Jaiswal_2019}, the coefficient is written as $(4/3+\lambda)$, and $\lambda=a_1^{\rm IS}-a_0=10/21$.   DNMR version of hydrodynamics \cite{Denicol:2012cn}  for Bjorken flow  is   identical to the two-moment equations (\ref{eq:Lequ}), namely, $a_1^{\rm IS} = a_1 = 38/21$.

The same analysis can be performed for BRSSS hydrodynamics \cite{Baier:2007ix}, a general approach based on conformal symmetry. As being presented in Appendix~\ref{sec:BRSSS}, the net result is that the equation for $\L_1$ has the same form as Eq.~(\ref{eq:Lequb}) with,
\beq\label{coefprime}
a_1^{\rm BRSSS}=a_0+\frac{2C_{\lambda_1}}{3C_\tau}, \qquad b_1^{\rm BRSSS}=2a_0\frac{C_\eta}{C_\tau},\qquad \tau_R=\tau_\pi=\frac{C_\tau}{T}.
\eeq
We have used here the notation of \cite{Baier:2007ix}, namely we have set 
\beq\label{confcoef}
\tau_\pi={C_\tau}/{T},\qquad \lambda_1=C_{\lambda_1} \,{\eta}/{T} ,\qquad \eta=C_\eta\, s,
\eeq
where the parameters $C_\eta$, $C_{\lambda_1}$ and $C_\tau$ depend on the underlying microscopic theory.  For the ${\cal N}=4$ super Yang-Mills theory, we have \cite{Baier:2007ix,Heller:2007qt,Heller:2015dha}
\beq\label{N4SYM}
C_\eta=\frac{1}{4\pi},\qquad C_\tau=\frac{2-\ln 2}{2\pi},\qquad C_{\lambda_1}=\frac{1}{2\pi}.
\eeq
Choosing the time scale so that $\tau_\pi=\tau_R$, one finds $a_1^{\rm BRSSS}\simeq 1.843$ and $b_1^{\rm BRSSS}\simeq 1.02$.  
A summary of the various versions of the second order visous hydrodynamics and their relations to the two-moment truncation via the constant coefficients $a_1$ and $b_1$ is given in \Tab{table:1}.

\begin{table}[]
\begin{tabular}{|c|c|c|c|}
\hline
            & $a_1$                                 & $b_1$                         & $\tau_\pi$            \\ \hline
Two moments/DNMR hydro& $38/21$                               & $8/15$                        & $\tau_R$            \\ \hline
Navier-Stoke Hydro    & undetermined                          & $\quad 2\eta/(\varepsilon \tau_R) \quad$  & $ 0^+$            \\ \hline
Isreal-Stewart Hydro    & $\quad a_0 \mbox{ or }a_0 + 21/10\quad$                                 & $2\eta/(\varepsilon \tau_R)=8/15$  & $\tau_\pi$          \\ \hline
BRSSS Hydro & $a_0+\dfrac{2C_{\lambda_1}}{3C_\tau}$ & $2a_0 \dfrac{C_\eta}{C_\tau}$ & $\dfrac{C_\tau}{T}$ \\ \hline
Kinetic-Hydro  & $31/15$ & $8/15$ & $\quad \tau_R \quad$ \\ \hline
\end{tabular}
\caption{Coefficients $a_1$, $b_1$ and $\tau_\pi$ in various versions of second order viscous hydrodynamics.}
\label{table:1}
\end{table}

We see therefore that in the late time hydrodynamic regime, second order viscous hydrodynamics reduce to the simple linear system (\ref{eq:Lequ}), with suitably adjusted  coefficients $a_1$ and $b_1$ (as well as $\tau_R$). There is however an important point that needs to be underlined here. The system (\ref{eq:Lequ})  originates from kinetic theory, which provides a definite description of the short time behavior:  when $\tau\ll \tau_R$,  the system is collisionless and particles are free streaming, irrespective of the microscopic dynamics. Even though within the two-moment truncation the description of the collisionless regime is only approximate, the physical motivation for the existence of this regime is well motivated. The time derivative of the viscous pressure, introduced via a relaxation equation by Israel and Steward, or emerging naturally in the BRSSS approach, confers to second order hydrodynamics a  mathematical structure similar to that of the two-moment truncation of the kinetic equations. If one takes hydrodynamics as an effective theory for long wavelength modes, the short time regime can be viewed as a so-called ``UV-completion'' \cite{Heller:2015dha} of this effective theory. Thus, what second order viscous hydrodynamics does is to provide an UV completion which  has the same structure as that given by the collisionless regime of kinetic theory. In simpler terms, what the present analysis suggests is that, in the present context of Bjorken flow, the various versions of second order hydrodynamics differ solely in how (quantitatively, i.e. through the values of the parameters $a_1$ ans $b_1$), they mimic the collisionless regime at short time. In trying to extend second order hydrodynamics towards regimes or larger gradient, i.e. here shorter times, one is implicitly using  information about this short time regime which has, a priori little to do with hydrodynamics (this is particularly clear in strong coupling  approaches based on holography, where the short time behavior is obtained by solving Einstein equations). In fact, as we shall show at the end of this paper, a simple adjustment of the parameter $a_1$ allows us to reproduce accurately the collisionless regime, and hence to reproduce within second order hydrodynamics (referred to as kinetic-hydrodynamics) essentially the entire evolution obtained from the full solution of the kinetic equation. 

 
We now turn to the explicit solution of the coupled system (\ref{eq:Lequ}) for $\L_0$ and $\L_1$. However, one should bear in mind that the following discussion applies to all the second order viscous hydrodynamics as well, with suitable substitutions of the constant coefficients list in \Tab{table:1}.

\subsection{Solving the linear system for $\L_0$ and $\L_1$}\label{sec:solvinglienar}

We shall treat the linear system (\ref{eq:Lequ}) using standard techniques of linear algebra \cite{Blaizot:2019scw}. We write  Eqs.~(\ref{eq:Lequ}) in a matrix form
\beq\label{FSb}
\tau \frac{\partial}{\partial \tau}
\left(
\begin{array}{c}
 {\cal L}_0 \\
  {\cal L}_1
\end{array}
\right)
= -M(\tau)\left(
\begin{array}{c}
 {\cal L}_0(\tau) \\
  {\cal L}_1(\tau)
\end{array}
\right),\qquad 
M(\tau)=
\left(
\begin{array}{cc}
 a_0 & c_0 \\
b_1 &  a_1 +\frac{\tau}{\tau_R} \end{array}
\right)
,
\eeq
and  consider the two moments $\L_0$ and $\L_1$ as the components of a two-dimensional vector $\ket{\L}$. Using a bra-ket notation, we rewrite Eqs.~(\ref{FSb}) as
\beq\label{eomL0L1M}
\tau\del_\tau \ket{\L(\tau)}=-M(\tau) \ket{\L(\tau)}.
\eeq
We also introduce the natural basis of the two-dimensional  vector space, with the two basis vectors  $(1,0)\to \bra{e_0}$, $(0,1)\to \bra{e_1}$, 
so that $\L_0(\tau)=\bra{e_0}\L(\tau) \rangle$ and $ \L_1(\tau)=\bra{e_1}\L(\tau) \rangle$.

 One may also write Eq.~(\ref{eomL0L1M}) in terms of the variable $w$ defined in Eq.~(\ref{wdef}),
 \beq\label{FSbw}
\Delta w \frac{\partial}{\partial w}
\left(
\begin{array}{c}
 {\cal L}_0 \\
  {\cal L}_1
\end{array}
\right)
= -M(w)\left(
\begin{array}{c}
 {\cal L}_0 \\
  {\cal L}_1
\end{array}
\right),\qquad 
M(w)=
\left(
\begin{array}{cc}
 a_0 & c_0 \\
b_1 &  a_1 +w\end{array}
\right)
,
\eeq
where we have used $\rmd \ln w=\Delta\rmd\tau$. Recall that $w=(\tau/\tau_1)^\Delta$, so that as long as $\Delta>0$ we can use $w$ as a measure of time. Note that $\Delta$ can be eliminated  by a simple rescaling:   $w\to \bar w\equiv w/\Delta$, $a_0\to\bar a_0\equiv a_0/\Delta$,  and similarly for the other matrix elements. Written in terms of the barred quantities, the explicit dependence on $\Delta$ disappears, and the solution is a function $\L(\bar w)$ which coincides formally with the solution of Eqs.~(\ref{FSbw}) for $\Delta=1$.  The solution for arbitrary $\Delta$ can then be obtained from $\L(\bar w)$  by undoing the scaling (whenever possible, i.e. whenever we have a sufficient analytic control of the solution). 

In the following, we shall write 
\beq\label{M0M1pert}   
M(w)=M_0 + M_1\,,\qquad
M_0=
\left(
\begin{array}{cc}
 a_0 & c_0 \\
b_1 &  a_1 \end{array}
\right), \qquad M_1=
\left(
\begin{array}{cc}
0 & 0 \\
0 & w \end{array}
\right),
\eeq
and consider first the collisionless regime with $M=M_0$. Then we shall  treat $M_1$ using time-dependent perturbation theory in order to analyze the departure from free streaming caused by collisions at short time, i.e., $w\ll1$. Next we look at the large time behavior and  the emergence of hydrodynamics.  We then consider an approximation akin to the adiabatic approximation of quantum mechanics, and which appears to be quite accurate when $0\lesssim\Delta\lesssim 1$. We end this section with a discussion of the particular case $\Delta=0$. Details on some of the calculations involved are given in Appendices~\ref{app:pert} and \ref{app:adiabatic}.

\subsubsection{The collisionless regime as a coupled mode problem}

The collisionless regime corresponds to the solution of Eqs.~(\ref{FSbw}) in the absence of the collision term. 
To proceed, we  expand the vector $\ket{\L}$ on the basis of the eigenvectors of $M_0$
\beq
\ket{\L(w)}=\sum_{n=1,2}C_n(w)\ket{\phi_n},\qquad M_0\ket{\phi_n}=\lambda_n\ket{\phi_n}.
\eeq
The solution of the equation of motion  is then easily obtained
\beq
\ket{\L(w)}=C_0\left( \frac{w_0}{w} \right)^{\lambda_0} \ket{\phi_0}+C_1\left( \frac{w_0}{w} \right)^{\lambda_1} \ket{\phi_1},
\eeq
where $C_0\equiv C_0(w_0)$ and $C_1\equiv C_1(w_0)$ are fixed by the initial condition\footnote{ Note that since the equations are linear, one may rescale the two moments by the energy density at some particular time $\tau_0$. The solution then depends on a single parameter, e.g. the initial pressure asymmetry $\L_1(\tau_0)/\L_0(\tau_0)$.}.
By projecting on $\bra{e_0}$ and $\bra{e_1}$, one gets the moments $\L_0$ and $\L_1$:
\beq\label{2modesfs}
\L_k(w)=C_0\langle e_k\ket{\phi_0}\left( \frac{w_0}{w} \right)^{\lambda_0}+C_1 \langle e_k\ket{\phi_1}\left( \frac{w_0}{w} \right)^{\lambda_1}, \qquad (k=1,2).
\eeq

The two moments $\L_0$ and $\L_1$ appear then as superpositions of the two eigenmodes $\ket{\phi_0}$ and $\ket{\phi_1}$, which are damped as $w$ increases. Since $\lambda_1>\lambda_0$, the mode $\ket{\phi_1}$ is damped faster than the mode $\ket{\phi_0}$. Consequently the latter dominates at late times, and the pressure asymmetry is given at large $w$ by \beq\label{L1overL0fs}
\frac{\L_1}{\L_0}\to \frac{\langle e_1\ket{\phi_0}}{\langle e_0\ket{\phi_0}}=-\frac{1}{c_0} (a_0-\lambda_0),
\eeq
independently of the initial condition (see Appendix~\ref{app:pert} for the values of $\bra{e_i}\phi_n\rangle$ used here).



The mixing of modes is responsible for the transient regime that one may observe for arbitrary initial conditions. Such a transient regime disappears indeed if the initial condition is chosen such that either $C_0$ or $C_1$ vanishes. If for instance, $C_1\simeq 0$, $\ket{\L(w_0)}\sim \ket{\phi_0}$, and a small initial admixture of the mode $\ket{\phi_1}$ will be damped, leaving eventually the system in the pure eigenmode $\ket{\phi_0}$ at late time. To see that, rewrite the first of Eqs.~(\ref{2modesfs}) as follows 
\beq
\L_0(w)=C_0\left( \frac{w_0}{w} \right)^{\lambda_0} \frac{c_0}{\lambda_0-a_0}\left\{  1+\frac{C_1}{C_0} \frac{\lambda_0-a_0}{\lambda_1-a_0}\left( \frac{w_0}{w} \right)^G \right\},
\eeq
where 
\beq
G\equiv \lambda_1-\lambda_0
\eeq
denotes the gap, i.e., the difference between the two eigenvalues.  Since $G>0$ the contribution of the term proportional to $C_1$ is indeed damped at late time  and, after a transient regime, $\L_0$ is  proportional to  $\langle e_0\ket{\phi_0}$.  In contrast to the case $C_1\simeq 0$, a small initial admixture of  $\ket{\phi_0}$ completely changes the evolution of the system, since only  $\ket{\phi_0}$ survives at late time. We shall see later that this behavior can also be understood from the existence of a stable and an unstable fixed points in the non linear equation that governs the evolution of the pressure asymmetry (see Sect.~\ref{sec:nonlinear}).

Thus, in the absence of collisions,  the moments evolve as powers laws, with exponents given by the eigenvalues of the matrix $M$. This provides an approximate description of the collisionless regime. Approximate because the change of the momentum distribution caused by the expansion requires many moments for its accurate description\footnote{Recall indeed that the modification of the momentum distribution caused by the free streaming of particles is accounted for by a simple rescaling of the longitudinal momentum, $p_z\to  p_z(\tau_0/\tau)$. Thus the surfaces of constant $p_0$ are ellipsoids $p_0^2=\p_\perp^2+p_z^2(\tau_0/\tau)^2$, which becomes flat oblate spheroids at large time. Note that in the opposite limit where $\tau\ll \tau_0$, the distribution becomes a prolate spheroid along the $p_z$ axis. These two shapes of the momentum distribution can be associated to the two eigenmodes of the linear system (\ref{FSb}), or equivalently, as we shall see later, to the corresponding fixed points in the non-linear equation for the pressure asymmetry.\label{note:ellispoid}}.  However, as we already emphasized, these higher moments do not play a major role in the late time dynamics of $\L_0$ and $\L_1$.\footnote{The complete eigenvalue problem is analyzed in detail in \cite{Blaizot:2019scw}. It is shown there that the two real eigenvalues of the two-moment truncation dominates the dynamics of the lowest moments and they are only moderately afffected as one includes more moments.} Thus, for the exact free streaming, the eigenvalues of the general (infinite dimensional) matrix $M$ (with all moments included) are respectively  1 and 2. In the two moment truncations these are instead 0.929 and 2.213.  These small deviations lead to some unphysical features, like possible excursions into regimes of small negative longitudinal pressure, as mentioned in the introduction, and discussed in \cite{Blaizot:2019scw}. Such unphysical features can be seen here as violations of the bound $\L_1/\L_0>-1/2$ (see after Eq.~(\ref{pressureasym})): here we have $\L_1/\L_0=-0.6060$. Note that in the  IS theory the violation is even more severe: with the choice often made $a_1^{\rm IS}=a_0$, the eigenvalues are respectively 0.737 and 1.93, leading  to $\L_1/\L_0=-0.895$. For BRSSS, with the values given above for $a_1^{\rm BRSSS}$ and $b_1^{\rm BRSSS}$ (see after Eq.~(\ref{N4SYM})), we get a similar value, $\L_1/\L_0=-0.912$. In the last part of this paper, we shall show how a simple adjustment of the parameter $a_1$ allows us to circumvent this difficulty. 

\subsubsection{Perturbation theory at small times}\label{tdpt}

When the collision rate is not too large, and for a small interval of time, one may treat the effect of the collisions using time-dependent perturbation theory. This is what we do now,  treating  $M_1$ as a perturbation. The perturbative correction depends on the initial state that is being perturbed.  We consider first the solution that coincides initially with the eigenmode $\ket{\phi_0}$\footnote{Recall that the mode $\ket{\phi_0}$ corresponds to the ratio of moments $\L_1/\L_0\simeq -0.6$, with $\L_0=\bra{e_0}\phi_0\rangle$ and $\L_1=\bra{e_1}\phi_0\rangle=1$. }, that is, we assume that $\ket{\L(w_0)}=\ket{\phi_0}$,
and we expand the solution on the eigenstates of the constant matrix $M_0$. We set
\beq
\ket{\L(w)}=C_0(w)(1+ a_{00}(w))\ket{\phi_0}+ a_{01}(w) C_1(w) \ket{\phi_1},
\eeq
where $a_{00}$ and $a_{01}$ are small numbers chosen so that 
$
a_{00}(w_0)=a_{01}(w_0)=0.$ The functions 
$ C_0(w)=\left( {w_0}/{w}  \right)^{\lambda_0}$ and  $C_1(w)=\left( {w_0}/{w}  \right)^{\lambda_1}$ encode the natural $w$-dependence of the eigenmodes (i.e. that induced by the free streaming).
A simple calculation  yields (see Appendix~\ref{app:pert})
\beq
a_{00}(w)= -\frac{a_0-\lambda_0}{G} (w-w_0),
\qquad
a_{01}(w)= \frac{a_0-\lambda_1}{G}\frac{w_0}{1+G}\left[ \left(\frac{w}{w_0}  \right)^{G+1}-1 \right].
\eeq
From this, one can deduce in particular the  pressure asymmetry
\beq\label{presasymphi0}
\frac{\L_1}{\L_0}=\frac{\lambda_0-a_0}{c_0}  \left\{1-  \frac{w}{1+G}\left[1-\left(\frac{w_0}{w}  \right)^{G+1} \right]   \right\}.
\eeq
The presence of the term $(w_0/w)^{G+1}$ limits a priori the validity of this result to not too small values of $w$. However, the limit $w_0\to 0$ is perfectly smooth and can thus be taken. This eliminates the ``dangerous'' term, leaving us with a well defined expansion down to $w=0$. We shall see later that the linear contribution in Eq.~(\ref{presasymphi0}) is in fact the leading term of a convergent expansion, which corresponds to the small $w$ expansion of the analytic attractor solution obtained in the next section (see Sect.~\ref{sec:analyticsol}).    

We can repeat the same analysis starting from the state $\ket{\L(w_0)}=\ket{\phi_1}$:
\beq
\ket{\L(w)}=C_0(w) a_{10}(w))\ket{\phi_0}+ (1+a_{11}(w)) C_1(w) \ket{\phi_1},
\eeq
where (see Appendix~\ref{app:pert})
\beq
a_{10}(w)= -\frac{a_0-\lambda_0}{G} \frac{w_0}{1-G}\left[ \left(\frac{w}{w_0}  \right)^{1-G}-1 \right],
\qquad
a_{11}(w)= \frac{a_0-\lambda_1}{G}(w-w_0).
\eeq
The pressure asymmetry now reads
\beq\label{presasymphi1}
\frac{\L_1}{\L_0}=\frac{\lambda_1-a_0}{c_0}  \left\{1-  \frac{w}{1-G}\left[ 1-\left(\frac{w}{w_0}  \right)^{G-1} \right]   \right\}.
\eeq
In contrast to Eq.~(\ref{presasymphi0}),  now the expansion remains valid all the way down to $w= 0$  for any  $w_0>0$. However the limit $w_0$ cannot be taken. Thus, the small $w$ expansion will contain non analytic contributions of the form $(w/w_0)^{G-1}$, that depend explicitly on the value $w_0$ where the  initial condition is fixed. This is the origin of  the trans-series representation of the small $w$ expansion around this particular solution (see Appendix~\ref{trans}).\\

This analysis reveals important features of the solution of the linear system. The behavior of the solution at small  $w$ is very sensitive to the initial condition. In one particular case, the initial condition can be set at  $w_0=0$. This is possible only if $\ket{\L(w_0)}=\ket{\phi_0}$ and it corresponds to the attractor solution. For all other solutions, the initial condition needs to be fixed at some finite $w_0$. However, all such solutions have a perfectly well defined limit as $w\to 0$. For the pressure asymmetry, this limit is $(\lambda_1-a_0)/c_0$, which, as we shall see in the next section, coincides with the unstable free streaming fixed point (see also the discussion in Sect.~\ref{sec:initialcond}).

\subsubsection{Viscous hydrodynamics at late times}

When the collision rate overcomes the expansion rate, which eventually occurs if $\Delta>0$,  hydrodynamics sets in. The way this occurs is easy to see: the off-diagonal matrix elements $c_0$ and $b_1$ that couple $\L_0$ and $\L_1$ become negligible as compared to the matrix element $M_{22}=a_1+\tau/\tau_R$ which increases linearly with $\tau/\tau_R$ , and the matrix $M$ becomes essentially diagonal. One expects then, ignoring the term $a_1$ as compared to $\tau/\tau_R$, and for constant $\tau_R$, 
\beq\label{idealhydro}
\L_0\sim \tau^{-a_0},\qquad \L_1\sim \rme^{-\tau/\tau_R}.
\eeq
That is, at late time $\tau\gg\tau_R$, the moment $\L_1$ is exponentially damped, and the energy density evolves according to ideal hydrodynamics.  
In fact, things are more subtle. Indeed, in the hydrodynamic regime, the exponential damping of $\L_1$ is hidden by a power law induced by the coupling $b_1$ of $\L_1$ to $\L_0$. That is, $\L_1$ is fed by its coupling to $\L_0$, which hinders its potential exponential damping.  A regime then emerges where an exact  cancellation takes place in Eq.~(\ref{eq:Lequb}) among the dominant terms at late time 
\beq\label{cancelL0L1}
\frac{\L_1}{\tau_R}\simeq - b_1\frac{\L_0}{\tau}, 
\eeq
so that eventually  
\beq\label{L1overL0andg}
\frac{\L_1}{\L_0}\simeq -\frac{b_1 }{w}+\cdots
\eeq
This regime is that in which deviations from ideal hydrodynamics are well accounted for by viscous corrections. 
 The right-hand side of Eq.~(\ref{L1overL0andg}) is the first term in the hydrodynamic gradient expansion of the pressure asymmetry.  As for Eq.~(\ref{cancelL0L1}), it can be seen as the leading constitutive equation for $\L_1$ relating it to the viscosity, $-\L_1=b_1 \varepsilon \tau_R/\tau=2\eta/\tau$. Thus, subsituting this relation (\ref{cancelL0L1}) into Eq.~(\ref{eq:Lequa}) yields the Navier Stokes equation (\ref{viscosity}) in the form
\beq
\frac{\del \L_0}{\del \tau}=-a_0 \frac{\L_0 }{\tau} -c_0 b_1 \tau_R \frac{\L_0 }{\tau^2},
\eeq
while $\L_1$ completely decouples. 

 Note that the cancellation  (\ref{cancelL0L1}) is  independent of the specific time dependence of $\tau_R$, that is, it occurs independently of the value of $\Delta>0$, as can be verified by considering the equation (\ref{FSbw}) in terms of $w$ (note that $b_1/w$ in Eq.~(\ref{L1overL0andg})  is invariant under the rescaling to the barred quantities discussed after Eq.~(\ref{FSbw})). A similar cancellation takes place in the equations for the higher moments, and it was used in \cite{Blaizot:2017ucy} in order to determine the leading behaviors of  high order transport coefficients. 
 
 Another perspective on this cancellation comes from the observation that the parameter that controls the coupling of  $\L_1$  to $\L_0$ is $b_1$, directly proportional to the viscosity (the parameter $c_0$ is fixed by energy-momentum conservation, as mentioned earlier). In the (formal) limit of vanishing viscosity, $\L_1$ completely decouples and is given by $\L_1\sim \tau^{-a_1} \rme^{-\tau/\tau_R}$, leading to ideal hydrodynamics when $\tau\gtrsim \tau_R$ without going through a viscous phase.
 

\subsubsection{The adiabatic approximation}\label{sec:adiapprox}

The writing of Eq.~(\ref{FSbw}) suggests that, when $0<\Delta\ll 1$, the derivative term plays a minor role.  This invites us to consider  an approximation akin  to the adiabatic approximation of quantum mechanics (see e.g. \cite{messiah1962quantum}), similar also to what is referred to  in other contexts as the slow roll approximation (see \cite{Heller:2015dha} and references therein). As we shall see in the next section, such an approximation turns out to be a good approximation for $0<\Delta \lesssim 1$, and within that range of physically relevant values of $\Delta$, it provides an accurate view of the entire $w$-dependence of the moments. There are  subtle issues involved in the implementation of this approximation in the present setting, which we shall return to in the next section. We should note in particular that, in contrast to the more familiar situation in quantum mechanics, here we are dealing with a non hermitian problem, in ``imaginary time'', and standard arguments based on phases and transition probabilities are not immediately applicable. Further details  are given in Appendix~\ref{app:adiabatic} (see also \cite{Brewer:2019oha}). We  consider here the adiabatic approximation in its leading order. 

We denote by $\ket{\phi_n(w)}$ the (right)  instantaneous eigenvectors of the (non hermitian) matrix $M(w)$, and by $\lambda_n(w)$ the corresponding eigenvalues:
\beq
M(w)\ket{\phi_n(w)}=\lambda_n(w)\ket{\phi_n(w)}, \qquad (n=0,1).
\eeq
The instantaneous eigenstates provide a convenient basis to expand $\ket{\L(w)}$. Writing $\ket{\L(w)}=\sum_{n=0,1}C_n(w) \ket{\phi_n(w)}$, one obtains  
\beq\label{expanLphi0phi1}
\L_0(w)=C_0(w) \bra{e_0}\phi_0(w)\rangle+C_1(w)\bra{e_0}\phi_1(w)\rangle,
\eeq
and similarly for $\L_1(w)$, with $\bra{e_0}$ substituted by $\bra{e_1}$. The instantaneous eigenvalues are given by 
\beq\label{lambda0r0}
\lambda_n(w)&=& \frac{1}{2}\left[a_0 + a_1 + w -(-1)^n \sqrt{(a_0-a_1-w)^2 +4 b_1 c_0 }\right],\qquad (n=0,1).
\eeq
As $w$ goes from 0 to $\infty$, $\lambda_0(w)$ evolves from its free streaming value to the hydrodynamical value $\lambda_0=a_0$. As for the  largest eigenvalue, $\lambda_1(w)$, it  evolves toward $+\infty$. More precisely, as $w\to\infty$, 
\beq
  \lambda_0(w)\simeq a_0-\frac{b_1c_0}{w},\quad  \lambda_1(w)\simeq w+a_1+\frac{b_1 c_0}{w}, \quad  G(w) \simeq w+ a_1 - a_0 .
\eeq 

A simple calculation presented in Appendix~\ref{app:adiabatic} shows that, in the leading order of he adiabatic approximation,   the eigenvalue $\lambda_0(w)$ captures accurately the $w$ behavior of the pressure asymmetry, with the  ratio $\L_1/\L_0$ being given by the generalization of Eq.~(\ref{L1overL0fs}):
  \beq\label{L1overL0fsr0}
\frac{\L_1}{\L_0}\to \frac{\langle e_1\ket{\phi_0(w)}}{\langle e_0\ket{\phi_0(w)}}=-\frac{1}{c_0} (a_0-\lambda_0(w)).
\eeq
In particular, when $w\to\infty$, ${\L_1}/{\L_0}\sim -{b_1}/{w}$, in agreement with Eq.~(\ref{L1overL0andg}). This is as expected since, as we have already observed,  this leading order is independent of the value of $\Delta$, so it holds for $\Delta\to 0$ where the adiabatic approximation  becomes exact. 
%


\subsubsection{The special case $\Delta=0$}\label{sec:constantDelta}

While one expects the adiabatic approximation to be exact in the limit $\Delta\to 0$, the  limit is singular and the case $\Delta=0$ requires a special treatment. In particular, when $\Delta=0$,  the relaxation time grows linearly with $\tau$ and $w=\tau/\tau_R$ is constant: one cannot use any longer $w$ as a measure of time. However, the time evolution can be deduced from Eq.~(\ref{FSb}). Starting the evolution at some time $\tau_0$ with some arbitrary initial condition, one observes a transient regime before the solution becomes ``stationary": it is then given by a power law whose exponent is the constant eigenvalue $\lambda_0(w)$, i.e., $\L_0(\tau)\sim\L_1(\tau)\sim (\tau_0/\tau)^{\lambda_0(w)}$, with the ratio $\L_1/\L_0$ given by Eq.~(\ref{L1overL0fsr0}). As is the case in the collisionless case, the transient regime occurs if initially the system is not exactly in the lowest eigenmode. An explicit example will be discussed in Sect.~\ref{sec:smallDelta} (see in particular Fig.~\ref{fig:stationarymode}).

In fact, in the  particular case $\Delta=0$, the full hierarchy of equations for the moments reduces to a linear problem with constant coefficients
\begin{align}
\label{eq:eomL2}
\frac{\partial \L_n}{\partial \tau} =& -\frac{1}{\tau}\left[a'_n\L_n
+b_n\L_{n-1}+c_n\L_{n+1}\right],\qquad a'_n=a_n+w(1-\delta_{n0}).
\end{align}
 It follows that the generalization of the matrix $M(w)$ that includes all moments is a simple tridiagonal constant matrix.  The pattern of eigenvalues is very similar to that found in the free streaming case \cite{Blaizot:2019scw}. There are two (and only two) real eigenvalues, accurately given by the two-moment truncation, the accuracy actually  increasing rapidly as $w$ increases. Besides these two real eigenvalues there are pairs of complex conjugate ones whose real parts are nearly equal and slightly less than $\lambda_1$. In the case of odd truncations, only the lowest real eigenvalue is present in the spectrum,  the second real eigenvalue  appearing only in even truncations. 
 
 By keeping $N$ moments ($N$ can be infinite), and expanding the moments $\L_n$ on the eigenstates of the corresponding generalized $N\times N$ matrix $M(w)$, we get
 \beq
 \L_k(\tau)=\sum_{n=0}^N C_n\left( \frac{\tau_0}{\tau} \right)^{\lambda_n(w)} \,\bra{e_k}\phi_n(w)\rangle,
 \eeq
 where $\lambda_n(w)$ is an eigenvalue of the matrix $M(w)$, $\ket{\phi_n(w)}$ the corresponding (constant) eigenvector, $\bra{e_k}$ a vector of the natural basis of the $N$-dimensional vector space spanned by the moments. The constants $C_n$ are determined by the initial conditions. 
 
 Clearly, the late time behavior is dominated by the mode associated to the lowest eigenvalue, i.e.  $\lambda_0(w)$, which remains separated from the next eigenvalue by a gap. The corresponding eigenmode plays the role of attractor, as in the two-moment truncation. Thus at late time all the moments have the same time dependence, like in free streaming, and their ratios to the lowest one are constants given by 
 \beq
 \frac{\L_k(\tau)}{\L_0(\tau)}=\frac{\bra{e_k}\phi_0(w)\rangle}{\bra{e_0}\phi_0(w)\rangle}=A_k(w),
 \eeq
 with the number $A_k(w)$ solution of the recursion relation
 \beq
 (a'_k-\lambda_0)A_k(w)+b_k A_{k-1}(w)+c_k A_{k+1}(w)=0.
 \eeq
 These numbers $A_k(w)$ characterize the distorsion of the momentum distribution as a function of $w$. 
 Since $b_0=0$, the recursion relation can be solved iteratively starting with $A_0=1$. 
 For  $w=0$ the  numbers $A_k(0)=P_{2n}(0)$ coincide with those characterizing the flat oblate distribution obtained in the late stage of the free streaming (see Eq.~(3.3) in \cite{Blaizot:2019scw}). The moments evolve then as $\L_k(\tau)\sim 1/\tau$.  At large $w$, the solution of the recursion relation above yields $A_{k\ne 0}\to 0$, characterizing an isotropic momentum distribution.  In this large $w$ limit, the dynamics is then entirely captured by $\L_0(\tau)$ which evolves in time as in ideal hydrodynamics, $\L_0(\tau)\sim 1/\tau^{4/3}$, with all other moments vanishing. In the intermediate regimes, the momentum distribution acquires a stationary spheroidal shape determined by the value of $w$, and isotropy is never reached.
 
 The physical interpretation of this regime is developed further in Sect.~\ref{typbehav}.

\section{The equation for the pressure anisotropy}\label{sec:nonlinear}

The seemingly simple system of equations (\ref{eq:Lequ}) hides a rather rich mathematical structure, which can be further unveiled by  rewriting this system as a single differential equation for the quantity 
\beq\label{gPL}
g(\tau) \equiv \frac{\tau}{\L_0}\frac{\partial \L_0}{\partial \tau} .
\eeq
Formally, $g$  may be understood as the exponent of  the power laws that govern the evolution of  the energy density at early or late times (where, in both cases, $g(\tau)$ becomes constant).  It also  generalizes, as we shall see,  the notion of instantaneous eigenvalue that we considered in the previous section. More physically, $g$ is related to the pressure asymmetry. To see that, we use Eq.~(\ref{eq:Lequa}) to obtain 
\beq\label{relg}
\frac{\P_L-\P_T}{\varepsilon}=\frac{\L_1}{\L_0}=-\frac{1}{c_0} (a_0+g)
\overset{w\to\infty}{=}-2  \frac{\pi}{\varepsilon+{\cal P}} ,
\eeq
where, in the last step, we have used the expression $-c_0\L_1=\pi$ of the viscous pressure~\cite{Blaizot:2019scw}, and $\P=\varepsilon/3$  is the local equilibrium pressure. The relations  (\ref{relg}) are independent of the specific time dependence of $\tau_R$. 

 
 The function $g(\tau)$ contains essentially the same information as the moments $\L_0$ and $\L_1$, and indeed the time dependence of both these moments can be reconstructed from $g(\tau)$, if desired. We shall therefore, in this section, be led to  revisit some of the results that we have obtained in the previous section. However, the equation obeyed by $g(\tau)$ allows us to get results that are not so easily obtained with the methods used in the previous section, and it provides much additional insight. In particular it can be solved analytically, which allows us to test precisely approximations that can be used more generally when no exact solution is available, for instance the adiabatic approximation, or perturbation theory. Furthermore, as a non linear equation,  it is  amenable to a  fixed point analysis \cite{Blaizot:2019scw} which provides a simple physical picture for the attractor solution.

\subsection{First order non linear ODE for $g(w)$ and fixed point analysis}

The equation for $g(\tau)$  is  obtained by transforming  \Eqs{eq:Lequ} into a first order nonlinear ODE, and  reads \cite{Blaizot:2019scw}
\beq\label{eqforbg0} 
\tau\frac{\rmd g}{\rmd \tau}+g^2+\left(a_0+a_1+\frac{\tau}{\tau_R}\right)g+a_1a_0-c_0b_1+a_0 \frac{\tau}{\tau_R} =0.
\eeq
It is valid for time-dependent $\tau_R$. It can also be written in terms of $w=\tau/\tau_R$ (see Eq.~(\ref{FSbw})) 
\beq\label{eqforbg0w} 
\Delta\frac{\rmd g}{\rmd \ln w}+g^2+\left(a_0+a_1+w \right)g+a_1a_0-c_0b_1+a_0 w =0.
\eeq
This equation\footnote{A similar equation holds in the conformal setting ($\tau_R T={\rm cste})$, with however a Jacobian (linearly dependent on $g$)  multiplying the derivative term~\cite{Blaizot:2019scw}.}  is valid as long as the mapping between $w$ and $\tau$ is well defined, that is as long as $\Delta >0$. As we did for the linear system (see after Eq.~(\ref{FSbw})), one can rescale variables so as to absorb the factor $\Delta$ into barred variables, together with redefining $  g= \Delta\,\bar g $. Keeping $\Delta$ explicit is however convenient, in particular when discussing the regimes of small and large $\Delta$, which we shall do in this section.


\subsubsection{Fixed point analysis}

To proceed with the fixed point analysis,  we write Eq.~(\ref{eqforbg0w}) as 
\beq\label{betafunct}
\Delta w\frac{\rmd g}{\rmd w}=\beta(g,w),\qquad  \beta(g,w)=-g^2-\left(a_0+a_1+w\right)g-a_1a_0+c_0b_1-a_0 w.
\eeq
It is easily verified that the zeros of the function $\beta(g,w)$ are, to within a sign, the instantaneous eigenvalues of the matrix $M(w)$ (the function $\beta(g,w)$ is in fact nothing but the characteristic polynomial of the matrix $M(w)$). When $w\to 0$, one recovers the collisionless regime and  the zeroes of the  function $\beta_0(g)=\beta(g,w=0)$ are true fixed points. As will be shown shortly, one is stable, $g_+=-\lambda_0\simeq -1$, the other is unstable, $g_-=-\lambda_1\simeq -2$. In fact, the free streaming solution of Eq.~(\ref{eqforbg0}) is easily obtained:  
\beq\label{freestreamg}
g(u)=\frac{A g_-+g_+ u^G}{A+u^G},\qquad A=\frac{C_1}{C_0} \frac{\lambda_0-a_0}{\lambda_1-a_0},\qquad G=g_+-g_-,
\eeq
where we have set $u=\tau/\tau_0$ and $A$, which fixes the initial condition at $u=1$,  is expressed in the right hand side using the notation of Sect.~\ref{sec:solvinglienar}.   Note that, irrespective of the value of $A$, the solution goes to $g_+$ at late time. The solution can also be evolved backward in time and converges then to the unstable fixed point \footnote{Note, however,  that if $A<0$ one may have to cross a pole to reach this limit $w\to 0$. This pole is associated to an unphysical situation where the energy density vanishes.\label{note:poleA}} $g_-$ as $u\to 0$  (see the discussion in  Sect.~\ref{sec:solvinglienar}).

  In the regime of large $w$, which is collision dominated, $\beta(g,w)\simeq (g+a_0)w$, and  another (attractive) fixed point emerges,  $g_*=-a_0$, corresponding to ideal hydrodynamics. Away from these limiting cases,  the values of $g$ for which $\beta(g,w)=0$ are $w$ dependent and are no longer true fixed points. They were referred to as ``pseudo fixed points'' in \cite{Blaizot:2019scw}. That these are useful to understand the solution is because their ``motion'' is slow on the time scale provided by the eigenvalues. Consider for instance the moment $\L_0$. After some transient, its time dependence (measured by $w$) is of the form $\L_0(w)\sim (w_0/w)^{\lambda_0}$. In a time interval $w$ of order one, $\lambda_0$ changes into $\lambda_0+\delta\lambda_0$ with $\delta\lambda_0=\rmd\lambda_0/\rmd w$.  The statement then is that $\delta\lambda_0\ll \lambda_0$, which indeed holds as soon as $w\gtrsim 1$, and it becomes increasingly accurate as $w$ increases (as one can see for instance by using the explicit expression of $\lambda_0(w)$ given in Eq.~(\ref{lambda0r0})). This observation is of course in line with the adiabatic approximation discussed in the previous section (see also Appendix~\ref{app:adiabatic}).\footnote{Eq.~(\ref{eqforbg0w} ) is written in a way that is well suited for a ``slow roll'' expansion, with the parameter $\Delta$ multiplying the derivative playing the role of expansion parameter.  The slow roll approximation in the present context has been analyzed in detail in the literature (see e.g. \cite{Denicol:2017lxn}, and references therein). Note that the expansion for small values of $\Delta$ is singular (See Appendix~\ref{app:adiabatic} for details). } In the leading order of this adiabatic approximation, the solution of the differential equation is simply given by the zero of the beta function (as a function of $w$), which, to within a sign, coincides with the instantaneous eigenvalues of the linear system (\ref{eq:Lequ}).  For the stable fixed point, we have 
    \begin{align}\label{FPgpm}
g_+(w)= -\lambda_0(w)=\frac{1}{2}\left[-a_0 - a_1 - w +  \sqrt{(a_0-a_1-w)^2 +4 b_1 c_0 }\right]\,.
\end{align}
This expression is an approximate solution of the differential equation (\ref{eqforbg0w}). It provides a simple picture of the continuous evolution from the collisionless to the collision dominated regimes, as captured by the adiabatic evolution of the pseudo fixed point $g_+(w)$ from $g_+=g_+(w=0)$ to $g_*$ at large $w$. All these pseudo fixed points are attractive, meaning that all solutions are  locally attracted to them as $w$ increases. In fact, $g_+(w)$ is an approximation to what we shall refer to in this paper as the attractor solution: the (unique) solution of Eq.~(\ref{eqforbg0w}), denoted $g_{\rm att}(w)$, that joins the free streaming fixed point $g_+$ at $w=0$ to the hydrodynamic fixed point $g_*$ as $w\to\infty$. Thus the adiabatic approximation states that $g_{\rm att}(w)\simeq g_+(w)$. The approximation is  exact when $\Delta=0$ and, as we shall see later in this section,  it remains an excellent approximation for $0\lesssim\Delta\lesssim 1$ (see e.g. Fig.~\ref{fig:AttSmallDelta} below). For further reference, we give  here the explicit expression of the unstable fixed point 
\beq\label{FPgm}
g_-(w) = -\lambda_1(w) = \frac{1}{2}\left[-a_0 - a_1 - w -  \sqrt{(a_0-a_1-w)^2 +4 b_1 c_0 }\right]\,.
\eeq

The fixed point analysis also helps to understand another limit, that of large $\Delta$, and fixed $w/\Delta$. In this regime,  the dominant terms in the beta function in Eq.~(\ref{betafunct})  are those  linear in $w$. The equation reduces then to
\beq
\bar w \frac{\rmd g}{\rmd \bar w}=-(g+a_0)\bar w,
\eeq
with the attractor (defined by $g(\bar w=0)=g_+$) given by
\beq\label{glargeD}
g(\bar w)+a_0=(g_++a_0)\rme^{-\bar w}.
\eeq
This result could have been anticipated from a simple analysis of the two mode problem in the previous section: as $\Delta\to\infty$, the equations for $\L_0$ and $\L_1$ decouple, and yield  $\L_0={\rm cste}$ and $\L_1\sim \rme^{-\bar w}$.
To better appreciate physically what this solution corresponds to, it is useful to express it in terms of the physical time $\tau$. Recalling that $w=(\tau/\tau_1)^\Delta$, one sees that in the limit $\Delta\to\infty$, the exponential contribution becomes a step function, and 
\beq\label{steplargeD}
g(\tau)=g_+ \theta(\tau_1-\tau)+g_* \theta(\tau-\tau_1).
\eeq
This approximation of large $\Delta$ is thus akin to the sudden (or diabatic) approximation of quantum mechanics. It represents a quasi-instantaneous transition from the free-streaming to hydrodynamics.\footnote{
Note that the value of $\tau_1$ remains of order $1$ as $\Delta\to\infty$. Indeed the transition region occurs when $\bar w\sim 1$, i.e.,  $w\sim \Delta$ at large $\Delta$. Since $w=(\tau/\tau_1)^\Delta$, one sees that the transition takes place at time $\tau/\tau_1\sim \Delta^{1/\Delta}\sim 1+\frac{1}{\Delta}\ln\Delta\simeq 1$.}
Such a rapid transition has been envisaged in some hydrodynamical studies, see e.g.  Ref.~\cite{Broniowski_2009}.

As a final observation here note that the fixed points $g_\pm$ and $g_*$ do not depend on $\Delta$. The parameter $\Delta$ only controls the shape of the attractor, i.e. how fast the transition between $g_+$ and $g_*$ occurs, not its end points $g_+$ and $g_*$. 

\subsubsection{Stability analysis}

The nature of the fixed points that we have identified above can be determined via a simple stability analysis\footnote{In the literature on dynamical systems, the behaviors near the various fixed points discussed here is analyzed in terms of ``forward'' and  ``pullback'' attractors  \cite{Behtash:2019txb,behtash2020transasymptotics}}. Consider first small fluctuations around $g_+$.  Setting $g(w)=g_++\delta g(w)$, and linearizing the $\beta$-function near $g_+$, we get 
\beq\label{deltagplus}
\delta g_+(w)\propto \left( \frac{w_0}{w}    \right)^{G/\Delta}.
\eeq
The fluctuation is damped as $w$ grows beyond $w_0$, confirming the stability of $g_+$ for increasing $w$. This stability becomes ``extreme'' as $\Delta\to 0$ (adiabatic regime), the damping being then essentially instantaneous. Note also that fluctuations are suppressed as $w_0\to 0$: this is connected with the fact there is only one solution such as $g(w=0)=g_+$, this is the attractor solution (see also the discussion in Sect.~\ref{tdpt}). \\   

The fluctuations around $g_-$ behaves as
\beq\label{deltagminus}
\delta g_-(w)\propto \left( \frac{w_0}{w}    \right)^{-G/\Delta}=\left( \frac{w}{w_0}    \right)^{G/\Delta}.
\eeq
This fluctuation grows as $w$ increases, justifying the qualification of unstable fixed point for $g_-$. Note however that, as we move backward in time, $g_-$ becomes attractive, and indeed we shall see later that all solutions (but the attractor) start as $g(w=0)=g_-$ (see also the comment after Eq.~(\ref{freestreamg})). Power laws such as (\ref{deltagminus}) also appears within perturbation theory (see Eq.~(\ref{presasymphi1})). They are regular as $w\to 0$, and are parts of the small $w$ expansion. However,  they are not analytic and this expansion is not a simple Taylor expansion, but rather a trans-series \cite{Behtash:2019txb} (see Appendix~\ref{trans}). \\

At late times  the system evolves to the hydrodynamic fixed point $g_*=-a_0$. When approaching this fixed point, $g(w)$ deviates slightly form $g_*$, the deviation being given by the hydrodynamic gradient expansion. The solution which includes the first gradient correction reads $g(w)=g_*+ \frac{b_1c_0}{w}$, and  a small deviation  $\delta g_*$ about this  solution  obeys  the equation
\beq\label{dampeddeltag0}
\Delta \frac{\rmd \delta g_*}{\rmd w}+(a_1-a_0)\frac{\delta g_*}{w}+\delta g_*=0,
\eeq
whose solution reads ($\bar w\equiv w/\Delta$)
\beq\label{dampeddeltag}
\delta g_*(w)\propto \rme^{-\bar w} \bar w^{\bar a_0-\bar a_1},
\eeq
where $\bar a_0=a_0/\Delta$, $\bar a_1=a_1/\Delta$. Thus the fluctuations about the hydrodynamic solution are, as expected,  exponentially damped by the collisions. The power laws that multiply the exponential factor involve the diagonal matrix elements of $M_0$ (in contrast for instance to the fluctuations near the stable fixed point (see Eq.~(\ref{deltagplus})), where the power law is given  rather by the gap $G=g_+-g_-$). In fact, the off-diagonal matrix elements $b_1$ and $c_0$ which play a dominant role in the gradient expansion do not play any role here (the terms $b_1 c_0$ cancel out in the linearization which yields (\ref{dampeddeltag0})). The form (\ref{dampeddeltag}) of the exponential correction plays an important role in the representation of the solution as a trans-series  (see Appendix~\ref{trans}). 

As shown by Eq.~(\ref{dampeddeltag}), the exponential damping occurs on a time scale of order $\Delta\tau_R$, which goes to zero as $\Delta\to 0$. This fast damping  also occurs  for the fluctuations around $g_+$ and $g_-$ (see Eqs.~(\ref{deltagplus}) and (\ref{deltagminus})). This may be seen as a hallmark of the adiabatic regime: when $\Delta\to 0$, the (pseudo) fixed point moves so slowly with increasing $w$ that the solution relaxes to it essentially instantaneously. \\

\subsection{Analytic solution for $g(w)$}\label{sec:analyticsol}

As was shown in \cite{Blaizot:2020gql}, an analytic solution of Eq.~(\ref{eqforbg0w}) exists  in terms of confluent geometric functions{\footnote{In \cite{Denicol:2017lxn}, a similar solution is given in terms of Wittaker functions. It can be easily verified that for the parameters appropriate to IS hydrodynamics, and constant relaxation time, the two solutions are identical owing to the relations between Wittaker functions and the confluent hypergeometrical functions (see Appendix~\ref{sec:analsol}).}. 
The general solution for $g(w)$ can be written in the form (see Appendix~\ref{sec:analsol} for details)
\begin{align}
\label{eq:gsol2}
g(w) = g_+ 
-w 
+ a w
\frac{\frac{1}{b}{M\left(1+a,1+b,\frac{w}{\Delta}\right)}- A U\left(1+a,1+b,\frac{w}{\Delta}\right)}
{M\left(a,b,\frac{w}{\Delta}\right)+ A U\left(a,b,\frac{w}{\Delta}\right)}\,,
\end{align}
where $A$ is a constant to be fixed by the initial condition.
The functions $M\left(a,b,\frac{w}{\Delta}\right)$ and $ U\left(a,b,\frac{w}{\Delta}\right)$ are confluent hypergeometric functions, 
 with parameters $a$ and $b$ given by (see Appendix~\ref{sec:analsol})
\be
a= 1 - \frac{g_- -g_* }{\Delta}\,,\qquad
b = 1+\frac{G}{\Delta}, \qquad G=g_+-g_-\,.
\ee
The attractor solution is   obtained for $A=0$. It is given by   
\be
\label{eq:expgatt}
g_{\rm att}(w)=g_+ - w+w
\frac{a M(1+a,1+b,w/\Delta)}{b M(a,b,w/\Delta)}\,.
\ee
 By using the expansions at small and large $w$ given in Appendix~\ref{sec:analsol}  one can verify that  $g_{\rm att}(w)$  indeed connects  $g_+$ at $w=0$ to   $g_*$ at large $w$. A plot of this attractor solution for various values of $\Delta$ is given in Fig.~\ref{fig:AttSmallDelta}. The curves displayed in this figure illustrate perfectly the definition that we are using for the attractor as the solution that joins two fixed points. Such a solution exhibits a logarithmic behavior in the vicinity of each fixed point and a transition region. 
\begin{figure}[h]
\begin{center} 
\includegraphics[angle=0,scale=0.62]{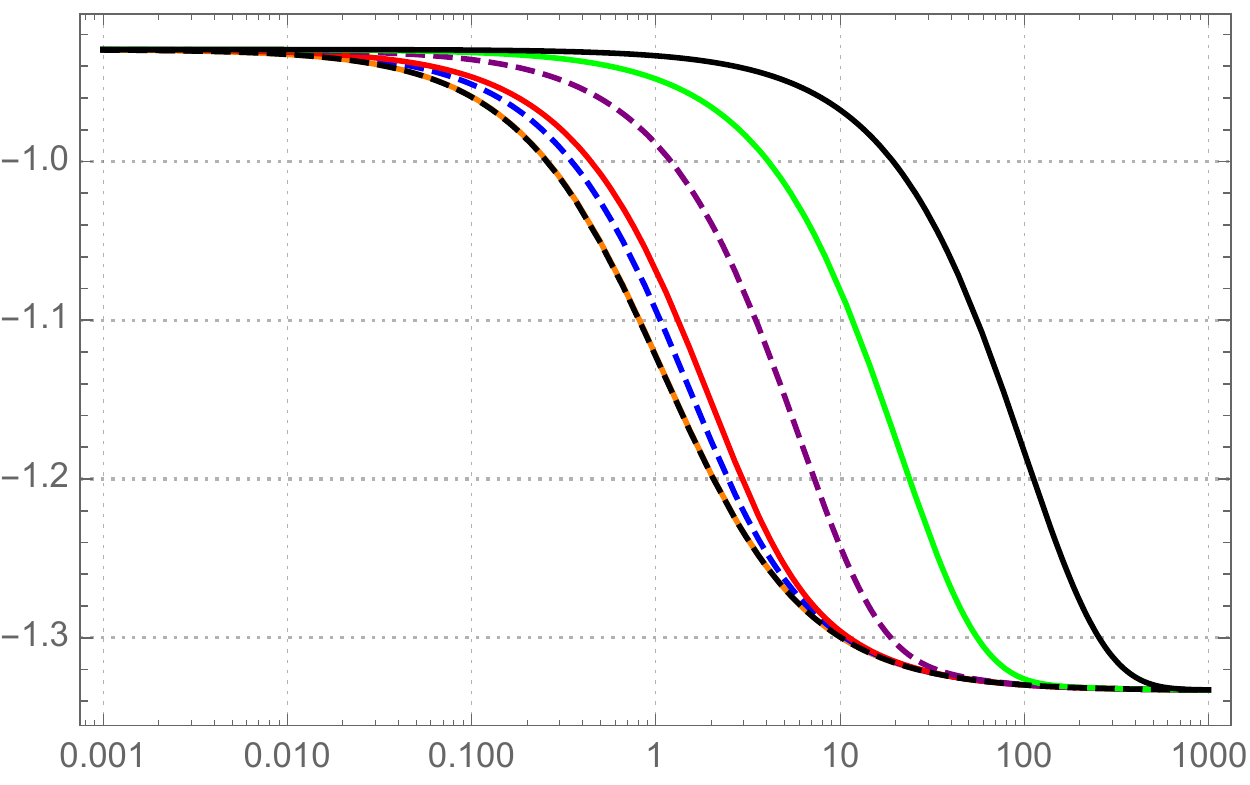} \includegraphics[angle=0,scale=0.62]{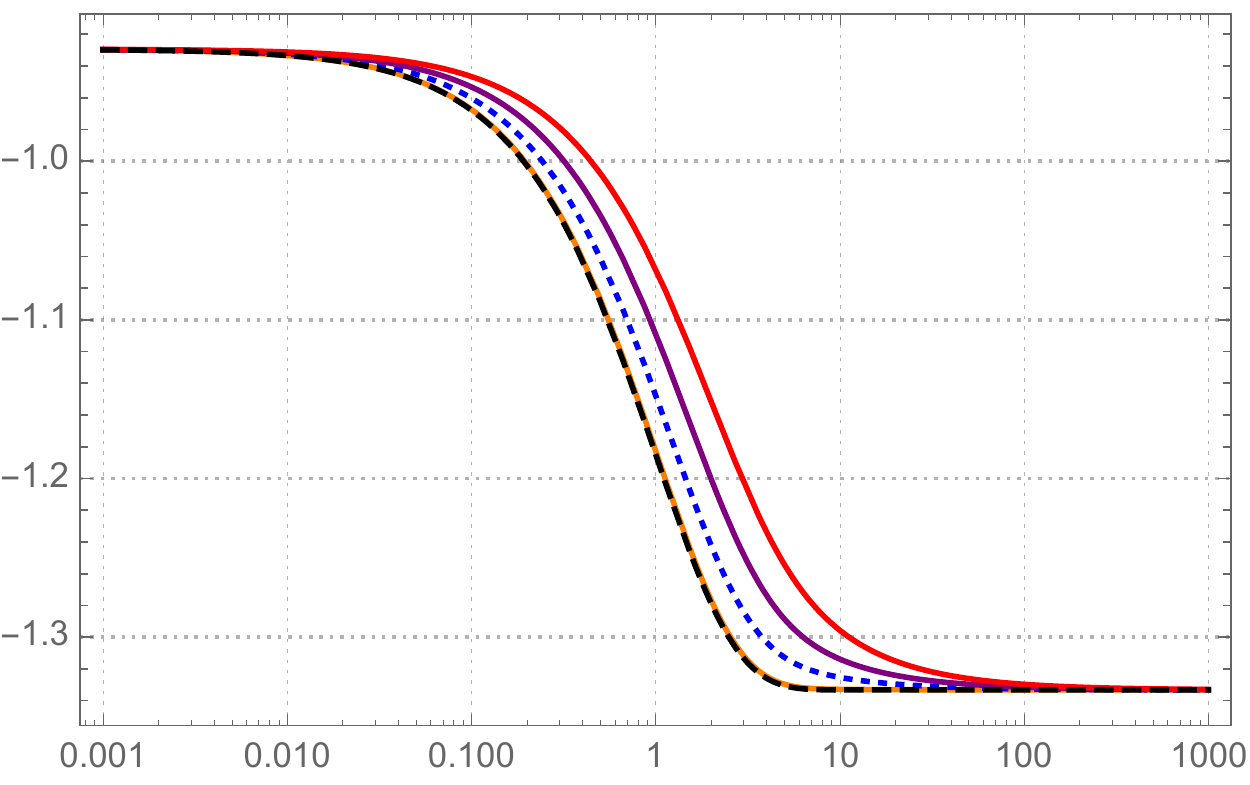} 
\end{center}
\caption{Color online. Left: The attractor solution as a function of $w$ for increasing values of $\Delta$: from left to right, $\Delta=0.01$ (orange),  0.5 (blue, dashed),  1(red), 5 (purple, dashed), 20 (green), 100 (black). The black dashed line represents the function $g_+(w)$ ($\Delta=0$), Eq.~(\ref{FPgpm}), and is indistinguishable from the orange curve. Right: The same as a function of $\bar w$ for large values of $\Delta$: from right to left,  $\Delta=1$ (red), 2 (purple), 5 (blue dotted),  100 (orange). The black dashed line is the exact large $\Delta$ limit in Eq.~(\ref{glargeD}) \label{fig:AttSmallDelta} and is indistinguishable from the orange curve corresponding to $\Delta=100$. }
\end{figure}

In the rest of this section we shall examine various properties of the explicit solution (\ref{eq:gsol2}), starting with the limiting cases of small and large values of $\Delta$.

\subsubsection{The limiting cases $\Delta\to 0$, and $\Delta\to\infty$}
 
We consider here the attractor solution and look first at the limit $\Delta\to 0$. In this case, the function $g_{\rm att}(w)$ converges to the function $g_+(w)$ of Eq.~(\ref{FPgpm}), as we have seen in the previous subsection (see also Appendix~\ref{sec:expansions}). This is illustrated in the left panel of Fig.~\ref{fig:AttSmallDelta}.

  As the left panel of  Fig.~\ref{fig:AttSmallDelta} also reveals, as $\Delta$ increases, the solution develops a travelling wave structure, with the location of the transition region scaling proportionally to $\Delta$ (for large $\Delta$, the solution remains approximately constant until $w\sim \Delta$). In the limit $\Delta\to\infty$  with $w/\Delta$ fixed, as shown in Appendix~\ref{sec:expansions}, the attractor becomes a function of $\bar w$, that is the entire $\Delta $ dependence is in the scaling of $w$. One recovers the result already given in Eq.~(\ref{glargeD}). This behavior is illustrated in the right panel of Fig.~\ref{fig:AttSmallDelta}. 
 
The hydrodynamic limit corresponds to the large $w$ limit at fixed $\Delta$. In this limit, we recover the leading terms of the gradient expansion (see Eq.~(\ref{hydroexp}))
\beq\label{attw2D}
g_{\rm att}(w)\simeq-a_0+\frac{b_1 c_0}{w}+\frac{b_1c_0(a_0-a_1)}{w^2}+\frac{ \Delta b_1 c_0}{w^2}+\cdots
\eeq
As already mentioned, the leading term ($\sim w^{-1}$) is independent of $\Delta$. At next to leading order ($\sim w^{-2}$), one finds a term that is independent of $\Delta$ and which agrees with the corresponding term in the expansion of the adiabatic solution $g_+(w)$. There is in addition a $\Delta$ dependent term which signals a departure from the adiabatic result.  
  
\subsubsection{Small $w$ behavior}

Staying with the attractor solution we revisit now its small $w$ expansion, and complete the discussion of the previous section (see Sect.~\ref{tdpt}) with remarks on the $\Delta$ dependence. 
In leading order, we have
\beq\label{gsmallw}
g_{\rm att}(w)\simeq g_++\frac{1}{2}\frac{a_1-a_0-G}{\Delta+G} \,w.
\eeq
This result of perturbation theory (see Eq.~(\ref{presasymphi0}) after taking the limit $w_0\to 0$) can be easily verified from the exact solution (see Appendix~\ref{sec:expansions}). In the denominator, the factor $\Delta$ next to gap $G=g_+-g_-$ represents a correction to the adiabatic approximation. The  small $w$ expansion of $g_+(w)$ in Eq.~(\ref{FPgpm}) yields indeed the same linear term as Eq.~(\ref{gsmallw}) with $\Delta=0$. Incidentally, we observe that for large $\Delta$ the expression (\ref{gsmallw}) becomes a linear function of $\bar w$ which agrees with the leading order of the expansion of $g_{\rm att}(\bar w)$ in Eq.~(\ref{glargeD}).

Going beyond this leading order result, it is not difficult to show that perturbation theory yields, for the attractor,  a convergent series in $w$. This series has a finite radius of convergence, which in the case of vanishing $\Delta$  is simply the gap $G=g_+-g_- \simeq 1$ (the same as that of the small $w$ expansion of $g_+(w)$ in Eq.~(\ref{FPgpm})). For general $\Delta$ the convergence of the small $w$ expansion of the attractor solution (\ref{eq:expgatt}) follows from the analyticity of the function $M(a,b,w/\Delta)$, the convergence being limited by the (complex) zeroes in the denominator.
\begin{figure}[h]
\begin{center}
\includegraphics[angle=0,scale=0.62]{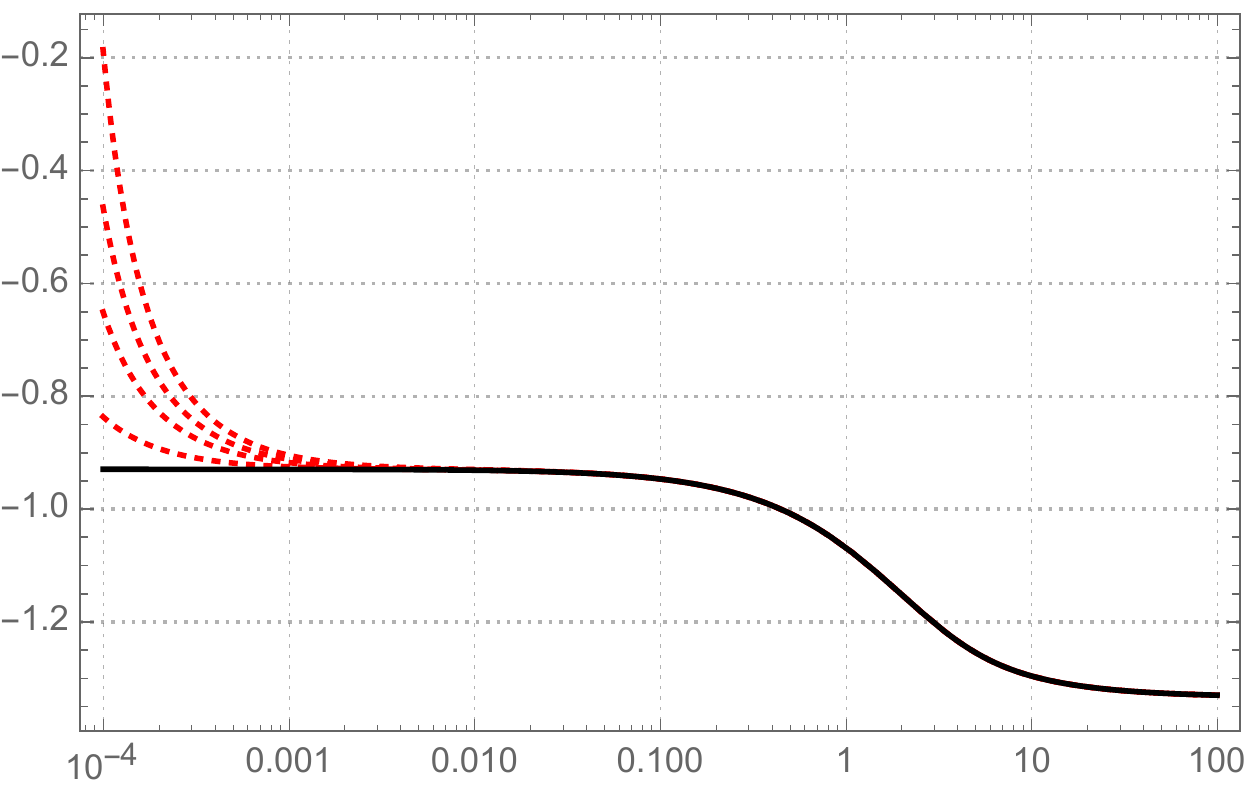} \includegraphics[angle=0,scale=0.62]{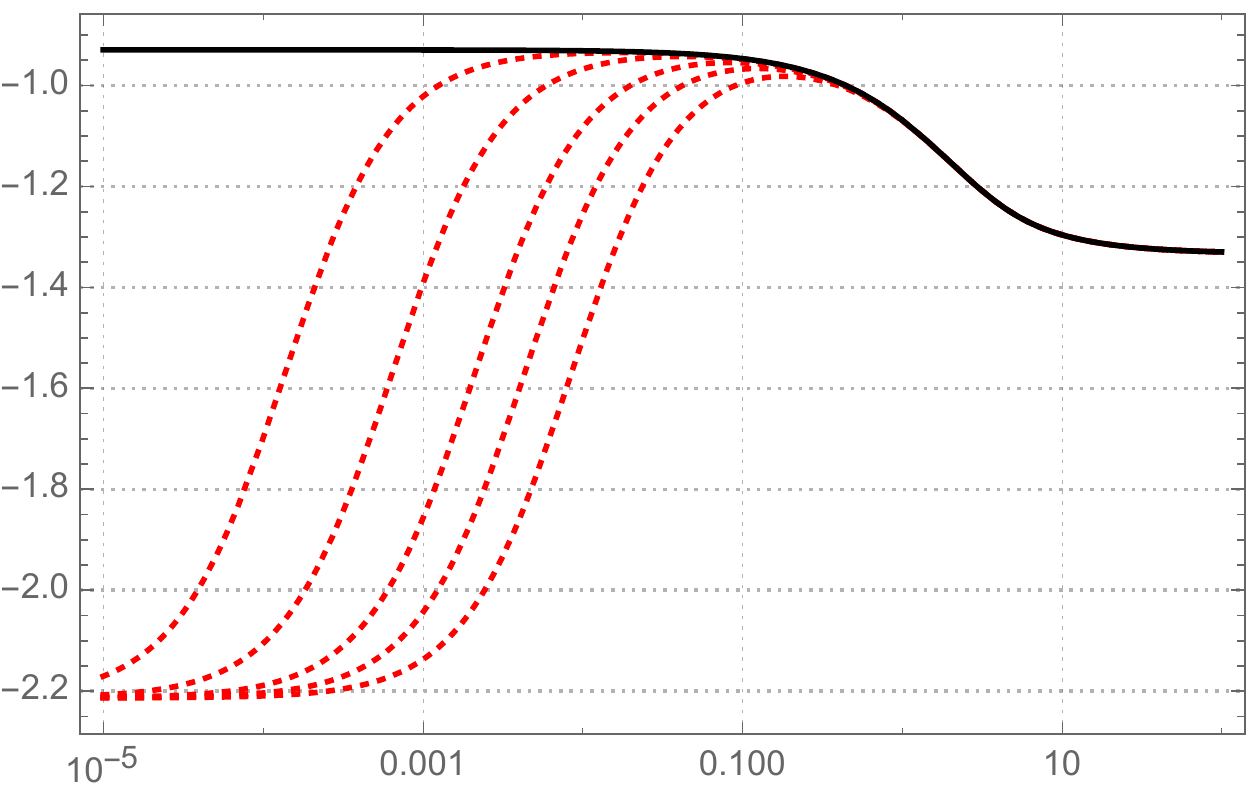} 
\end{center}
\caption{Solutions for various initial conditions (dotted lines). The full black line represents the attractor solution joining  $g_+$ at $w=0$ to  $g_*$ at large $w$. Left: the parameter $A<0$ is adjusted so that, at $w_0=10^{-4}$, $g(w_0) >g_+$. As $A<0$ the solution exhibits a pole located at a value of $w<w_0$. Right: the initial conditions are set at $w_0=10^{-3}$ such that $g_-<g(w_0)<g_+$. In this case, $A>0$ and the solution can be continued to $w=0$ where it reaches the unstable fixed point $g_-$. When the initial condition is chosen such that $g(w_0)<g_-$, the solution blows up (has a pole) at some value of $w>w_0$ but eventually reaches $g_*$ at large $w$.  \label{fig:initialcond} }
\end{figure}

\subsubsection{Remarks on the initial conditions}\label{sec:initialcond}

We turn now to the general solution with $A\ne 0$. As shown in Appendix~\ref{trans}, such solutions, when extrapolated backward to $w=0$, converge to the free streaming fixed point $g_-$, i.e.,  $g(w)\to g_-$ as $w\to 0$. However, since $g_-$ is a repulsive fixed point, the solution quickly deviates from $g_-$ as $w$ increases. It is then possible to adjust $A$ so as to satisfy any initial condition at a finite $w_0$. Note however that, in some cases, the solution may not be smooth  all the way from $w_0$ down to $w=0$. Indeed it may happen that the value of $A$ required to satisfy the initial condition at $w_0$ is negative, in which case  the denominator in Eq.~(\ref{eq:gsol2}) may vanish (see also footnote~\ref{note:poleA}). Typical solutions are illustrated in Fig.~\ref{fig:initialcond} for the two cases, $g(w_0)>g_+$ ($A<0$) and $g_-<g(w_0)<g_+$ ($A>0$).
\begin{figure}[h]
\begin{center}
\includegraphics[angle=0,scale=0.6]{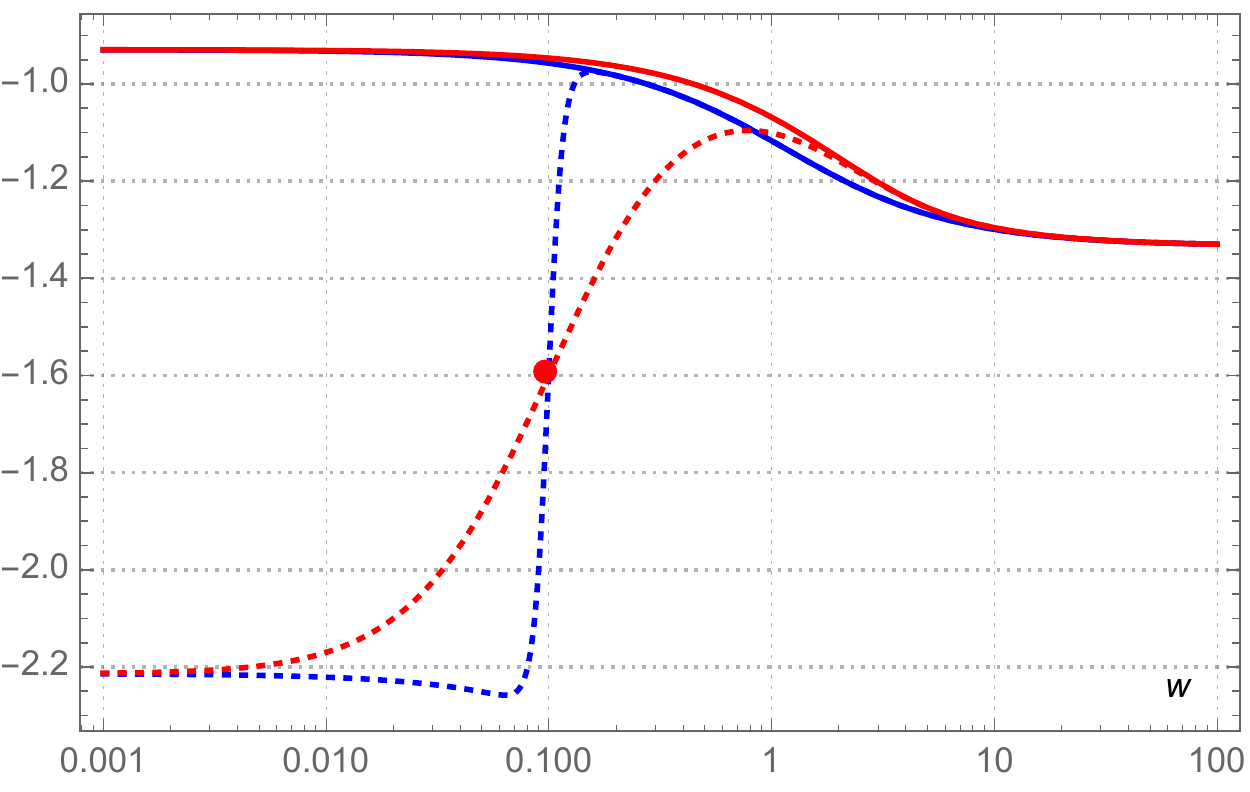} 
\end{center}
\caption{Typical behavior of a general solution extrapolated backwards to $w=0$. The initial condition is fixed at $w_0=0.1$ with $g(w_0)=-1.6$ (the initial condition is indicated by the  red dot). The two sets of curves correspond to $\Delta=1$ (red) and $\Delta=0.1$ (blue). The solid lines are the attractors for the corresponding values of $\Delta$, and the dotted lines are the actual solutions.  When $w$ increases from $w_0$, the solution quickly reaches its attractor. As one moves backward, to smaller and smaller values of $w$, the solutions converge to $g_-$. From small $\Delta$, the transition is very rapid, and the solution evolves quickly to $g_-(w)$ for $w<0.1$ and $g_+(w)$ for $w>0.1$. \label{fig:initialcondD} }
\end{figure}

The plots in Fig.~\ref{fig:initialcond} correspond to $\Delta=1$. As $\Delta\to 0$, one observes that the transition towards the (adiabatic) attractor $g_+(w)$ occurs very rapidly beyond the point $w_0$ where the initial condition is set, as illustrated in Fig.~\ref{fig:initialcondD}. This is in line with the remarks made earlier concerning the fast relaxation of fluctuations around the attractor in the adiabatic regime. In this regime, there is a rapid loss of the memory of the initial conditions. 

%
%
%
\subsubsection{Large $w$ behavior}\label{sec:largewex}

At large $w$ the exact solution can be represented by a trans-series (see Appendix~\ref{trans}): 
\beq
\label{eq:aexp}
g(w)= \sum_{m=0} (\sigma \zeta(w))^m \sum_{n=0} f_n^{(m)} w^{-n} 
\eeq
where $\sigma$ is a complex number (see Eq.~(\ref{sigmaA})), and 
\be\label{expocorr}
\zeta(w) = e^{-w}\,w^{b-2a+1}=e^{-w}\,w^{a_1-a_0},
\ee  
which we recognize as the fluctuation around the hydrodynamic solution (see Eq.~(\ref{dampeddeltag})). The presence of such exponential corrections is familiar in this context (see e.g. \cite{Basar:2015ava}).      
The leading term in the trans-series ($m=0$) is the  hydrodynamic gradient expansion, 
\be
g_{\rm hydro}= \sum_{n=0} f_n^{(0)} w^{-n},   
\ee
and is independent of the initial condition: the parameter $A$ which determines the initial condition is hidden in the real part of $\sigma$ and the hydrodynamic gradient expansion does not depend on $\sigma$. The first two orders of the gradient expansion are recalled in Eq.~(\ref{attw2D}) (see also Eq.~(\ref{coeff0})).

\begin{figure}[h]
\begin{center}
\includegraphics[angle=0,scale=0.5]{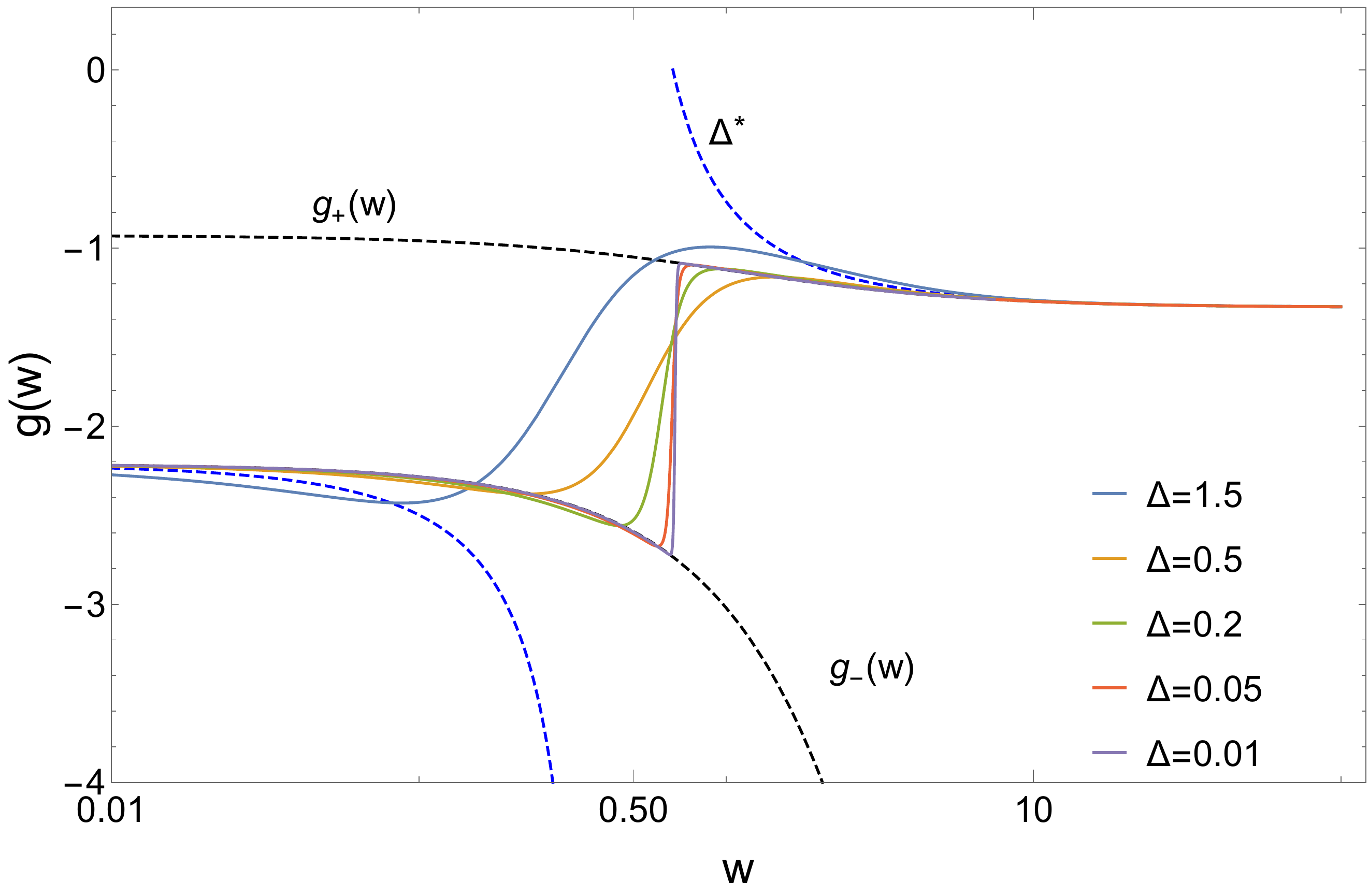} 
\end{center}
\caption{The continuation of the hydrodynamic gradient expansion via its Borel sum. The black dashed lines are the function $g_+(w)$ and the function $g_-(w)$. As $\Delta\to 0$, the Borel sum matches the function $g_+(w)$ at large $w$ and the function $g_-(w)$ at small $w$, and jumps from one to the other for $w\simeq 0.67$. Note that the jump from $g_+(w)$ to $g_-(w)$ turns into a singularity for $\Delta^*\approx0.873$, which corresponds to $a=2$, a value for which the function $\tilde{\cal F}$ in the Borel sum of  Eq.~(\ref{eq:ghydro}) is singular. \label{fig:gradexpBS} }
\end{figure}

The hydrodynamic gradient expansion can be resummed by the standard Borel summation technique. Here the Borel sum is known analytically \cite{Blaizot:2020gql} (see Appendix~\ref{sec:translargew}), which allows us to explore its property to arbitrarily small $w$.  This Borel sum has an interesting behavior illustrated in Fig.~\ref{fig:gradexpBS}. At large $w$, its real part follows the $g_+(w)$, then gradually deviates from it 
towards 
$g_-(w)$ when $w$ decreases. 
This feature indicates that hydrodynamic gradient expansion ``knows'' about the early time dynamics, and in particular about the two free streaming fixed points. 

It is interesting to analyze how this Borel sum evolves as a function of $\Delta$. It is easily verified that the branch point singularity of the Borel transform,  which is located at $w=1$ for $\Delta=1$, moves as  $w=\Delta^{-1}$ for arbitrary $\Delta$. Thus, when $\Delta\to0$, the branch cut on the real axis of the Borel transform is pushed to infinity. In this limit the trans-series collapses. As we already observed, when $\Delta\to 0$, a typical solution is represented at large $w$ by $g_+(w)$ and at small $w$ by $g_-(w)$, with a jump at an intermediate value of $w$  that depends on the initial condition. Such features are clearly illustrated in Fig~\ref{fig:gradexpBS}, although here the location of the jump does not depend on any initial condition.  

By taking into account the exponential corrections (\ref{expocorr}), one can reconstruct arbitrary solutions from the trans-series (\ref{eq:aexp}), the various subseries being given exactly by their corresponding Borel sums \cite{Blaizot:2020gql}. By choosing appropriately the value of the real part of $\sigma$ ($A=0$) one can in particular reconstruct the attractor, and verify the resurgence relations between the coefficients of the various subseries, as discussed in \cite{Blaizot:2020gql}.  

%

  As $w$ decreases, the exponential corrections become comparable to higher contribution of the gradient expansion. One can then rearrange the trans-series as discussed in Appendix~\ref{sec:translargew}:
  \beq\label{tranasymb}
\chi(w)=\sum_k  \frac{1}{w^k} F_k(\sigma\zeta) ,\qquad F_k(\sigma\zeta)=\sum_{m=0}^\infty \sigma^m \zeta^m f^{(m)}_k.
\eeq
The first two terms in the expansion in powers of $1/w$ are   
\beq\label{effectiveeta2}
\chi(w)= \sigma \zeta f_0^{(1)}+\frac{1}{w} \left( f_1^{(0)}+\sigma \zeta  f_1^{(1)}+ \sigma^2 \zeta^2 f_1^{(2)}\right)+\cdots
\eeq
Folowing \cite{Behtash_2019a}, we may interpret the coefficient of the $1/w$ term as an effective viscosity. Normalizing to the leading order viscosity, and taking the real part, we get
\beq\label{effectiveeta2a}
\frac{\eta_{\rm eff}}{\eta}=1-2\sigma_R \zeta(w)+\frac{1}{c_0 b_1}(\sigma_R^2-\sigma_I^2)\zeta(w)^2,
\eeq
where $\eta$ is explicitly given in footnote~\ref{note:eta}. Note that since the coefficients in Eq.~(\ref{effectiveeta2}) depend on $\sigma$, the effective viscosity (\ref{effectiveeta2a}) depends on the initial condition. Denoting by $\sigma_R$ the real part of $\sigma$ for $A=0$, one can appreciate the effect of changing the initial condition by  correcting $\sigma_R$ by a small amount, keeping the imaginary part $\sigma_I$ constant. The result is illustrated in Fig.~\ref{fig:effectivevisc2}. One observes a sizeable reduction of the effective viscosity in the vicinity of $w=1$. Note that this picture of effective viscosity makes sense only for not too small values of $w$, as the convergence of the expansion in powers of $\sigma\zeta$ appears to be poor. 
\begin{figure}[h]
\begin{center}
\includegraphics[angle=0,scale=0.6]{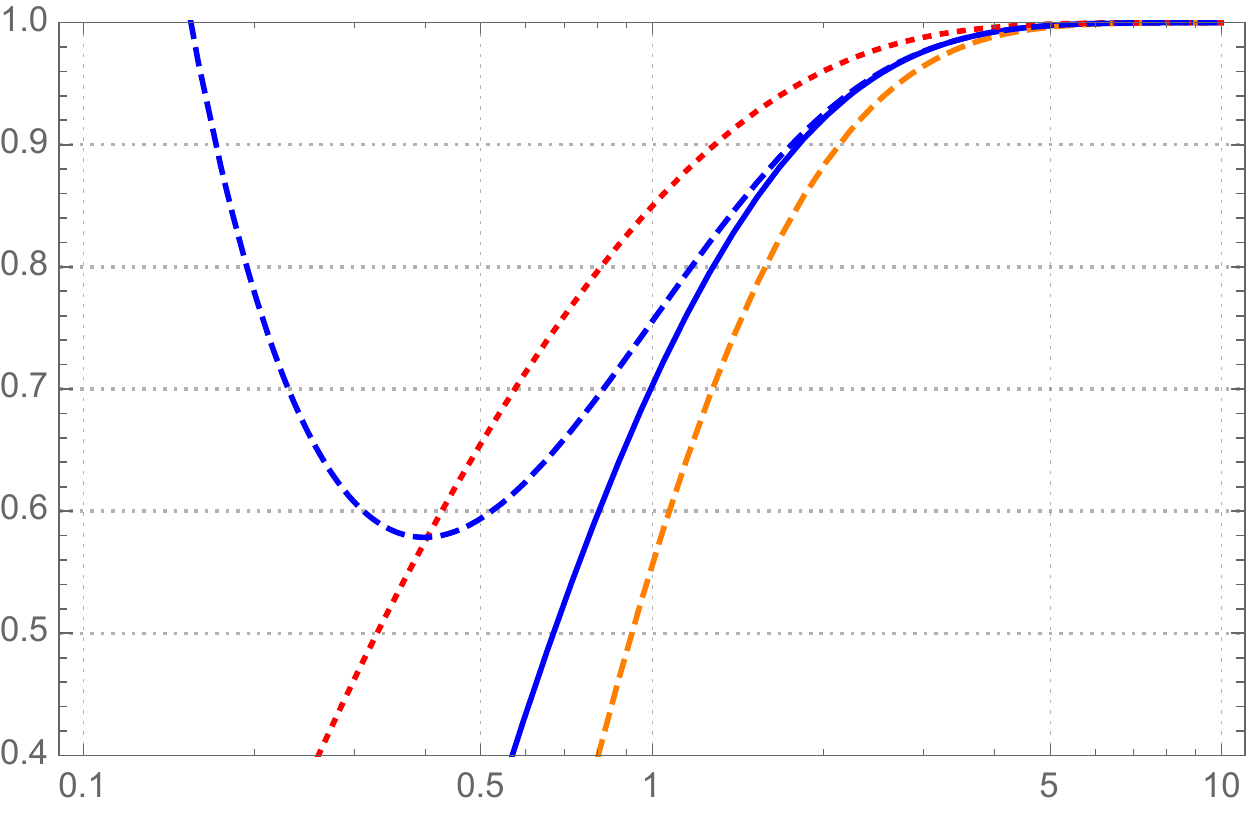}\includegraphics[angle=0,scale=0.6]{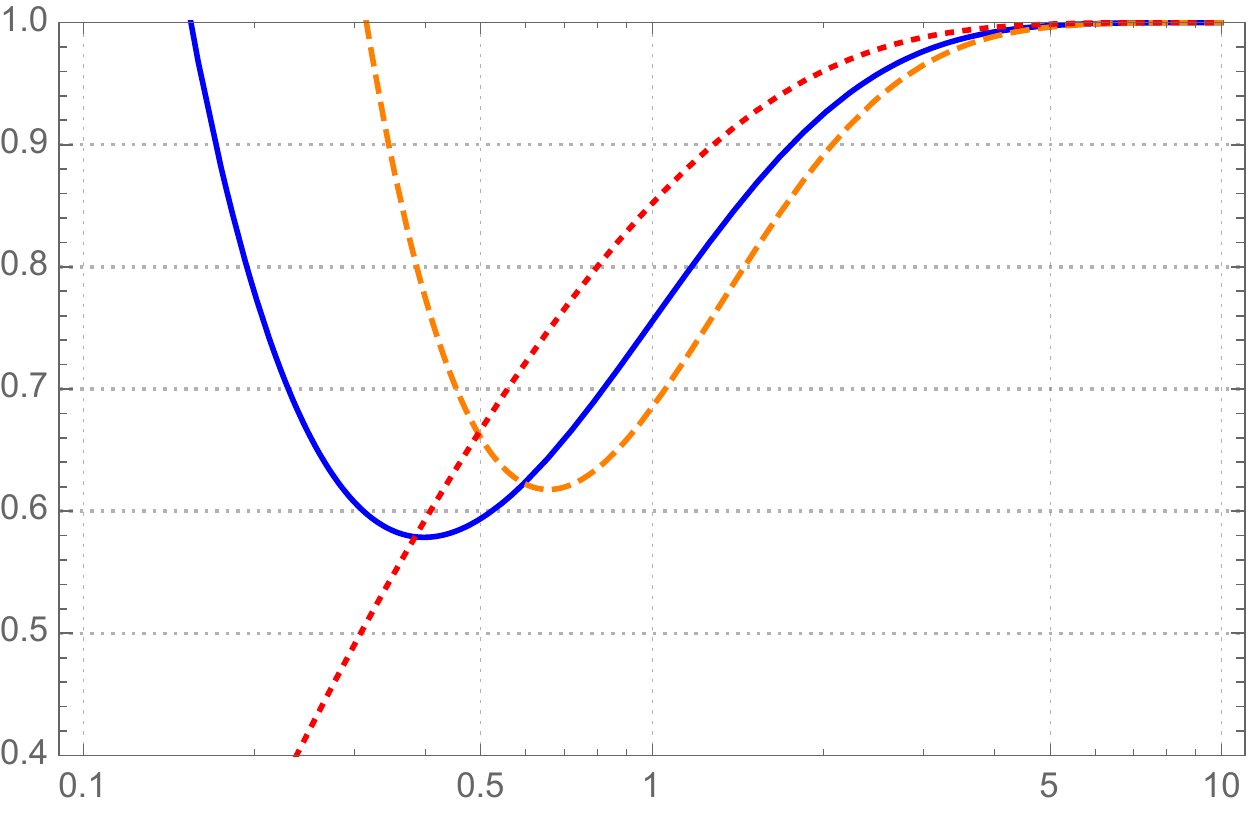} 
\end{center}
\caption{The effective viscosity normalized to the leading order one (Eq.~(\ref{effectiveeta2a})). The blue curve corresponds to $\sigma_R$, the attractor value, the dashed (orange) and dotted (red) curves correspond respectively to $\sigma_R+0.2$ and $\sigma_R-0.2$. The left panel corresponds to the first order contribution in $\zeta(w)$. The dashed blue line there includes the contribution in $\zeta^2$. Right panel, same for the full contribution, including the $\zeta^2$ contribution. \label{fig:effectivevisc2} }
\end{figure}

The renormalization of the viscosity discussed here bears some similarity with that introduced in \cite{Blaizot:2017ucy} (see also \cite{Lublinsky:2007mm}). The corrections have however different origins. In \cite{Blaizot:2017ucy}, the effective viscosity was introduced to account for the effects of the moments $\L_n$ that are left out in the two-moment truncation. It involves higher orders in the gradient expansion, and is somewhat similar to that discussed in \cite{Romatschke:2017vte}. Here, we are dealing with exponential corrections to leading order of the hydrodynamic gradient expansion. Both corrections act in the same direction: they tend to tame the growth of $g(w)$ predicted by the gradient expansion as $w$ decreases, that is, as one approaches the free streaming fixed point. That different dynamical effects can contribute to reduce the effective viscosity points to an ambiguity that should be kept in mind when discussing the viscosity extracted from heavy ion data.

\section{Additional physics remarks}\label{typbehav}

The previous two sections were mostly concerned with the mathematical properties of the approximate solution of the kinetic equation via a two moment truncation, taking two different perspectives, with each one bringing its own insights: solving a coupled set of linear equations in Sect.~\ref{sec:coupledmodes} or a single non linear differential equation in Sect.~\ref{sec:nonlinear}. In this section we shall rather focus on more physical considerations, and  also comment on some recent calculations in the literature. In the last part of this section, we shall show that a simple renormalization of a second order transport coefficient (involving the coefficient $a_1$) allows us to bring the two-moment truncation in good agreement with the exact solution of the kinetic equation.

So far the discussion has been carried out mostly in terms of the variable $w$. In this section we shall discuss the solution of the two-mode problem in terms of the physical time. We use Eq.~(\ref{eqforbg0})
that we rewrite here for convenience
\beq\label{eqforbg02} 
\tau\frac{\rmd g}{\rmd \tau}+g^2+\left(a_0+a_1+\frac{\tau}{\tau_R}\right)g+a_1a_0-c_0b_1+a_0 \frac{\tau}{\tau_R} =0.
\eeq
We measure time in units of $\tau_1$ and set $u\equiv \tau/\tau_1$, so that $w=u^\Delta$. Recall that $\tau_1$ is defined such that $w=1$ for $\tau=\tau_1$, that is, $\tau_1$ marks the time of the transition to hydrodynamics. In the case $\Delta=0$, $\tau_1$ is undefined, and we shall set $u=\tau/\tau_0$, with $\tau_0$ the initial time, i.e., the time at which the evolution starts.


\subsection{Generic behaviors as a function of $\Delta$}

\begin{figure}[h]
\begin{center}
\includegraphics[angle=0,scale=0.85]{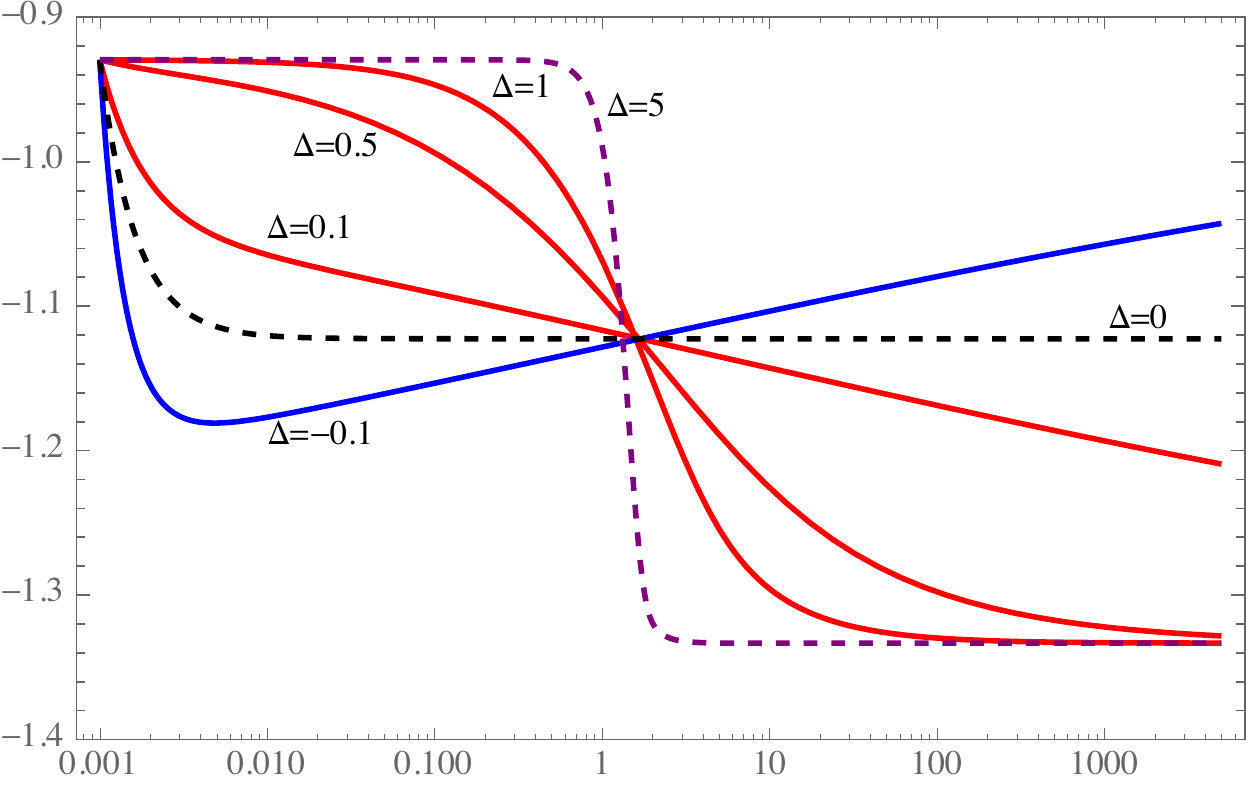} 
\end{center}
\caption{The function $g(u)$ as a function of $u=\tau/\tau_1$ for $\Delta=5,1,0.5,0.1,0,-0.1$. The initial condition corresponds to $g=-0.929$ (the free streaming fixed point) at $u_0=10^{-3}$.  The corresponding values of $w_0=u_0^\Delta$, the ratio of the collision rate to the expansion rate at time $\tau_0/\tau_1=10^{-3}$  are respectively $w_0=10^{-15}$,  0.001, 0.03, 0.5, 1, 2. \label{fig:varDelta} }
\end{figure}

We  start by summarizing the generic  behaviors that are expected as a function of $\Delta$. These are displayed in  Fig.~\ref{fig:varDelta}. The various curves in this figure represent   solutions that start from $g_+$ at  time $\tau_0=10^{-3}\,\tau_1$. The shape of the curves near $u_0$  can be understood by recalling the relation $w=u^{\Delta}$, so that $\del g/\del u=(\del g/\del w)\Delta u^{\Delta-1}$, with  $\del g/\del w<0$ at $w=0$ (see e.g. Eq.~(\ref{gsmallw})). The curves for $\Delta\gtrsim 1$ represent (approximately) the attractor. However, when $\Delta<1$ the attractor obviously extends to lower values of $u$, so that the portions of the corresponding curves near $u_0$ should be interpreted as transients. This is particularly obvious for the case $\Delta=0$.

 All the curves cross nearly at the same point, i.e. at $u\gtrsim 1$, the point at which collision rate and expansion rate balance each other. As one varies $\Delta$, keeping  $u_0$ fixed, one varies the initial collision rate, given by $w_0=u_0^\Delta$.  A large value of $\Delta$ corresponds  to a small initial collision rate. This is why the curve corresponding to $\Delta=5$ stays initially almost constant: the system spends a lot of time in the collisionless regime before making a rapid transition to hydrodynamics. Such a regime, already identified earlier (see Eq.~(\ref{steplargeD})),  has been considered for instance in \cite{Broniowski_2009}. As $\Delta $ decreases, the transition smoothens, and eventually the initial collision rate is large enough for the collisions to be effective very early on. This is visible already for $\Delta=0.5$ where the solution exhibits a visible negative slope at this initial time. 
The solution  $\Delta=0$ is special: after a transient regime, a stationary state is reached where the expansion exactly balances the effects of the collisions. For small, but non vanishing $\Delta$ (for instance  $\Delta=0.01$ in Fig.~\ref{fig:varDelta}), the stationary state observed for $\Delta=0$ is replaced by a a very long (several orders of magnitude in $u$) regime where $g(u)$ is a linear function of $\ln u$. This regime will be discussed later in this section, together with the special case $\Delta=0$. 

Note that when $\Delta>0$, $\tau_R$ increases with time. This increase is slower than the increase of the expansion time as long as $\Delta>0$, i.e., $w$ grows with time, so that the collision rate always overcomes the expansion rate at sufficiently late times in which case the system eventually isotropizes.  When $\Delta<0$, on the contrary,  after a transient regime during which the collisions can compete with the expansion, the expansion rate eventually exceeds the collision rate, and the system ends up being  collisionless. Such a behavior has been observed in \cite{Dash:2020zqx} in a study of Gubser flow\cite{Gubser:2010ze}, and was more thoroughly discussed in  \cite{Chattopadhyay:2019jqj}.

\begin{figure}[h]
\begin{center}
\includegraphics[angle=0,scale=0.6]{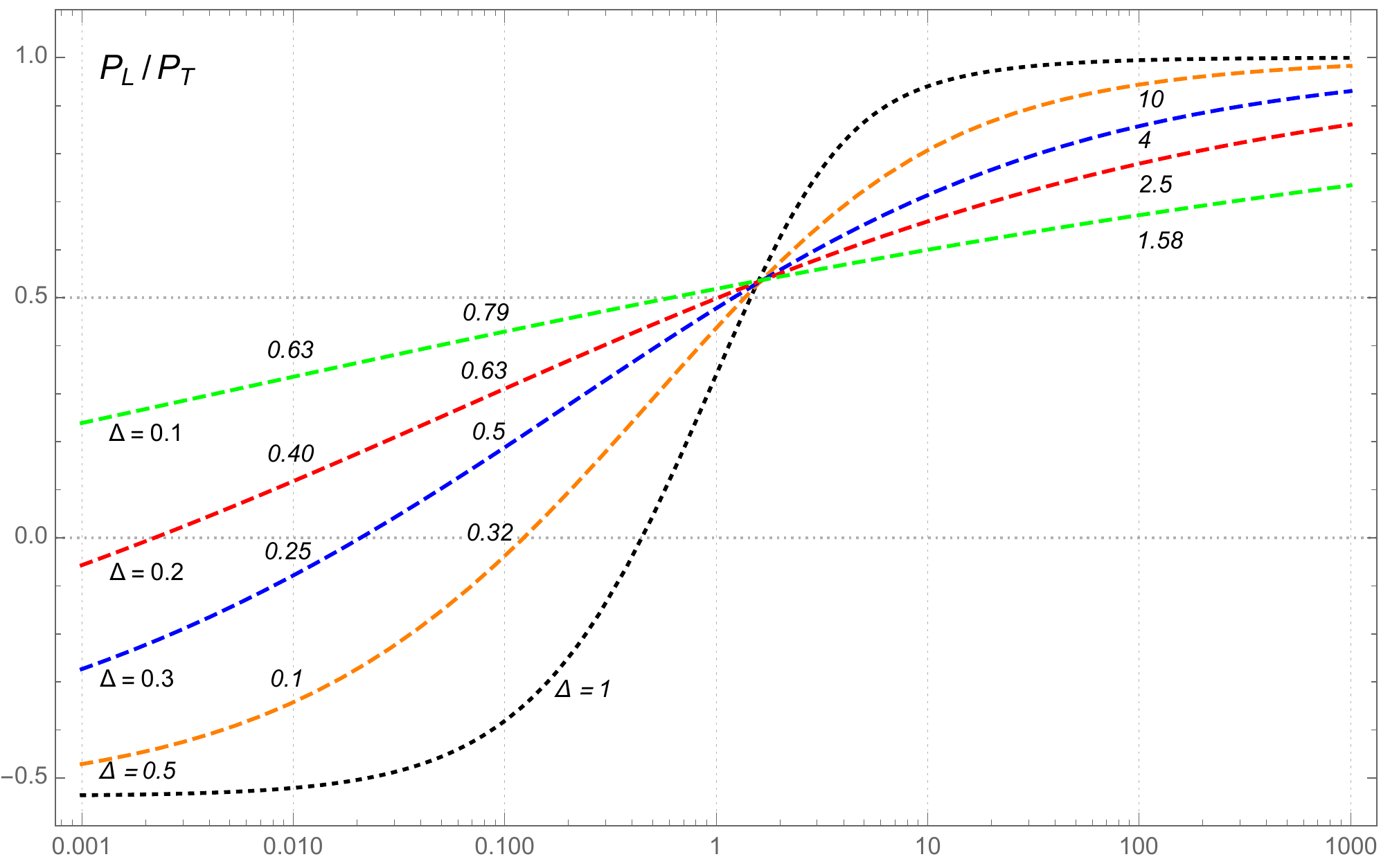} 
\end{center}
\caption{The ratio $\P_L/\P_T$ as a function of $\tau/\tau_1$, and for various values of $\Delta$. The ratio of the collision rate over expansion rate, $w$, is indicated on each curve for $\tau/\tau_1=0.01, 0.1, 100$. For $\Delta=1$ this ratio is simply $\tau/\tau_1$.  \label{fig:pLpTDelta} }
\end{figure}

Another view of the various solutions is provided in Fig.~\ref{fig:pLpTDelta} which displays  the attractor for the pressure ratio $\P_L/\P_T$, for various values of $\Delta$, as a function of $\tau/\tau_1$. This plot exhibits the well known unphysical feature of the two-moment truncation, and of most second order viscous hydrodynamic equations as well, namely a region where the longitudinal pressure becomes negative. Although, in general,  this concerns only a small part of the evolution, and only some initial conditions, this is clearly an unpleasant feature of the approximation. We shall see in the last part of this section how this issue can be overcome, without altering the qualitative overall picture. One can see in Fig.~\ref{fig:pLpTDelta} how the ratio of the collision rate to the expansion rate varies along a given attractor depending on the value of $\Delta$: this ratio varies rapidly for $\Delta=1$, where the transition region is clearly visible, but slowly for small $\Delta$, such as $\Delta=0.1$, in which case the transition region is just a long linear (in $\ln u$) regime extending over several decades. In the next subsection we explore further this region of small $\Delta$.

\subsection{Solutions for small $\Delta$} \label{sec:smallDelta}

To understand better  the case of small values of $\Delta$ we start with the case $\Delta=0$, already addressed in Sect.~\ref{sec:constantDelta} from a different perspective. Then we move to the case of small $\Delta$. 

\subsubsection{Constant attractors for $\Delta=0$}
 
 The case $\Delta=0$ corresponds to the case where $w=\tau/\tau_R$ is a constant, the collision time $\tau_R$ growing linearly with $\tau$. In this case  we measure time with respect to the initial time $\tau_0$ and set $
 u\equiv \tau/\tau_0.
$   
The solution of Eq.~(\ref{eqforbg02}) reads 
 \beq\label{solutionDelta0}
g(u)=\frac{Ag_-(w)+g_+(w)u^{G(w)} }{A+u^{G(w)}},\qquad \L_0(u)=\frac{u^{g_+(w)}+Au^{g_-(w)}}{1+A},
\eeq
with $A$ a constant determined by the initial condition (we have set $\L_0(u=1)=\L_0(\tau_0)=1$). Here $g_+(w)$ denote the fixed point given as a function of $w$ in Eq.~(\ref{FPgpm}), and similarly for $g_-(w)$, Eq.~(\ref{FPgm}), or the gap  $G(w)=g_+(w)-g_-(w)$.   
\begin{figure}[h]
\begin{center}
\includegraphics[angle=0,scale=0.9]{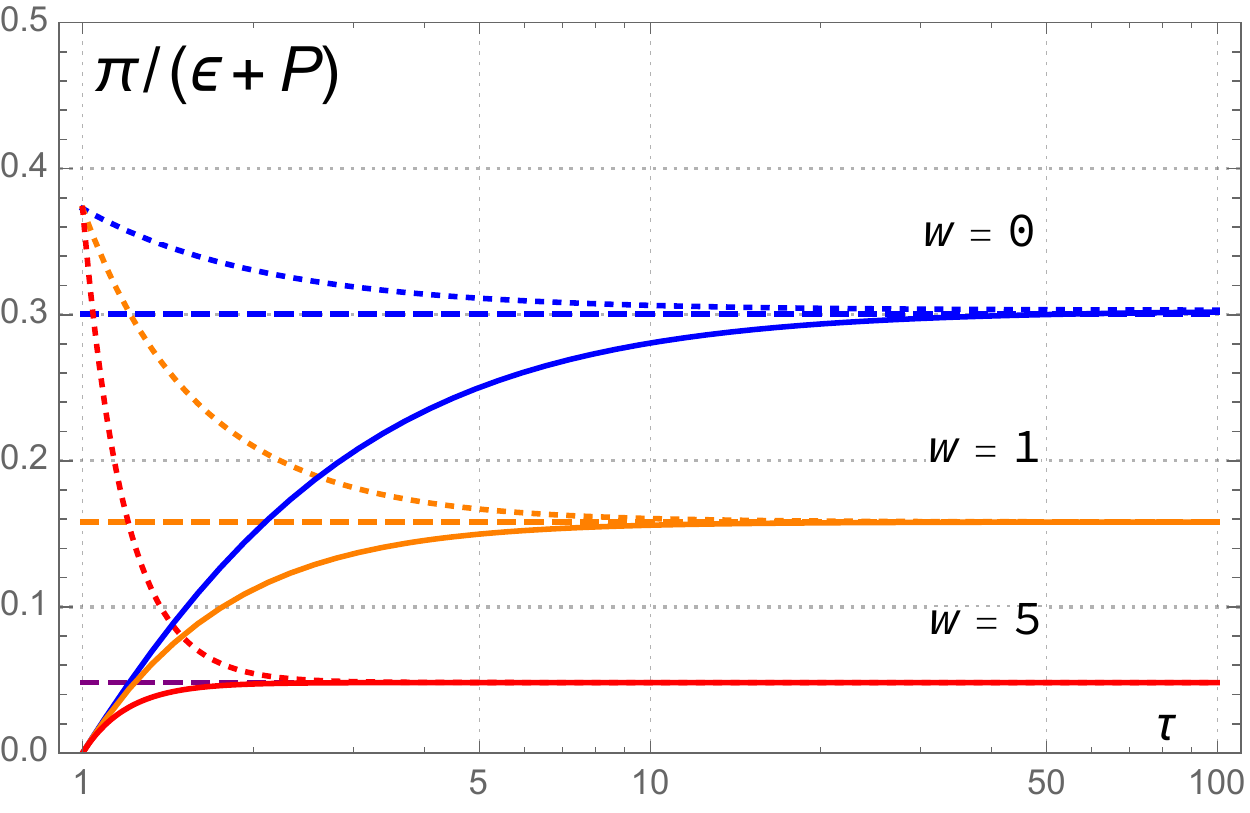} 
\end{center}
\caption{The function $g(u)$ for $\Delta=0$ as a function of $u=\tau/\tau_0$, for various values of $w=0,1,5$. The dashed lines correspond to the (constant) attractor solutions. The initial condition is set to $g(u=1)=0.9*g_+(w=0)$ for the dotted lines, while the full lines correspond to $g(u=1)=g_*$. \label{fig:stationarymode}}
\end{figure}
Irrespective of the value of $A$, the solution (\ref{solutionDelta0}) is driven to the stable fixed point as $u\to \infty$, that is $g(u\to\infty)=g_+(w)$. Similarly, the energy density behaves as $\L_0(u)\sim u^{g_+(w)}$ at late time. Note that it is only when $w\to\infty$ that $g_+(w)$ coincides with $g_*$ characterizing ideal hydrodynamical behavior. For all other values, the evolution of the system is towards a stationary state determined by a perfect balance between collisions and expansion, controlled by the (constant) value of $w$. The attractor in this case is a constant that depends only on $w$, or equivalently the Knudsen number. It is obtained by setting $A=0$ in Eq.~(\ref{solutionDelta0}).  An illustration of the solutions for different initial conditions ($A\ne 0$) is given in Fig.~\ref{fig:stationarymode}  for the pressure asymmetry measured by the quantity $\pi/(\varepsilon+\P)$ (see Eq.~(\ref{relg})). The transient regime and the approach of the solutions to their corresponding attractors are clearly visible. The picture illustrated here is of course compatible with that in Fig.~\ref{fig:varDelta}. We note that in the latter figure, the value of the constant attractor is determined by the initial value $w_0$.


\begin{figure}[h]
\begin{center}
\includegraphics[angle=0,scale=0.5]{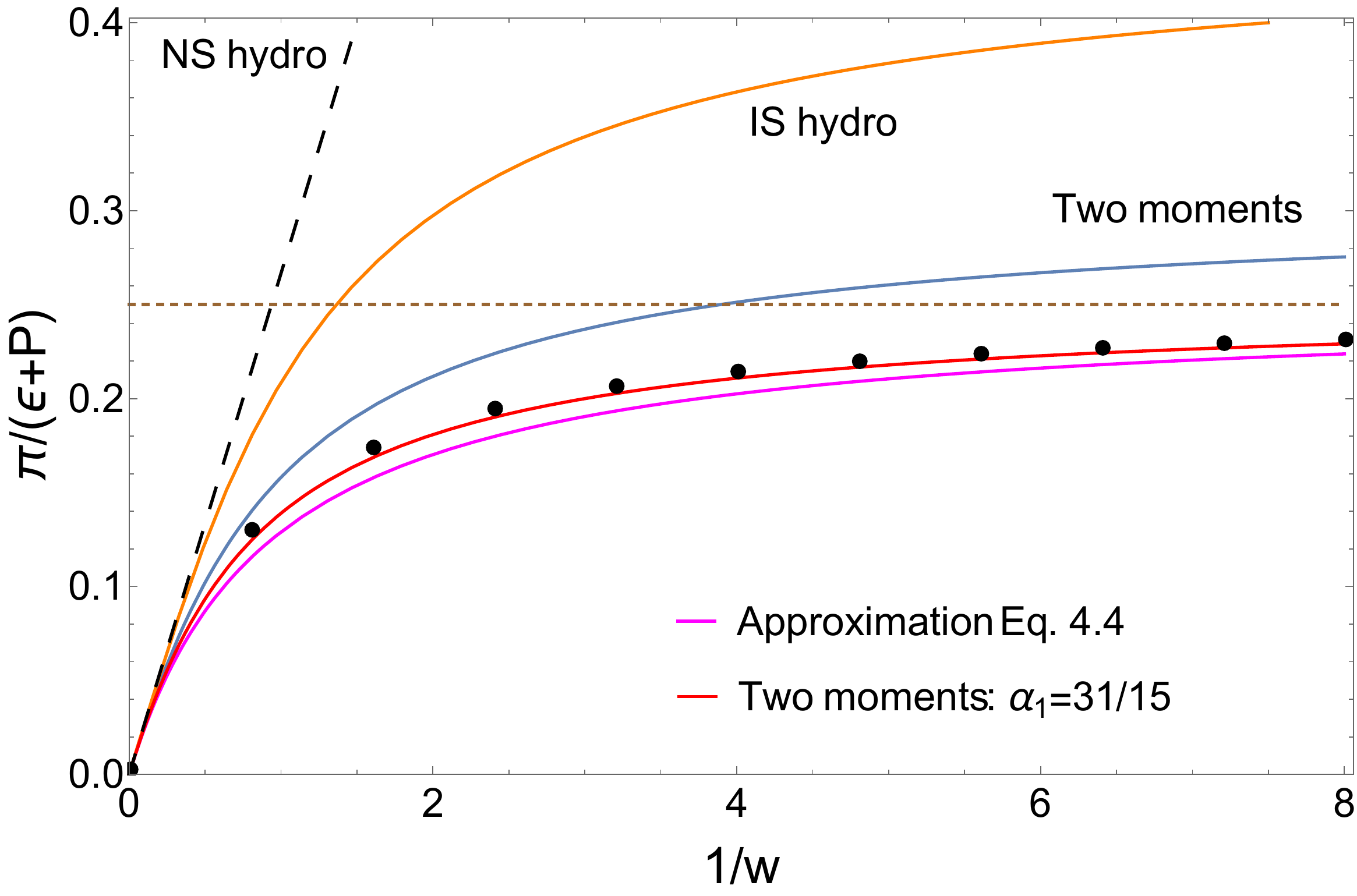} 
\end{center}
\caption{The quantity $\pi/(\varepsilon +P)$ as a function of $1/w$. The black dashed line  is the Navier-Stokes approximation  $\pi/(\varepsilon +P)=b_1/(2w)$. The exact result  (red) converges slowly  to the exact (free streaming) result 0.25 at large values of the Knudsen number. The result of the two moment equation  (blue line), lies slightly above the exact result. It is equivalent to the solution of the IS equations with $a_1=38/21$. The purple line is the solution of the IS equations with $a_1=a_0$. The black points  represent the results of the renormalized second order hydro ($a_1=31/15$) and converge towards the exact value, $0.25$, at large Knudsen number.  \label{fig:DNplot2b}}
\end{figure}

The behavior just described is indeed 
 reminiscent of that observed in \cite{Denicol:2019lio} where the Boltzmann equation is solved for a gas of hard spheres (binary collisions with constant cross section) in an expanding system. In that context one can relate the collision rate to the constant cross section $\sigma$
\beq
\frac{1}{\tau_R}=\sigma n(\tau)=\sigma n_0 \frac{\tau_0}{\tau},\qquad w=\frac{\tau}{\tau_R}=\sigma n_0\tau_0,
\eeq
where $n(\tau) $ is the particle density and we have used the fact that in the expanding system, $n(\tau)$ decreases as $1/\tau$.\footnote{There might be a slight inconsistency here since we do not implement particle number conservation when solving the kinetic equation.}
 To make the comparison with \cite{Denicol:2019lio} tighter, we have plotted in Fig.~\ref{fig:DNplot2b} the quantity 
${\pi}/({\varepsilon+\P})=-{\L_1}/{(2\L_0)}$ 
as a function of the Knudsen number, taken here to be simply $1/w$. The resulting plot is  similar to that presented in Fig.~1 of \cite{Denicol:2019lio}.  The black curve is the Navier-Stokes approximation, with the slope at small Knudsen number (large $w$) proportional to the viscosity.  
 At large values of the Knudsen number (small $w$), the exact solution of the kinetic equation converges to the free streaming value, which is  $1/4$, and this is the value to which the Boltzmann result in \cite{Denicol:2019lio} appears to converge. The curve representing the solution of the two-moment equations is equivalent  to the solution of the IS equations in \cite{Denicol:2019lio}. It only matches approximately the free streaming result. One can also solve the IS equations for different values of $a_1$. As seen in Fig.~\ref{fig:DNplot2b}, decreasing $a_1$, e.g. choosing $a_1=a_0$, increases the deviation from the exact solution. Conversely, choosing a smaller value allows us to reproduce the correct limit at large Knudsen number. We return to this in the next subsection.   
 The red curve representing the exact solution has universal features: its slope at the origin is fixed by the viscosity, and at large Knudsen number ($K\!n \ge 1$) it is dominated by the collisionless regime to which all solutions of the Boltzmann equation should eventually converge.  To emphasize this universal character, we may consider the following ansatz \cite{Bhalerao:2005mm} for the ratio $\P_L/\P_T$ as a function of the Knudsen number $K\!n$
 \beq
 \frac{\P_L}{\P_T}\simeq \frac{b_1}{2}\frac{K\!n}{1+2b_1K\!n},
 \eeq
 which is just a simple interpolation formula between the Navier-Stoke regime and the collisionless regime.  As one can see  in Fig.~\ref{fig:DNplot2b}, this simple formula captures rather well the overall behavior of the exact solution.

%
 

\subsubsection{Scaling solutions near $\Delta=0$}

As we have mentioned earlier, for small $\Delta$, the transition region between the two fixed points is  linear in $\ln u$. In order to analyze this regime, we use the adiabatic approximation, which, as we have seen, is accurate at small $\Delta$. The attractor solution is then approximately given, as a function of $w$, by $g_+(w)$. To get the solution as a function of the physical time we simply substitute $w\to u^\Delta$ in the expression (\ref{FPgpm}) of $g_+(w)$, with $u=\tau/\tau_1$. That is, $g_+(w)$ plays here the role of a scaling function, all the time dependence being contained in the relation $w\to u^\Delta$. Note that this is only approximate since some $\Delta$ dependence remains hidden in the coefficients $a$ and $b$ of the hypergeometric functions (see Eqs.~(\ref{abamb})). This is a small effect however, which does not affect the present discussion in any significant way. 
By calculating the derivative of $g_+(u^\Delta)$ at $u=1$, one gets
\beq\label{slopegamma}
\left.\frac{\del g_+}{\del \ln u}\right|_{u=1}=-\frac{\Delta}{2} \left(1+\frac{a_0-a_1-1}{  G(1)}   \right)\equiv\gamma.
\eeq
The slope proportional to $\Delta$ agrees qualitatively (and quantitatively for small $\Delta$) with that of the attractor solutions displayed in Fig.~\ref{fig:pLpTDelta}. As $\Delta\to 0$, $\gamma$  decreases and the regime linear in $\ln u$  stretches over several decades. In that regime of small $\gamma$, 
\beq
g_+(u)-g_+(1) \sim \gamma \ln u \sim u^\gamma-1.
\eeq
In other words, there is a long time interval where $g_+(u)$ behaves as a power law.

What is responsible for this particular regime is exactly what is causing the flat behavior of the solution in the case $\Delta=0$: the almost exact balance between collisions and expansion. This near  equilibrium holds the system in a quasi stationary state,  which  makes the transition between the collisionless regime and hydrodynamics extremely slow: for a long time, the system is stuck in a regime where the influences of two fixed points nearly annihilate each other. It is tempting to speculate that a similar mechanism is responsible for the so-called ``non-thermal fixed point'' phenomenon leading to the scaling laws observed in the solutions of the  QCD kinetic equations at (very) weak coupling (see e.g. the recent work  \cite{Mazeliauskas_2019} and references therein).

\subsection{Kinetic-hydrodynamics}
\label{sec:kinhyd}

\begin{figure}[h]
\begin{center}
\includegraphics[angle=0,scale=0.8]{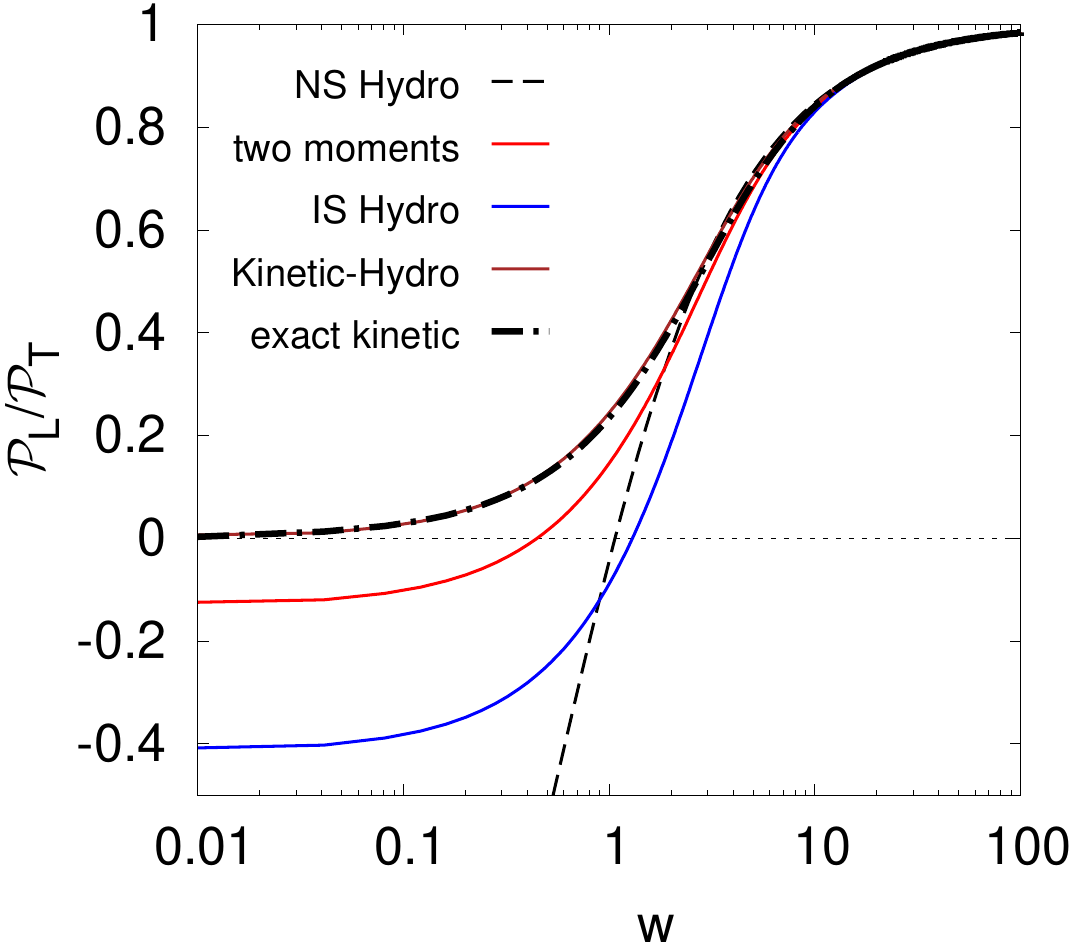} 
\end{center}
\caption{\label{fig:plpta1}
Attractor solution fro the ratio $\P_L/\P_T$, as calculated from NS hydrodynamcis (black dashed), the two-moment truncation of the $\L$-moments (red solid), IS hydrodynamics (blue solid) and the kinetic-hydrodynamics (brown solid), and compared to the exact solution of the full kinetic theory (black dash-dotted line) for $\Delta=1$.
}
\end{figure}

As mentioned in the previous subsection, when discussing the results presented in \Fig{fig:DNplot2b}, it is possible to modify the behavior of the pressure asymmetry at large Knudsen number (small $w$) by tuning the value of $a_1$. Indeed it is a simple matter to verify that for $a_1=31/15$, the lowest eigenvalue of the matrix $M(w=0)$ in Eq.~(\ref{FSbw}) is $\lambda_0=1$, and  that, correspondingly, the exact value of $\L_1/\L_0=1/4$ is reached at small $w$ as shown in \Fig{fig:DNplot2b} (red solid line). 

As was shown in \Sect{sec:2ndhyd}, the equations for the moments  $\L_0, \L_1$  are common to all variants of second order viscous hydrodynamics, these variants being characteirized by specific  values of the   coefficients $a_1$ and $b_1$. We may then regard the case of $a_1=31/15$  as a novel set of second order viscous hydrodynamic equations.  This particular version of second order hydrodynamics which, for lack of a better name, one may call ``kinetic-hydrodynamics'', reproduces accurately the exact attractor solution of the kinetic theory for the Bjorken flow. This is illustrated in  \Fig{fig:plpta1} which displays the attractor solution of the ratio $\P_L/\P_T$: the agreement of  the kinetic-hydrodynamics  with the exact kinetic solution is remarkable, through the whole $w$ region, including both extremes,  hydrodynamics as $w\to\infty$ and collisionless regime as $w\to 0^+$. In particular, the region of negative pressure appearing in NS hydro, IS hydro, or the two-moment truncation (DNMR hydro) at small $w$, has now disappeared. Although \Fig{fig:plpta1} is obtained with $\Delta=1$, similar comparisons hold for arbitrary values of $\Delta$.

To appreciate the physics of the kinetic-hydrodynamics with $a_1=31/15$ it is useful to recall how hydrodynamics emerge from kinetic theory. Within the present moment approach, there are two basic ingredients:  the truncation of the $\L$-moments to $n\le 1$ and the hydrodynamic fixed point at $w\to \infty$. By construction, the truncation of the tower of equations for the $\L_n$ moments to the lowest two moments results in equations for the independent components in the energy-momentum tensor $T^{\mu\nu}$ (see Eq.~(\ref{Txxyyzz})).
It is only in the hydrodynamic limit $w\to\infty$ that the lost information caused by the truncation becomes negligible, and that  the evolution of the energy-moment tensor, i.e., second order viscous hydrodynamics   become accurate. For large $w$ (or small Knudsen number), all versions of viscous hydrodynamcis converge to the to the simple Navier-Stokes approximation which describes accurately the evolution in the vicinity of the hydrodynnamic fixed point, as shown in \Fig{fig:plpta1}. 
However, away from the small Knudsen number regime ($w\to \infty$), the truncation cannot be exact. In particular, as the Knudsen number increase ($w\to 0^+$), the effects from higher moments  $\L_n$ become more  an more important. We know from previous works \cite{Blaizot:2017ucy,Blaizot:2019scw} that these higher moments do not change the overall fixed point structure, they mainly affect the location of the collisionless fixed point. This can be taken into account by a simple renormalization of the coefficients $a_1$ and/or $b_1$. The coefficient $b_1$ is fixed by the viscosity and cannot be changed without affecting the Navier-Stokes regime. However $a_1$ can be tuned, and as we have seen, the value $a_1=31/15$ puts the collisionless fixed point at the right place. The coefficient $a_1$ is related to the transport coefficient $\lambda_1$ that appears in second order hydrodynamics \cite{Baier:2007ix}, with  
 $\lambda_1\propto a_1-a_0$.  The choice $a_1=31/15$ of kinetic-hydrodynamics leads to the value $\lambda_1/\eta\tau_\pi=11/10$. This is not too different from those obtained   in kinetic theory \cite{York:2008rr,Teaney:2013gca}, respectively $\lambda_1/\eta\tau_\pi=5/7$ or $1$, or of that obtained in ${\cal N}=4$ super Yang-Mills theory~\cite{Baier:2007ix}, namely $1/(2-\log 2)$.

\section{Conclusions}


In this paper, we have first looked at second order viscous hydrodynamics as a coupled mode problem, using techniques of linear algebra. The two eigenmodes are associated to simple angular moments of the momentum distribution, $\L_0$ and $\L_1$, and the two independent components of the energy momentum tensor, the energy density $\varepsilon=\L_0$ and the difference between the longitudinal and the transverse pressures, $\L_1=\P_L-\P_T$. In the collisionless regimes, the two coupled modes are damped, one faster than the other, so that at late time, after some transient regime,  only one mode survives. The collisions change gradually the nature of this eigenmode until, at late time, it describes hydrodynamics. 

Then we turned the  coupled equations into a single non linear differential equation for the pressure asymmetry, measured by the ratio of the two moments $\L_0$ and $\L_1$. The two modes of the linear problem are then associated to fixed points of this non linear equation, with the  stable collisionless fixed point evolving slowly under the effects of the collisions into the hydrodynamic fixed point. The attractor appears then as the solution that connects two distinct physical regimes,  the early time collisionless regime and the late time hydrodynamic, collison dominated, regime. In contrast to the notion of fixed points, which may be considered as ``local'' concepts, the  attractor is a non local object. It emerges here as  a generic feature of the competition between two fixed points, between expansion and hydrodynamics.  

The non linear solution is amenable to an analytic solution in terms of special functions. This allowed us to verify a number of properties of the solution that can be derived by elementary means, as well as to test various approximations. The solution depends on a parameter $\Delta$ that controls the speed of the transition between the collisionless regime and hydrodynamics. Varying this parameter allowed us to reveal a number of interesting features of the solution. It also allowed us to reproduce easily a number of regimes identified in more sophisticated simulations.  A particularly interesting regime is that of slow transitions well accounted for by an adiabatic approximation. In this regime, expansion and collisions nearly balance each other and a phenomenon reminiscent of the so-called non-thermal fixed point is observed. The collisionless fixed point explains the universality of the attractor at small time, a property which has been recently exploited in \cite{Giacalone:2019ldn}.

  When analyzing how hydrodynamics emerges, we see no real mystery: it emerges when the collision rate becomes comparable to the expansion rate. The ``success'' of second order hydrodynamics to match kinetic theory at early times, when gradients are large, is essentially connected to the choice of a second order transport coefficient,  its value determining how well  the collisionless fixed point is approximated. In other words, moving backward in time, one does not ``improve'' hydrodynamics (since the Knudsen number increases), one just get deeper into the collisionless regime, that is, closer to the free streaming fixed point, present in all versions of second order hydrodynamics of IS type\footnote{It is the time derivative of the viscous tensor, that is introduced in an ad hoc fashion in Israel-Stewart theory, or that emerges naturally in BRSSS analysis, that is responsible for the presence of the collisionless fixed point in second order hydrodynamics.}.  We have demonstrated this mechanism by implementing a simple renormalization of the second order transport coefficient $\lambda_1$. This renormalization puts the free streaming stable fixed point at the right place, and allows us to reproduce with great accuracy  the exact solution of the kinetic equation within second order viscous hydrodynamics. From that point of view, the fact that hydrodynamics matches kinetic theory while the pressure anisotropy is still ``large", as measured by the ratio $\P_l/\P_T\sim 0.5$, is not so surprising. In the context of Bjorken flow at least, the   success of hydrodynamics is perhaps not so  ``unreasonable" once one realizes that its extension to large gradients, or equivalently to early times, just involves a correct treatment of the early time collisionless regime, which, as we have seen, can be achieved in a simple fashion.

\acknowledgements 

L.Y. is supported in part by National Natural Science Foundation of China (NSFC) under Grant No. 11975079.

\appendix

\section{BRSSS hydrodynamics}\label{sec:BRSSS}

In this Appendix, we provide details on the derivation of the generalization of Eq.~(\ref{eq:Lequb}) in the case of  BRSSS hydrodynamics~\cite{Baier:2007ix}. We note first that this equation can be written as follows \cite{Blaizot:2019scw}
 \beq
 \pi=\frac{4\eta}{3\tau}-\tau_\pi\left( \del_\tau \pi +a_0 \frac{\pi}{\tau}  \right)-\frac{\lambda_1}{2\eta^2}\pi^2,
\eeq
which coincides for instance (to within trivial notation changes) with Eq.~(4) in Ref.~\cite{Heller:2015dha}. 
At this order of the gradient expansion,  one can substitute $\pi/\eta$ by $4/3\tau$ in the term quadratic in $\pi$, and obtain the linear equation
\beq\label{A2}
\pi=\frac{4\eta}{3\tau}-\tau_\pi\left( \del_\tau \pi +a_0 \frac{\pi}{\tau}  \right)-\frac{2\lambda_1}{3\eta}\frac{\pi}{\tau}.
\eeq
Note that, in contrast to the IS approach, where it is introduced as a relaxation term, here  the derivative of the viscous pressure appears naturally among the various terms linear in gradients.
We can rewrite Eq.~(\ref{A2}) in the following way\footnote{In kinetic theory for massless particles, $\lambda_1/(\eta\tau_\pi)=5/7$, so that the coefficient of $\pi/\tau$ is just $a_1=38/21$. }
\beq\label{eqforpi2}
\frac{\rmd\pi}{\rmd\tau}+\frac{\pi}{\tau}\left(\frac{4}{3}+\frac{2\lambda_1}{3\eta  \tau_\pi}  \right)=-\frac{1}{\tau_\pi}\left( \pi-\frac{4}{3}\frac{\eta}{\tau}\right),
\eeq
or, equivalently, as 
\beq\label{eq:Lequconf}
\frac{\rmd\L_1}{\rmd \tau}+a_1'\frac{\L_1}{\tau}+b_1'\frac{ \L_0}{\tau}=-\frac{\L_1}{\tau_R} ,
\eeq
where the coefficients $a_1'$ and $b_1'$ are given in Eq.~(\ref{coefprime}) of the main text.

\section{Perturbation theory}\label{app:pert}

In this appendix, we consider the solution of Eqs.~(\ref{FSb}) at small $w$, i.e. for small (relative) collision rates, where one can expect time-dependent perturbation theory to be valid. We set $\Delta=1$, and 
write the matrix $M$ as $M=M_0+M_1$ (cf. Eq.~(\ref{M0M1pert})), with $M_1$ the perturbation. We shall obtain the solution to Eqs.~(\ref{FSb}) in leading order in $M_1$.

We call $\ket{\phi_0}$ and $\ket{\phi_1}$ the (constant) eigenstates of $M_0$, and $\lambda_0,\lambda_1$ the corresponding eigenvalues:
\beq
M_0\ket{\phi_n}=\lambda_n\ket{\phi_0},\qquad (n=0,1).
\eeq
We normalize the eigenstates so that 
$\bra{e_1}\phi_n\rangle=1.$
It is then easy to show that 
\beq\label{eigenM0}
\bra{e_0}\phi_n\rangle=\frac{c_0}{\lambda_n-a_0}=\frac{\lambda_n- a_1}{b_1} ,\qquad \bra{e_1}\phi_n\rangle=1,\qquad (n=0,1).
\eeq

   Let us consider first the solution to Eqs.~(\ref{FSb}) that corresponds to the initial condition $\ket{\L(w_0)}=\ket{\phi_0}$ for some finite $w_0$. 
We expand this solution on the eigenstates of $M_0$ as follows
\beq
\ket{\L(w)}=C_0(w)(1+ a_{00}(w))\ket{\phi_0}+ a_{01}(w) C_1(w) \ket{\phi_1},
\eeq
where the (small) coefficients $a_{00}$ and $a_{01}$ are chosen such that 
$a_{00}(w_0)=0, \, a_{01}(w_0)=0,$ while 
\beq\label{trivialC0C1}
C_n(w)=\left( \frac{w_0}{w}  \right)^{\lambda_n},\qquad (n=0,1),
\eeq
encode the ``natural''  time dependence of the eigenstates of $M_0$ (that induced by $M_0$ alone). 
A simple calculation, using the equations of motion (\ref{FSb}), leads to 
\beq
\left[w\del_w +M_0\right]\ket{\L(w)}=C_0(w) w\frac{\del a_{00}}{\del w} \ket{\phi_0} +C_1(w) w\frac{\del a_{01}}{\del w} \ket{\phi_1}=-M_1\ket{\L(w)}.
\eeq
To determine the  action of $M_1$ on $\ket{\L(w)}$, we note that, in leading order, we need only consider the action of $M_1$ on $\ket{\phi_0}$ (since $\ket{\phi_1}$ is multiplied by the small quantity $a_{01}$). We have, with $\alpha$ and $\beta$ two constants to be determined, 	
\beq
M_1\ket{\phi_0}=\alpha \ket{\phi_0}+\beta \ket{\phi_1}.
\eeq
By projecting on the natural basis, one gets
\beq
&& \bra{e_0} M_1\ket{\phi_0}=0=\alpha \bra{e_0}\phi_0\rangle+\beta  \bra{e_0}\phi_1\rangle,\nn
&& \bra{e_1} M_1\ket{\phi_0}=w \bra{e_1}\phi_0\rangle=\alpha \bra{e_1}\phi_0\rangle+\beta  \bra{e_1}\phi_1\rangle,
\eeq
from which one extracts the values of $\alpha$ and $\beta$:
\beq
\alpha= w \frac{a_0-\lambda_0}{G},\qquad   \beta= -w \frac{a_0-\lambda_1}{G},\qquad G\equiv \lambda_1-\lambda_0.
\eeq
The equation of motion (\ref{FSb}) becomes then 
\beq
\left[w\del_w +M_0\right]\ket{\L(w)}&=&-M_1\ket{\L(w)}\simeq - C_0(w) \left[  \alpha\ket{\phi_0}+\beta\ket{\phi_1} \right]\nn
&=& C_0(w) w\frac{\del a_{00}}{\del w} \ket{\phi_0} +C_1(w) w\frac{\del a_{01}}{\del w} \ket{\phi_1},
\eeq
from which one deduces
\beq
\frac{\del a_{00}}{\del w}=-\frac{a_0-\lambda_0}{G},\qquad \frac{\del a_{01}}{\del w} =\frac{C_0(w)}{C_1(w)} \frac{a_0-\lambda_1}{G}.
\eeq
These equations are easily integrated. Taking into account the initial condition, one gets
\beq
a_{00}(w)= -\frac{a_0-\lambda_0}{G} (w-w_0),
\qquad
a_{01}(w)= \frac{a_0-\lambda_1}{G}\frac{w_0}{1+G}\left[ \left(\frac{w}{w_0}  \right)^{G+1}-1 \right].
\eeq
The expressions of the moments, in first order perturbation theory, are then  
\beq\label{L0Phi0}
\L_0=C_0(w)\langle e_0\ket{\phi_0} \left\{1-\frac{a_0-\lambda_0}{G} (w-w_0) +\frac{a_0-\lambda_0}{G}\frac{w}{1+G}\left[ 1 -\left(\frac{w_0}{w}  \right)^{G+1} \right]\right\},
\eeq
and
\beq\label{L1Phi0}
\L_1=C_0(w)\left\{1- \frac{a_0-\lambda_0}{G} (w-w_0)+\frac{a_0-\lambda_1}{G}\frac{w}{1+G}\left[ 1 -\left(\frac{w_0}{w}  \right)^{G+1} \right]\right\}.
\eeq
One deduces from these expressions that of the pressure asymmetry
%
\beq\label{L1L0pert}
\frac{\L_1}{\L_0}=\frac{1}{\langle e_0\ket{\phi_0}}  \left\{1-  \frac{w}{1+G}\left[1-\left(\frac{w_0}{w}  \right)^{G+1} \right]   \right\}.
\eeq
Note that,  in Eqs.~(\ref{L0Phi0}) and (\ref{L1Phi0}), besides the terms linear in $w$ coming from the first order in the perturbation $M_1(w)\sim w$,  there are additional, non analytic terms $\sim (w_0/w)^G$, whose origin lies in the free streaming coefficients $C_0(w)$ and $C_1(w)$ given in (\ref{trivialC0C1}). Such   terms prevent the small $w$ expansion of the moments to extend all the way to $w=0$, once the initial condition has been fixed at $w_0>0$. However, these terms cancel out in the pressure asymmetry (\ref{L1L0pert}). This  quantity is insensitive to the ``trivial'' short time behavior of the moments, and for it, one can fix the initial condition at $w_0=0$, leaving $\L_1/\L_0$ as an analytic function near $w=0$. One can easily verify that Eq.~(\ref{L1L0pert})  coincides with the corresponding expansion of the analytic solution. 
\\

The previous results depend crucially on the choice of the initial state on the which the perturbation is acting. Let us then repeat the same analysis starting from the mode $\ket{\phi_1}$, i.e.,  $\ket{\L(w_0)}=\ket{\phi_1}$. We set
\beq
\ket{\L(w)}=C_0(w) a_{10}(w))\ket{\phi_0}+ (1+a_{11}(w)) C_1(w) \ket{\phi_1},
\eeq
with $a_{10}(w_0)=0$ and $a_{11}(w_0)=0.$
A calculation similar to that done above yields the following expressions for the moments
\beq
\L_0=C_1(w)\langle e_0\ket{\phi_1} \left\{1+\frac{a_0-\lambda_1}{G} (w-w_0) -\frac{a_0-\lambda_1}{G}\frac{w_0}{1-G}\left[ \frac{w}{w_0} -\left(\frac{w}{w_0}  \right)^{G} \right]\right\},
\eeq
\beq
\L_1=C_1(w)\left\{1+\frac{a_0-\lambda_1}{G} (w-w_0)-\frac{a_0-\lambda_0}{G}\frac{w_0}{1-G}\left[ \frac{w}{w_0} -\left(\frac{w}{w_0}  \right)^{G} \right]\right\},
\eeq
and for the pressure asymmetry
\beq\label{presasymphi1b}
\frac{\L_1}{\L_0}=\frac{1}{\langle e_0\ket{\phi_1}}  \left\{1-  \frac{w}{1-G}\left[ 1-\left(\frac{w}{w_0}  \right)^{G-1} \right]   \right\}.
\eeq
In contrast to the previous case where $\ket{\L(w_0)}=\ket{\phi_0}$, here there is  no obstacle to let $w\to 0$. However, it is now not possible to fix the initial condition at $w_0=0$, even for the pressure asymmetry.  The small $w$ expansion remains modified by the presence of non analytic terms $\sim (w/w_0)^{G-1}$. Such terms are the origin of the trans-series structure for the corresponding all-order solution (see Appendix~\ref{trans}).

The behaviors that we have observed in the two cases that correspond respectively to the initial conditions  $\ket{\L(w_0)}=\ket{\phi_0}$ and $\ket{\L(w_0)}=\ket{\phi_1}$, can be also understood in terms of the fixed points of the non linear equation obeyed by the pressure asymmetry (see Sect.~\ref{sec:nonlinear}). These initial conditions correspond to what we have called the stable and the unstable fixed points in Sect.~\ref{sec:nonlinear}, ``stable'' or ``unstable'' referring to the behavior of the solution near these fixed points as $w$ is increasing: as $w$ increases, a generic solution is ``attracted'' toward the stable fixed point and ``repelled'' from the unstable one. When going backward, i.e. towards $w=0$ starting from some finite $w_0$, the attractive fixed point becomes repulsive and vice-versa. Thus, all solutions eventually go at small $w$ to the ``unstable'' fixed point, while the ``stable'' fixed point can only be reached for a single very specific initial condition (that corresponding to the attractor). 

\section{Adiabatic approximation}\label{app:adiabatic}

In this Appendix, we solve Eq.~(\ref{FSb}) in the adiabatic approximation introduced in Sect.~\ref{sec:adiapprox}. To do so, we expand the state of the system $\ket{\L}$ on the instantaneous eigenstates of the matrix $M(w)$. We call  $\ket{\phi_0(w)}$ and $\ket{\phi_1(w)}$ the right eigenvectors of the matrix $M(w)$ belonging respectively to the eigenvalues $\lambda_0(w)$ and $\lambda_1(w)$. That is 
\beq\label{eominstant}
M(w)\ket{\phi_n(w)}=\lambda_n(w)\ket{\phi_n(w)},\qquad (n=0,1).
\eeq
To each eigenvalue $\lambda_n(w)$, there corresponds a left eigenvector $\bra{\phi'_n(w)}$, whose transpose is an eigenvector of the transpose of the matrix $M$, that is 
\beq
\tilde M(w)\ket{\phi'_n(w)}=\lambda_n(w)\ket{\phi'_n(w)},\qquad (n=0,1).
\eeq
It is easy to show that 
\beq
\bra{\phi'_n(w) } \phi_m(w)\rangle=\delta_{n m} \bra{\phi'_n(w)}\phi_n(w)\rangle.
\eeq
The normalization of the eigenstates is fixed after projection on the natural basis as we did in  Appendix~\ref{app:pert}. We set
\beq
\bra{e_1} \phi_n(w)\rangle=1,\qquad (n=0,1)
\eeq
By using the relation
\beq
(a_0-\lambda_0(w)) \langle e_0\ket{\phi_0(w)}+c_0 \langle e_1\ket{\phi_0(w)}=0,
\eeq
and a similar one for $\ket{\phi_1(w)}$, one obtains then (see Eqs.~(\ref{eigenM0}))
\beq\label{pji0phi1natbasis}
\langle e_0\ket{\phi_n(w)}=\frac{c_0}{\lambda_n(w)-a_0},\qquad (n=0,1).
\eeq
Similarly, the non trivial components of $\ket{\phi_0'}$ and $\ket{\phi_1'}$ are
\beq\label{pji0phi1pnatbasis}
\langle e_0\ket{\phi_n'(w)}=\frac{b_1}{\lambda_n(w)-a_0},\qquad (n=0,1).
\eeq
The explicit expressions of $\lambda_0(w)$ and $\lambda_1(w)$ are given in Eqs.~(\ref{lambda0r0}) of the main text, from which we deduce in particular
\beq
\del_w \lambda_0(w)=\frac{a_0-\lambda_0(w)}{\lambda_1(w)-\lambda_0(w)},\qquad  \del_w \lambda_1(w)=\frac{\lambda_1(w)-a_0}{\lambda_1(w)-\lambda_0(w)}.
\eeq
These formulae will be useful later on.

In order to solve the equation of motion 
\beq\label{eomadiab0}
\Delta w\del_w\ket{\L(w)}=-M(w)\ket{\L(w)},
\eeq
 we expand $\ket{\L(w)}$ on the instantaneous eigenstates:
\beq\label{expinst}
\ket{\L(w)}=C_0(w)\ket{\phi_0(w)}+C_1(w)\ket{\phi_1(w)}.
\eeq
The equation of motion then reads
\beq\label{eqomwjo}
\Delta w\del_w \ket{\L(w)}&=&\dot C_0(w)\ket{\phi_0(w)}+ C_0(w)\ket{\dot\phi_0(w)} +\dot C_1(w)\ket{\phi_1(w)}+C_1(w)\ket{\dot\phi_1(w)}\nn
&=& -C_0(w)\lambda_0(w)\ket{\phi_0(w)}-C_1(w)\lambda_1(w)\ket{\phi_1(w)},
\eeq
where the dot denotes here $\Delta w\del_w$. To calculate the time derivative of the instantaneous eigenvectors,  we also expand these derivatives on the eigenstates:
\beq
&&\del_w \ket{\phi_0(w)}= \alpha_{00}(w) \ket{\phi_0(w)}+\alpha_{01}(w) \ket{\phi_1(w)},\nn && \del_w \ket{\phi_1(w)}= \alpha_{10}(w) \ket{\phi_0(w)}+\alpha_{11}(w) \ket{\phi_1(w)},
\eeq
with 
\beq
\alpha_{ij}(w) =\frac{\bra{\phi_i'}\del_w \ket{\phi_j}}{\bra{\phi_i'}{\phi_j}\rangle},\qquad i,j=1,2.
\eeq
A simple calculation, using for instance the explicit components of the eigenstates on a fixed basis  (see Eqs.~(\ref{pji0phi1natbasis}) and (\ref{pji0phi1pnatbasis})), 
yields  
\beq\label{alphaijw}
&&\alpha_{00}(w) =-\alpha_{01}(w)=\frac{\lambda_1(w)-a_0}{(\lambda_1(w)-\lambda_0(w))^2},\nn && \alpha_{10}(w) =-\alpha_{11}(w)=\frac{a_0-\lambda_0(w)}{(\lambda_1(w)-\lambda_0(w))^2}.
\eeq
One may also use the relations
\beq
&&\bra{\phi_i'}w\del_w \ket{\phi_j}=-\frac{ \bra{\phi_i'} w\del_w M \ket{\phi_j} }{\lambda_i(w)-\lambda_j(w)}=-\frac{ \bra{\phi_i'} M_1 \ket{\phi_j} }{\lambda_i(w)-\lambda_j(w)}= w \,\alpha_{ij}
(w), \qquad i\ne j,\nn
&&\frac{ \bra{\phi_i'}w\del_w \ket{\phi_i} }{\bra{\phi_i'}{\phi_i}\rangle}=w\del_w \lambda_i.
\eeq
As $w\to 0$ the coefficients (\ref{alphaijw}) go to constant values, the eigenvalues being then equal to those of the collisionless regime. As $w\to \infty$,  $\lambda_1(w)\sim w$ and $\lambda_0(w)-a_0\sim -b_1 c_0/w$, so that $\alpha_{00}(w)\sim 1/w$,  and $\alpha_{10}\sim b_1c_0/w^3$.

The adiabatic approximation requires the rate of change of the eigenvectors to be small as compared to the change induced by the eigenvalues.  To be more precise, we return to Eq.~(\ref{eqomwjo}), 
divide this equation  by $\Delta $ and separate the projections  on the two eigenstates. We get the following two coupled equations
\beq\label{eomcoupledCi}
&&w \del_w C_0(w)+\left(\frac{\lambda_0(w)}{\Delta } + w\alpha_{00}(w)\right)C_0(w) + w\alpha_{10}(w) C_1(w) =0,\nn
&&w\del_w C_1(w)+\left(\frac{\lambda_1(w)}{\Delta }-w\alpha_{10}(w)\right)C_1(w) - w\alpha_{00}(w) C_0(w) =0.
\eeq
 These equations (\ref{eomcoupledCi}) are an exact transcription of the equations of motion (\ref{eomadiab0}), obtained after projection on the instantaneous eigenstates, that is, no approximation has been done so far. 
We note now that, at small $w$ the $\alpha_{ij}$ play no role since they are multiplied by $w$. As we shall verify shortly,  by ignoring them one  just reproduces the free streaming regime. When $w$ becomes large, one can verify that the  coefficients  $\alpha_{ij}$ can still be ignored when $\Delta\to 0$. Indeed, as we have seen, $w a_{00}(w)$ goes to a constant and is therefore small compared to $\lambda_0/\Delta$ when $\Delta$ is small enough (recall that $\lambda_0\sim 1$ so that this implies $\Delta\lesssim 1$, a condition that we shall recover later). As for the term $w a_{10}(w)$, it decreases as $1/w^2$ and it can be safely ignored.  When all the terms proportional to the $\alpha_{ij}$'s are neglected, the equations (\ref{eomcoupledCi}) decouple and read
\beq\label{eomadiab02}
 w\del_w C^{(0)}_0(w)+\frac{\lambda_0(w)}{\Delta} C^{(0)}_0(w) =0,\qquad w\del_w C^{(0)}_1(w)+\frac{\lambda_1(w)}{\Delta} C^{(0)}_1(w) =0.
\eeq 
 This constitutes the leading adiabatic approximation. The coefficients $C^{(0)}_0(w)$ and $C^{(0)}_1(w)$ that are solutions of Eqs.~(\ref{eomadiab02}) are given by 
 \beq
C^{(0)}_i(w)=C^{(0)}_i(w_0)\,\exp\left( -\frac{1}{\Delta}\int_{w_0}^w \frac{\rmd w}{w} \lambda_i(w)\right), \qquad (i=0,1).
\eeq
At early time, the eigenvalues are constant, and these formulae yield
\beq
C^{(0)}_i(w)=C^{(0)}_i(w_0)\left( \frac{w_0}{w} \right)^{\frac{\lambda_i}{\Delta}}, \qquad (i=0,1).
\eeq
Note that the presence of the factor $\Delta$ in the exponent is just an artefact of our use of $w$ as a measure of time. Going back to the physical time $\tau$ eliminates this factor and leaves 
\beq
C^{(0)}_i(\tau)=C^{(0)}_i(\tau_0)\left( \frac{\tau_0}{\tau} \right)^{\lambda_i}, \qquad (i=0,1),
\eeq
which is the usual free streaming relation.
In this regime, 
\beq
\ket{\L(w)}\simeq C^{(0)}_0(w_0)\left( \frac{w_0}{w} \right)^{\frac{\lambda_0}{\Delta}}\left( \ket{\phi_0(w)}+ \frac{ C^{(0)}_1(w_0)}{ C^{(0)}_0(w_0)}  \left( \frac{w_0}{w} \right)^{\frac{G}{\Delta}}\ket{\phi_1(w)}\right).
\eeq
At late time on the other hand, $\lambda_1(w)-\lambda_0(w) \sim w$, and we get
\beq
\ket{\L(w)}\simeq C^{(0)}_0(w_0)\left( \frac{w_0}{w} \right)^{\frac{a_0}{\Delta}}\left( \ket{\phi_0(w)}+ \frac{ C^{(0)}_1(w_0)}{ C^{(0)}_0(w_0)}  \rme^{\frac{w_0-w}{\Delta}}\ket{\phi_1(w)}\right).
\eeq
In both cases, the dominant mode is the mode $\ket{\phi_0(w)}$, the other component being damped as $w$ increases. The attractor solution is obtained by starting the evolution in this particular state at some $w_0$. As was observed in several occasions, the moments do not have a well defined limit as we let $w_0\to 0$. However, let us consider 
\beq\label{gadiab}
g(w)=\frac{\dot\L_0}{\L_0}= -\frac{  C_0(w)\lambda_0(w)\langle e_0\ket{\phi_0}+C_1(w)\lambda_1(w)\langle e_0\ket{\phi_1}  }{  C_0(w)\langle e_0\ket{\phi_0}+C_1(w)\langle e_0\ket{\phi_1}  }.
\eeq
When we substitute in this expression $C_i(w)$ by $C_i^{(0)}(w)$ and furthermore choose the initial condition  $C_1^{(0)}(w_0)=0$, one obtains the simple result
\beq
g(w)=-\lambda_0(w).
\eeq
The rapidly varying functions $C_0^{(0)}(w)$ have cancelled out between numerator and denominator, leaving for $g(w)$ a simple result whose validity extends all the way to $w=0$: this result is nothing but the adiabatic attractor $g_+(w)$.  

As the comparison with the exact solution has shown, the adiabatic approximation turns out to be an excellent approximation, even for $\Delta\simeq 1$ (see the discussion in Sect.~\ref{sec:nonlinear} and Fig.~\ref{fig:AttSmallDelta}). This is in part due to the fact that the coefficients $\alpha_{ij}$ in Eq.~(\ref{eomcoupledCi}) decrease rapidly as $w$ gets large. We have seen for instance that the leading order in the gradient expansion is independent of $\Delta$, the corrections to the adiabatic approximations manifesting themselves only at order $1/w^2$  (see e.g  Eq.~(\ref{attw2D})). At small $w$, in order to see how the adiabatic approximation handles the effect of collisions, one may exploit the results of perturbation theory obtained in Sect.~\ref{app:pert}. 
We need to extend  the results obtained there to the general case $\Delta\ne 1$. As mentioned in the main text (see after Eq.~(\ref{FSbw})) this is achieved by rescaling $w\to \bar w=w/\Delta$, and $G\to \bar G=G/\Delta$. This rescaling leaves the eigenvectors of $M_0$ invariant. It follows that Eq.~(\ref{L1L0pert}) for instance becomes (dropping the non analytic piece, or assuming $w_0=0$)
\beq\label{L1L0pertDelta}
\frac{\L_1}{\L_0}=\frac{\lambda_0(w)-a_0}{c_0} \left\{1-  \frac{w}{\Delta+G}  \right\}.
\eeq
On sees that the term $\Delta$ enters as a correction to the gap, $G+1\to G+\Delta$. For small $\Delta$, this correction can be interpreted as a correction  to the adiabatic approximation. It vanishes when $\Delta\to 0$ and becomes significant only when $\Delta\sim G$. But the gap in the free streaming spectrum is $G\sim 1$. It follows that, at small $w$, the adiabatic approximation is expected to remain reasonably accurate in the whole range of $w$ values as long as  $\Delta\lesssim 1$. 

As a final remark, let us note that we can expand the instantaneous eigenstates on the eigenstates of $M_0$. At large $w$, these take simple forms. In particular the mode $\ket{\phi_0(w)}$ is given by
\beq
\ket{\phi_0(w)}\simeq \frac{w}{G}  \left(\ket{\phi_0}-\ket{\phi_1} \right).
\eeq
On can easily verify that $\ket{\phi_0(w)}$ is an eigenstate of $M$, as it should, with eigenvalue $a_0$ (to obtain this result, since $M_1(\ket{\phi_0}-\ket{\phi_1}$=0, one needs to consider the action of $M_1$ on the ``small'' component of $\ket{\phi_0(w)}$, i.e. on $\ket{\phi_1}$. Clearly, we have also $w\del_w \ket{\phi_0(w)}=a_0\ket{\phi_0(w)}$. It follows that the time variation of $\ket{\phi_0(w)}$ just cancels that coming from the coefficient $C_0(w)$ in Eq.~(\ref{expinst}), so that at late time, $\ket{\L(w)}$ is a stationary state. The time dependence of  the moments $\L_0$ and $\L_1$ can be extracted from the large component $\ket{\phi_0(w)}$, and one recovers the asymptotic  relation $\L_1/\L_0=-b_1/w$. 

\section{Analytical solution for $\Delta\ne 0$}\label{sec:analsol}
 
In this appendix, we provide details on the analytic solution of the equation (\ref{eqforbg0w}) for $g(w)$, namely
\be
\label{eq:dif_gw}
\frac{\rmd g}{\rmd \ln w}+g^2+\left(a_0+a_1+w\right)g+a_1a_0-c_0b_1+ a_0 w=0\,.
\ee
We have set here $\Delta=1$. As discussed after Eq.~(\ref{FSb}) the solution for a general value of $\Delta$ can be obtained from a simple rescaling of the parameters, which is easy to implement on the analytic solution. 

\subsubsection{Solution in terms of confluent geometric functions}
The first step towards the solution is to transform the first order, non linear differential equation (\ref{eq:dif_gw}) into a second order linear differential   equation. This is done with the help of an auxialiary function $y(w)$ related to $g(w) $ by 
\be
\label{eq:variable}
g(w) + a_0 = b-a -w + w \frac{y'( w)}{y( w)}
\ee
where the prime indicates a derivative with respect to $ w$. With the parameters $a$ and $b$  in \Eq {eq:variable} chosen to satisfy 
\begin{align}\label{solab}
2a-b =& 1 + a_1-a_0\,,\cr
(a-b)(a-1)  = & -b_1 c_0\,,
\end{align}
the equation (\ref{eq:dif_gw}) becomes 
the following second order ODE
\be
 w y'' + y'(b- w) - a y=0\,,
\ee
This is known as Kummer's equation, which is solved by the confluent hypergeometric functions $M(a,b, w)$ and $U(a,b, w)$\cite{10.5555/1098650}. Some properties of these functions are recalled in Appendix~\ref{app:sec1}. 
Eq.~(\ref{solab}) has two sets of solutions,
\be
a_\pm = 1 - (g_\mp + a_0 )\,,\qquad
b_{\pm} = 1\pm(g_+-g_-)\,.
\ee
However, a simple argument reveals that the physically meaningful solution  corresponds to the choice $a_+, b_+$. 
To see that, we note that in the late hydrodynamical regime the total entropy increases as $\delta S \propto \tau s$, where $\tau$ measures the proper volume and $s$ is the entropy density. Thus
\be
\frac{\tau}{S}\frac{\rmd S}{\rmd \tau}=1+ \frac{\rmd\ln s}{\rmd \ln\tau} =1+ 
\frac{3}{4} \frac{\rmd \ln \epsilon}{\rmd \ln \tau}\ge0\,, 
\ee
where in the last step we have used the ideal equation of state $\epsilon\sim s^{4/3}$.  It follows that 
\be
g(w) + a_0 \ge 0
\ee
Noting that $g_\pm+a_0 =b_\pm-a_\pm$, 
this condition translates into, 
\be
a_\pm-b_\pm\le 0\,,
\ee
which is only satisfied by $a_+$ and $b_+$.   Thus, from now on, we set $a=a_+$ and $b=b_+$. 

By using the relation of $y(  w)$ to $g(w)$ in \Eq{eq:variable}, and exploiting recursion relations and derivative properties of the confluent hypergeometric functions, \Eq{eq:recu_M}, one easily finds
\be
\label{eq:gsol2b}
g(w) = g_+-w + a w
\frac{\frac{1}{b}M\left(1+a,1+b,  w\right)- A U\left(1+a,1+b, w\right)}
{M\left(a,b,  w\right)+ A U\left(a,b,  w\right)}\,,
\ee
where $A$ is a constant  to be determined by the initial conditions. The attractor solution corresponds to $A=0$. It smoothly joins the free streaming fixed point at $w=0$  to the hydrodynamic fixed point at large $w$.



The solution (\ref{eq:gsol2b}}) holds for $\Delta>0$ only. Indeed, the  function $U\left(a,b, w\right)$ has a branch cut on the negative real axis. In order to extend the solution  to  negative $\Delta$ (and $w>0$), we need to avoid the corresponding singularity. This can be achieved via the transform $\Delta \to -\Delta$ in the solution \Eq{eq:gsol2}, namely,
\be
a=1-\frac{g_--g_*}{\Delta} \to \t a=1+\frac{g_--g_*}{\Delta}\,,\qquad
b=1+\frac{G}{\Delta} \to \t b = 1- \frac{G}{\Delta}
\ee
and $w/\Delta\to -w/\Delta$, so that for a negative $\Delta$
\begin{align}
\label{eq:sol_g0}
g(w) = g_+ 
-w 
+\t  a w
\frac{\frac{1}{\t b}{M\left(1+\t a,1+\t b,-\frac{w}{\Delta}\right)}- A U\left(1+\t a,1+\t b,-\frac{w}{\Delta}\right)}
{M\left(\t a,\t b,-\frac{w}{\Delta}\right)+ A U\left(\t a,\t b,-\frac{w}{\Delta}\right)}\,.
\end{align}
In terms of $w$, it is not difficult to show that the solutions in Eqs.~(\ref{eq:gsol2}) and (\ref{eq:sol_g0})  are identical, if the absolute value of $\Delta$ is the same in the two cases. In terms of $\tau$, Eq.~(\ref{eq:gsol2b}) characterizes the time evolution  toward local equilibrium, while \Eq{eq:sol_g0} does the opposite and describes the evolution toward the collisionless regime. 

%
%

\subsubsection{Relation to Wittaker's functions}\label{sec;Witt}

In Ref.~\cite{Denicol:2017lxn} a solution similar to that presented here was given, for the case of constant $\tau_R$, in terms of Wittaker's functions $M_{\kappa,\mu}(z)$. These functions are simply related to the confluent geometrical functions\cite{10.5555/1098650}:
\beq
M_{\kappa,\mu}(z)= \rme^{-z/2} z^{\mu+1/2} M(\mu+1/2-\kappa),1+2\mu,z). 
\eeq
The connection between the parameters in \cite{Denicol:2017lxn} and those of the present solution are as follows
\beq
\kappa=-\frac{1}{2} (\lambda+1),\qquad \mu=\frac{1}{2}\sqrt{4 a_{DN}+\lambda^2   },
\eeq
with
\beq
\lambda=a_1-a_0,\qquad  a_{DN}=c_0 b_1.
\eeq
It follows that 
\beq
1+2\mu= b_+=b,\qquad \frac{1}{2}+\mu-\kappa =a_+=a.
\eeq
The solution for the function $y$ introduced in \cite{Denicol:2017lxn} reads (to within an irrelevant multiplicative constant)
\beq\label{solDN}
y(w)=w^{-(1+\lambda)/2}\,\rme^{-w/2}\left( M_{\kappa,\mu}(w)+A W_{\kappa,\mu}(w)  \right),\qquad  \frac{w}{y}\frac{\rmd y}{\rmd w}=a_0+g_0.
\eeq
In terms of confluent geometrical functions, this is 
\beq
y(w)&=& w^{-(1+\lambda)/2}\,\rme^{-w/2}\, \rme^{-w/2} w^{\mu+1/2}\left[ M(a,b,w)+A U(a,b,w) \right]\nn
&=& w^\alpha \rme^{-w} \left[ M(a,b,w)+A U(a,b,w) \right],
\eeq
with $\alpha=b-a$. From there a simple change of variables allows one to identify the solution (\ref{solDN})  to that given in Eq.~(\ref{eq:gsol2}) above.


\section{Simple expansions for the attractor solution}\label{sec:expansions}

Let us first recall that the function $M(a,b,z)$ is an entire function of $z$ with the following expansion in powers of $z$ (see Eq.~(\ref{eq:1f1exp})):
\beq\label{seriesM}
M(a,b,z)=1+\frac{a}{b} z+\frac{(a)_2}{(b)_2} \frac{z^2}{2!}+\cdots +\frac{(a)_n}{(b)_n} \frac{z^n}{n!}+\cdots
\eeq
where
$(a)_n\equiv a(a+1)\cdots (a+n-1),\; (a)_0=1. 
$ 
For the forthcoming discussion, it is convenient to keep the factors $\Delta$ explicit. We have 
\beq\label{abamb}
&&a= 1 - \frac{g_- + a_0 }{\Delta}=1-\frac{1}{2\Delta}\left[ a_0-a_1-G \right]>1,\nn
&&b = 1+\frac{g_+-g_-}{\Delta}=1+\frac{G}{\Delta} >1,\nn
&&b-a=\frac{g_++a_0}{\Delta}=\frac{1}{2\Delta}\left[ a_0-a_1+G  \right]>0.
\eeq
These inequalities indicate that we are in the situation where the function $M(a,b,z)$ has no zero on the positive real $z$ axis \cite{DLMF}. 

 By keeping the first few terms of the expansion of $M(a+1,b+1,z)/M(a,b,z)$ in powers of $z$ and replacing  $z$ by $w/\Delta$, we get the following expansion for Eq.~(\ref{eq:expgatt})
\beq\label{attw3}
g_{\rm att}(w)\simeq g_+ -(g_++a_0)\frac{w}{b\Delta}\left[ 1   -\frac{ a}{ (b+1)}\frac{w}{b\Delta}+\frac{ a(2a-b)}{ (b+1) (b+2)}\frac{w^2}{b^2\Delta^2}\right].
\eeq
This expression suggests that, at least for the first few orders,  the expansion is in powers of $w/(b\Delta)$, with $b\Delta=\Delta+g_+-g_-$. This remark allows us to understand the limits of small and large $\Delta$ in simple terms. 

Consider first the limit $\Delta\to 0$. In this case, $b\Delta\simeq g_+-g_-=\sqrt{(a_0-a_1)^2+4b_1c_0  }$. Furthermore, in that limit, $a$ and $b$ are large, $a\sim -(g_-+a_0)/\Delta$ and $b\sim (g_+-g_-)\Delta$. 
One can then verify on the expression above that the terms of order $w^2$ and $w^3$ in $g_{\rm att}(w)$
coincide with the expansion of $g_+(w)$ (Eq.~(\ref{FPgpm})) up to order $w^3$, that is\footnote{A mismatch starts to occur at order $w^4$.}
\beq
g_{\rm att}(w)&\simeq& g_+-\frac{g_++a_0}{g_+-g_-}w+\frac{b_1 c_0}{(g_+-g_-)^3}w^2+\frac{(a_0-a_1)b_1 c_0}{(g_+-g_-)^5}w^3.
\eeq
Identifying the function $g_+(w)$ as an explicit limit from known analytic expressions of the function $M(a,b,z)$ turns out to be quite involved. However, we have checked that  when $\Delta\to 0$, the attractor is perfectly reproduced numerically   by the adiabatic approximation (\ref{E4}), as can be seen in Fig.~\ref{fig:AttSmallDelta}.  
 
 It is also interesting to consider the limit of a large $\Delta$. We have seen in the main text that when $\Delta$ is large, the function $g(w)$ has a travelling wave structure, the transition region evolving proportionally to $\Delta$.  It is easy to see how this emerges from the expansion (\ref{attw3}), by considering the  limit $\Delta\to\infty$, with $w/\Delta$ fixed. In this limit, we can substitute $a\to 1$,  $b\to 1$, except in the factor $a-b\to -\frac{g_++a_0}{\Delta}$. On sees then that the function becomes a function of $\bar w$, that is, the entire $\Delta $ dependence is in the scaling of $w$. Thus, in the limit $\Delta\to\infty$, with $w/\Delta$ fixed,  the attractor becomes a simple function of $\bar w$, whose first terms in the small $\bar w$ expansion read
 \beq
g_{\rm att}(w)\simeq g_+ -(g_++a_0)\bar w\left[ 1   -\frac{ 1}{ 2}\bar w+\frac{1}{ 6}\bar w^2\right],
\eeq
with the next term in the expansion easily shown to be $-\frac{\bar w^3}{4!}$ and 
$\frac{\bar w^4}{5!}$. One recognizes the expansion of the exponential and one recovers the result of the main text, Eq.~(\ref{glargeD}),

Finally consider the limit $w\to\infty$, at fixed $\Delta$. To study this regime, we use the following asymptotic expansion of $M(a,b,z)$ valid for fixed $a,b$ and large positive $z$ (see Eq.~(\ref{eq:1f1_asymp}))
\beq
M(a,b,z)\simeq \frac{\Gamma(b)}{\Gamma(a)}\rme^z z^{a-b} \sum_{n=0} \frac{(b-a)_n(1-a)_n}{n!} \frac{1}{z^n}.
\eeq
At large $z$, and  fixed $\Delta$ (that is, fixed $a$ and $b$) we have then
\beq\label{hydroexp}
\frac{a}{b}\frac{M(a+1,b+1,z)}{M(a,b,z)}\simeq  1+\frac{a-b}{z} \left[  1 +\frac{1-a}{z}\left[1 + \frac{2-2a+b}{z}\right]\right].
\eeq
The result quoted in the main text, Eq.~(\ref{attw2D}), follows immediately from this formula. 

%

\section{Asymptotic expansions and trans-series}\label{trans}

We now discuss the asymptotic expansions of the analytical solution in both the limit $w\to 0^+$ and $w\to \infty$. For simplicity, formulae will be written explicitly for $\Delta =1$, but we shall occasionally comment on their limits  for small or large values of $\Delta$. 

\subsection{Trans-series solution when $w\to 0^+$}


By using the  expansions of the confluent hypergeometric functions given  in Eqs.~(\ref{eq:1f1exp}) and (\ref{eq:Uexp}), one may expand the analytical solution (\ref{eq:gsol2})  for arbitrary small $w$. After some algebra the analytical solution can be rewritten as,
\begin{align}\label{eq:anaII}
g(w) = g_+ - w + a w \frac{R_1 M(1+a,1+b,w) + w^{-b} S_1 M(1+a-b,1-b,w)}{R_2 M(a,b,w) + w^{1-b} S_2 M(1+a-b,2-b,w)}\,,
\end{align}
where the four constants are
\begin{subequations}
\begin{align}
R_1 &= \frac{1}{b} R_2,\qquad 
R_2 = 1 + A \frac{\Gamma(1-b)}{\Gamma(1+a-b)} \\
S_1 &= -\frac{A}{a}\frac{\Gamma(b)}{\Gamma(a) },\qquad\quad\quad
S_2=-\frac{A}{1-b}\frac{\Gamma(b)}{\Gamma(a)}
\end{align}
\end{subequations}
The attractor solution is recovered for  $A=0$, i.e.,   $S_1=S_2=0$, $R_1=1/b$ and $R_2=1$. One recovers then Eq.~(\ref{eq:expgatt}).
However, for general initial conditions, with $A\ne 0$, the analytical solution has a singular contribution originating from the factor $w^{-b}$ present in both the denominator and the numerator in Eq.~(\ref{eq:anaII}). Because $b= 1 + G>1$, where $G=g_+-g_-$,  in the limit $w\to 0^+$, one can reorganize  \Eq{eq:anaII} as a double  expansion in powers of $w^{b-1}$ and $w$. This yields the following trans-series for $g(w)$
\begin{align}\label{transsmallw}
g(w)& = g_+ - w +  a \frac{S_1}{S_2} \frac{M(1+a-b,1-b,w)}{M(1+a-b,2-b,w)} \cr
&\times \left(1 + \frac{w R_1}{S_1} \frac{M(1+a,1+b,w)}{M(1+a-b,1-b,w)} w^{b-1}\right)
\left[1+ \frac{R_2}{S_2} \frac{M(a,b,w)}{M(1+a-b,2-b,w)} w^{b-1}+\cdots\right]\cr
& = \sum_{m=0} w^{m(b-1)} \sum_{n=0} \gamma^{(m)}_nw^n\,.
\end{align}
One recognizes  in this expansion  the typical non analytic contribution $\sim w^{b-1}=w^G$  identified in  perturbation theory (see Eq.~(\ref{presasymphi1b})).

Note that, since 
\be
\frac{S_1}{S_2} = \frac{1-b}{a},
\ee
we have
\be
g(w,A\ne 0) \xrightarrow{w\to 0} \gamma_0^{(0)}=g_+ + 1-b = g_- \,,
\ee
independently of the value of $A$.
It follows in particular that all the solutions, except the attractor,  start at the unstable fixed point $g_-$ at $w=0$. Note that the general solutions may present a pole singularity at small $w$ whenever the initial condition is such that $A<0$ (see the discussion at the end of Sect.~\ref{sec:nonlinear}). Note also that the first term in the trans-series, namely the first line of Eq.~(\ref{transsmallw}) has a finite radius of convergence: the denominator $M(1+a-b,2-b,w)$ indeed vanishes for $w\simeq 0.331$. 

 In fact, we can push the analysis a bit further, and look at the limit of small $\Delta$. 
Because $b$  becomes large when $\Delta$ is small (see Eqs.~(\ref{abamb})), we anticipate a collapse of the trans-series to its leading term, i.e. the first line of Eq.~(\ref{transsmallw}). As we have just argued, the convergence of the ratio of the two $M$ functions is limited by the zero of the denominator.  However, for $\Delta=0.642063$, corresponding to the value $2-b=-1$ where $M(1+a-b,2-b,w)$ has a simple pole, the zero of the denominator disappears, and the behavior of the function changes qualitatively.  There is then a delicate competition between the numerator and the denominator and for $w\gtrsim 0.6$ the solution jumps from $g_-(w)$ to $g_+(w)$, in very much the same way as the Borel sum of the hydrodynamic gradient expansion does, albeit at a slightly different value of $w$ (see Fig.~\ref{fig:gradexpBS} below). 

\subsection{Trans-series solution when $w\to +\infty$}\label{sec:translargew}

To perform this analysis, it is convenient to  write the equation for $g$ in terms of $\chi=g(w)+a_0$, that is
 \beq\label{eqforbg0b3bb3}
\frac{\rmd \chi}{\rmd \ln w}+\chi^2+(a_1-a_0)\chi -c_0 b_1+ \chi w =0.
\eeq
The advantage of this writing is that the coefficients $a_0,a_1, c_0, b_1$ enters in combinations $a_0-a_1$ and $c_0 b_1$ which have simple expressions    in terms of the parameters $a$ and $b $ of the $M$ function (see Eqs.~(\ref{solab})). The attractor solution takes then the from
\beq
\chi_{\rm att}=\chi_+ -w+\frac{w}{M}\frac{\rmd M}{\rmd w},\qquad \chi_+=b-a,
\eeq
where $\chi_+=g_++a_0$. The hydrodynamic fixed point corresponds here to $\chi_*=0$. 

Using the asymptotic expansion of the confluent hypergeometric functions (cf. Eqs.~(\ref{eq:1f1_asymp}) and (\ref{eq:asymp_U})),  one obtains the following asymptotic expansion of the analytical solution 
\begin{align}
\label{eq:aexp}
\chi(w)\to&
\chi_+ - w 
+   \frac{   w {\cal F}(-a,b-a,w) - a\sigma \frac{\zeta(w)}{w}  {\cal F}(1+a,1+a-b,-w)   }
{ {\cal F}(1-a,b-a,w) + \sigma \frac{\zeta(w)}{w} {\cal F}(a,1+a-b,-w)  },
\end{align}                         
where the function $\F(a,b,z)$ is given by the asymptotic series,
\be\label{calFdef}
\F(a,b,z) = \sum_{k=0} \frac{\Gamma(a+k)\Gamma(b+k)}{\Gamma(a)\Gamma(b)} \frac{z^{-k}}{k!}=\sum_{k=0} {\cal F}_k(a,b) \frac{1}{z^k}\,.
\ee
The expansion parameter $\sigma$ is a complex constant depending on the constant $A$,
\be\label{sigmaA}
\sigma = \frac{\Gamma(a)}
{\Gamma(b)}\left[e^{ i \pi a}  \frac{\Gamma(b)}{\Gamma(b-a)} + A\right]  
= \frac{A \Gamma(a)}{\Gamma(b)} + e^{ i\pi a}\frac{ \Gamma(a)}{\Gamma(b-a)}
\ee
and $\Im\sigma= \Gamma(a)/\Gamma(b-a) \sin (\pi a)$. In Eq.~(\ref{eq:aexp}), it is accompanied by the function
\be\label{zetaw}
\zeta(w) = e^{-w}\,w^{b-2a+1}=e^{-w}\,w^{a_0-a_1}\,,
\ee
which characterizes the  small (when  $w\to +\infty$) exponential corrections. The expansion with respect to $\sigma$ (or $\zeta(w))$ gives rise to  a trans-series 
\beq\label{translargew}
\chi(w)=\sum_{m=0}^\infty \sigma^m \chi^{(m)}(w),\qquad \chi^{(m)} (w)=\zeta^m\sum_k f_k^{(m)} \frac{1}{w^k}=\zeta^m f^{(m)}(w).
\eeq

The coefficients of the trans-series obtained by expanding Eq.~(\ref{eq:aexp}) in powers of $\sigma$ can be checked by a direct evaluation obtained by plugging the ansatz (\ref{translargew}) in Eq.~(\ref{eqforbg0b3bb3}), using for $\zeta(w)$ the general form 
\beq
\zeta(w)= \rme^{-Sw} w^\beta,\qquad  w\frac{\rmd \zeta^n}{\rmd w}=n(-Sw+\beta)\zeta^n.
\eeq
This yields the recursion relation 
\beq\label{recursionrel}
n(-Sw+\beta) f^{(n)}(w) +w\frac{\rmd  f^{(n)}(w)} {\rmd w}+\sum_{p=0}^n f^{(p)}(w)f^{(n-p)}(w)+(a_1-a_0+w)f^{(n)}-c_0 b_1 \delta^{n0}=0.\nn
\eeq
It is easily verified that the first terms in this recursion yields $S=1$ and $\beta=a_0-a_1$, in agreement with Eq.~(\ref{zetaw}). We shall exploit further this recursion relation in the rest of this section.

\subsubsection{The gradient expansion}

The leading order in the trans-series corresponds to $n=0$ 
\be
\label{eq:ghydro}
\chi_{\rm hydro}(w) \equiv \chi^{(0)}(w) = 
\chi_+ -  w\left(1-\frac{{\cal F}(-a,b-a,w)}{{\cal F}(1-a,b-a,w)}\right)\,,
\ee
and can be identified to the hydrodynamic gradient expansion:
\be
\chi_{\rm hydro}(w) = \sum_{n=0} f_n^{(0)} w^{-n}\,.
\ee
One may verify that the coeficients $f_n^{(0)}$ obtained either from expanding (\ref{eq:ghydro}) in powers of $1/w$, or by solving the recursion relation (\ref{recursionrel}) for $n=0$, yield identical results. The first few coefficients are recalled here for completeness  
\beq\label{coeff0}
&& f_0^{(0)}=0, \qquad f_1^{(0)}=c_0b_1,\qquad f_2^{(0)}=f_1^{(0)}(1+a_0-a_1),\nn
&& f_3^{(0)}=c_0b_1\left[  (1+a_0-a_1)(2+a_0-a_1)-(c_0b_1)\right].
\eeq


\subsubsection{Borel sums}

A standard tool in the analysis of asymptotic series is the Borel summation technique. In the present case, the Borel transform of the function ${\cal F}(a,b,w)$ is known analytically:
\beq
\sum_k {\cal F}_k(a,b)\frac{1}{k!}\frac{1}{z^k}= {}_2F_1(a,b,1,z),
\eeq
where ${}_2F_1(a,b,1,z)$ is a hypergeometric function, which  has a branch cut running from $z=1$ to $\infty$. 
The Borel sum  $\tilde {\cal F}$  is  the inverse Laplace transform of  ${}_2F_1(a,b,1,z)$, whose analytical expression reads~\cite{PhysRevA.32.1341}
\begin{align}\label{tildeF}
\frac{\t{\cal F}(a,b,z)}{\pi \csc[(a-b)\pi]} =& -\frac{e^{-i a \pi} z^a M(a,1+a-b,-z)}{\Gamma(b)\Gamma(1+a-b)}
 + \frac{e^{-i b \pi} z^b M(b,1-a+b,-z)}{\Gamma(a)\Gamma(1-a+b)}\,.
\end{align}
The Borel summation  is here quite efficient. One can indeed verify that the substitution of the function ${\cal F}$ by its Borel sum $\tilde {\cal F}$ in the asymptotic expression Eq.~(\ref{eq:aexp}) reconstructs the exact solution. It follows that the Borel sum of the hydrodynamic gradient expansion is given by Eq.~(\ref{eq:ghydro}) in which such a substitution has been made. 



\subsubsection{Trans-asymptotic matching}

We now return to the trans-series (\ref{translargew}) and reorder it as a series in powers of $1/w^n$, including at each order the complete set of exponential corrections. That is we write 
\beq\label{tranasym}
\chi(w)=\sum_k  \frac{1}{w^k} F_k(\sigma\zeta) ,\qquad F_k(\sigma\zeta)=\sum_{m=0}^\infty \sigma^m \zeta^m f^{(m)}_k.
\eeq
The coefficients $f^{(m)}_k$ can be obtained from the  recursion relation (\ref{recursionrel}). Keeping terms up to order $1/w^2$ one gets
\beq\label{transasymp0}
\chi(w)= \sigma \zeta f_0^{(1)}+\frac{1}{w} \left( f_1^{(0)}+\sigma \zeta  f_1^{(1)}+ \sigma^2 \zeta^2 f_1^{(2)}\right)+\frac{1}{w^2} \left( f_2^{(0)}+\sigma \zeta  f_2^{(1)}+\sigma^2 \zeta^2 f_2^{(2)}+\sigma^3 \zeta^3 f_2^{(3)  }\right)+\cdots\nn
\eeq
where coefficients not already given in Eq.~(\ref{coeff0}) are 
\beq\label{transasymp1}
&&f_0^{(1)}=-1,\quad f_1^{(1)}=-2c_0 b_1,\quad f_1^{(2)}=1,\quad f^{(1)}_2=-c_0 b_1 \left( 2c_0 b_1  +1+a_0-a_1  \right),\nn
&&    f^{(2)}_2=4c_0b_1+a_0-a_1-1, \quad f^{(3)}_2=-1. 
 \eeq
 
 The same expansion coefficients can be obtained by starting from the asymptotic expansion (\ref{eq:aexp}) and expanding in powers of $\sigma$. An important feature of this expansion is that $\zeta(w)$ always enters as the ratio $\zeta(w)/w$. It follows that each terms of the trans-series, when expanded in powers of $w^{-n}$ receives a finite number of exponential corrections (up to order $f_n^{(n+1)}$ for the term of order $1/w^n$). Thus the coefficients of the powers of $w^{-n}$ are, in this particular case, simple polynomials in $\sigma \zeta(w)$, instead of being themselves asymptotic series. The first polynomials are given in Eqs.~(\ref{transasymp0}) and (\ref{transasymp1}) above. 
 
 \begin{figure}[h]
\begin{center}
\includegraphics[angle=0,scale=0.75]{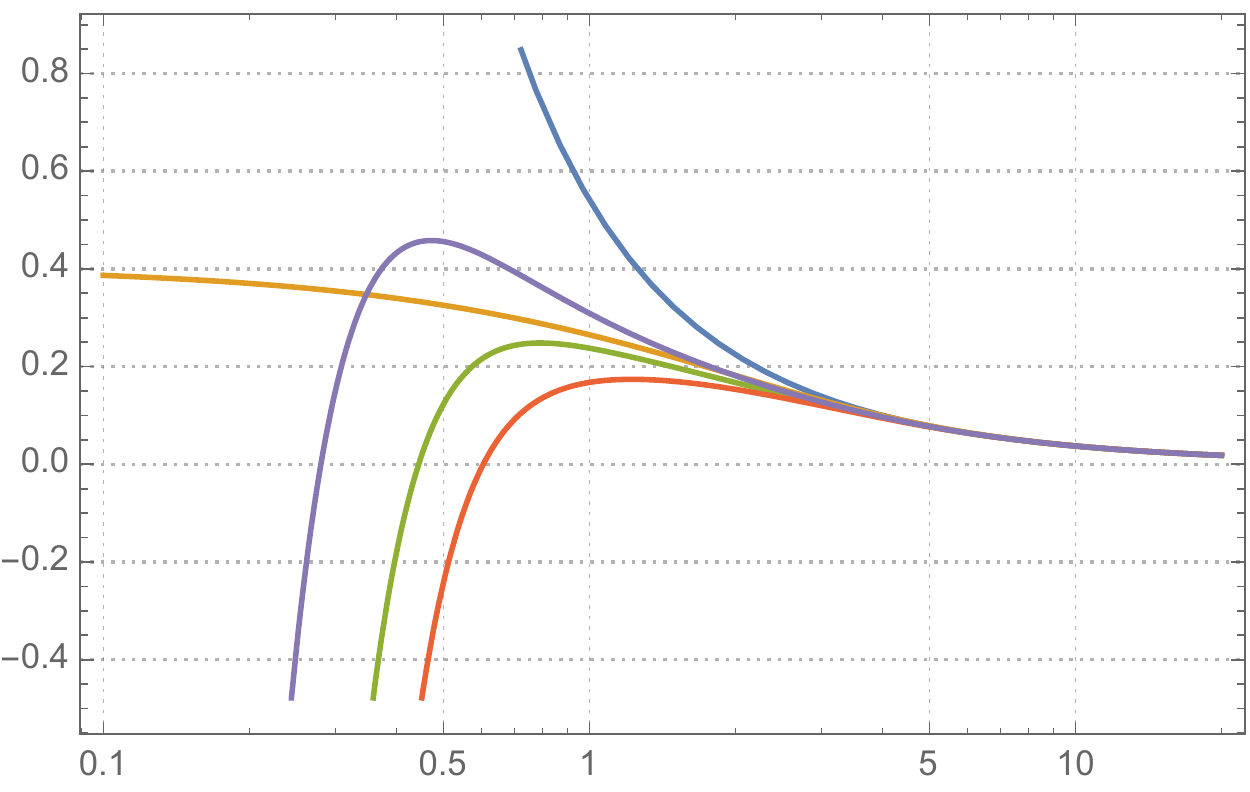} 
\end{center}
\caption{The transasymptic matching to order $1/w^2$ and $\sigma^2$. The orange curve is the exact attractor. The blue curve is the gradient expansion to order $1/w^2$. The green curve is obtained with $\sigma=\sigma_R +i \sigma_I$ after taking the real part. In the red curve we have chosen $\sigma_R+0.1$, in the purple curve $\sigma_R-0.1$, keeping the imaginary part to its value. Note that putting $\sigma_I=0$ does not change much the picture. \label{fig:transasymp} }
\end{figure}

 To appreciate the effects of these exponential corrections we have plotted some results in Fig.~\ref{fig:transasymp}. These curves are obtained by using the values of the real and imaginary parts of $\sigma$ deduced from Eq.~(\ref{sigmaA}) in which we set $A=0$. One can observe a sizeable improvement over the original gradient expansion for $1\lesssim w \lesssim 2$. We also see the effect of changing the initial condition by changing slightly the value of $\sigma_R$, with the curve moving above or below the (approximate) attractor depending on the sign of the correction to $\sigma_R$. Following the authors of \cite{Behtash_2019a} one may interpret the coefficients of the  powers of $w^{-n}$ as effective transport coefficients. This is dicussed in the main text (see Eq.~(\ref{effectiveeta2}) and the discussion in Sect.~\ref{sec:largewex}).

\section{Useful properties of the confluent hypergeometric functions}
\label{app:sec1}

%
%
General properties of the confluent geometric function can be found for instance in \cite{10.5555/1098650}. Here we gather a few relations that are used in the present paper. 

The confluent hypergeometric functions are solutions to the Kummer's differential equation, 
\be
z \frac{{\rmd}^2 f}{\rmd z^2}  + (b - z) \frac{\rmd f}{\rmd z} - a f = 0\,.
\ee
 The  confluent hypergeometric function of the first kind  is given as
\be
\label{eq:1f1exp}
M(a,b,z) \equiv {}_1F_1(a,b,z)= \sum_{n=0}^\infty \frac{(a)_n}{(b)_n} \frac{z^n}{n!}\,,
\ee
where the symbol $(a)_n$ stands for
\be
(a)_n = \prod_{k=0}^n(a-k+1)=\frac{\Gamma(a+n)}{\Gamma(a)}\,,\qquad
(a )_0 = 1\,.
\ee
 For large $|z|\rightarrow \infty$, the asymptotic expansion reads
\be
\label{eq:1f1_asymp}
\frac{M(a,b,z)}{\Gamma(b)}\sim \frac{e^z z^{a-b}}{\Gamma(a)} 
\sum_{n=0}^\infty \frac{(1-a)_n (b-a)_n}{n!} z^{-n}
+\frac{e^{ i\pi a} z^{-a}}{\Gamma(b-a)}
\sum_{n=0}^\infty \frac{(a)_n (a-b+1)_n}{n!} (-z)^{-n}
\ee
which holds for $-\frac{1}{2}\pi + \delta \leq {\rm arg}(z)\leq \frac{3}{2}\pi-\delta$, and when
$a \ne 0, -1, \ldots$ and $b-a\ne 0, -1,\ldots$. 
The first term is not needed when $\Gamma(b-a)$ is finite (that is, when $b-a$ differs from a non-positive integer) and the real part of $z$ goes to negative infinity, whereas the second term is not needed when $\Gamma(a)$ is finite (that is, when $a$ differs from a non-positive integer) and the real part of $z$ goes to positive infinity. 
The confluent hypergeometric function of the second kind, $U(a,b,z)$, is related to $M(a,b,z)$ by
\be
\label{eq:Uexp}
U(a,b,z)=\pi \csc (\pi b) \left[\frac{M(a,b,z)}{\Gamma(1+a-b)\Gamma(b)}-z^{1-b}\frac{M(1+a-b,2-b,z)}{\Gamma(a)\Gamma(2-b)}\right]\,,
\ee
and its asymptotic expansion reads
\be
\label{eq:asymp_U}
U(a,b,z)=z^{-a} \sum_n^\infty \frac{(a)_n(1+a-b)_n}{n!} (-z)^{-n}
\ee
The following   relations  are also useful 
\begin{align}
\label{eq:recu_M}
\frac{\rmd M(a,b,z)}{\rmd z} = \frac{a}{b} M(1+a,1+b,z), \qquad 
\frac{\rmd U(a,b,z)}{\rmd z} = -a U(1+a,1+b,z)\,.
\end{align}

\bibliography{refsbib2.bib}

\begin{thebibliography}{51}%
\makeatletter
\providecommand \@ifxundefined [1]{%
 \@ifx{#1\undefined}
}%
\providecommand \@ifnum [1]{%
 \ifnum #1\expandafter \@firstoftwo
 \else \expandafter \@secondoftwo
 \fi
}%
\providecommand \@ifx [1]{%
 \ifx #1\expandafter \@firstoftwo
 \else \expandafter \@secondoftwo
 \fi
}%
\providecommand \natexlab [1]{#1}%
\providecommand \enquote  [1]{``#1''}%
\providecommand \bibnamefont  [1]{#1}%
\providecommand \bibfnamefont [1]{#1}%
\providecommand \citenamefont [1]{#1}%
\providecommand \href@noop [0]{\@secondoftwo}%
\providecommand \href [0]{\begingroup \@sanitize@url \@href}%
\providecommand \@href[1]{\@@startlink{#1}\@@href}%
\providecommand \@@href[1]{\endgroup#1\@@endlink}%
\providecommand \@sanitize@url [0]{\catcode `\\12\catcode `\$12\catcode
  `\&12\catcode `\#12\catcode `\^12\catcode `\_12\catcode `\%12\relax}%
\providecommand \@@startlink[1]{}%
\providecommand \@@endlink[0]{}%
\providecommand \url  [0]{\begingroup\@sanitize@url \@url }%
\providecommand \@url [1]{\endgroup\@href {#1}{\urlprefix }}%
\providecommand \urlprefix  [0]{URL }%
\providecommand \Eprint [0]{\href }%
\providecommand \doibase [0]{http://dx.doi.org/}%
\providecommand \selectlanguage [0]{\@gobble}%
\providecommand \bibinfo  [0]{\@secondoftwo}%
\providecommand \bibfield  [0]{\@secondoftwo}%
\providecommand \translation [1]{[#1]}%
\providecommand \BibitemOpen [0]{}%
\providecommand \bibitemStop [0]{}%
\providecommand \bibitemNoStop [0]{.\EOS\space}%
\providecommand \EOS [0]{\spacefactor3000\relax}%
\providecommand \BibitemShut  [1]{\csname bibitem#1\endcsname}%
\let\auto@bib@innerbib\@empty
\bibitem [{\citenamefont {Heinz}\ and\ \citenamefont
  {Snellings}(2013)}]{Heinz_2013}%
  \BibitemOpen
  \bibfield  {author} {\bibinfo {author} {\bibfnamefont {Ulrich}\ \bibnamefont
  {Heinz}}\ and\ \bibinfo {author} {\bibfnamefont {Raimond}\ \bibnamefont
  {Snellings}},\ }\bibfield  {title} {\enquote {\bibinfo {title} {Collective
  flow and viscosity in relativistic heavy-ion collisions},}\ }\href {\doibase
  10.1146/annurev-nucl-102212-170540} {\bibfield  {journal} {\bibinfo
  {journal} {Annual Review of Nuclear and Particle Science}\ }\textbf {\bibinfo
  {volume} {63}},\ \bibinfo {pages} {123--151} (\bibinfo {year}
  {2013})}\BibitemShut {NoStop}%
\bibitem [{\citenamefont {Shen}\ and\ \citenamefont
  {Yan}(2020)}]{shen2020recent}%
  \BibitemOpen
  \bibfield  {author} {\bibinfo {author} {\bibfnamefont {Chun}\ \bibnamefont
  {Shen}}\ and\ \bibinfo {author} {\bibfnamefont {Li}~\bibnamefont {Yan}},\
  }\href@noop {} {\enquote {\bibinfo {title} {Recent development of
  hydrodynamic modeling in heavy-ion collisions},}\ } (\bibinfo {year}
  {2020}),\ \Eprint {http://arxiv.org/abs/2010.12377} {arXiv:2010.12377
  [nucl-th]} \BibitemShut {NoStop}%
\bibitem [{\citenamefont {Weller}\ and\ \citenamefont
  {Romatschke}(2017)}]{Weller:2017tsr}%
  \BibitemOpen
  \bibfield  {author} {\bibinfo {author} {\bibfnamefont {Ryan~D.}\ \bibnamefont
  {Weller}}\ and\ \bibinfo {author} {\bibfnamefont {Paul}\ \bibnamefont
  {Romatschke}},\ }\bibfield  {title} {\enquote {\bibinfo {title} {{One fluid
  to rule them all: viscous hydrodynamic description of event-by-event central
  p+p, p+Pb and Pb+Pb collisions at $\sqrt{s}=5.02$ TeV}},}\ }\href {\doibase
  10.1016/j.physletb.2017.09.077} {\bibfield  {journal} {\bibinfo  {journal}
  {Phys. Lett. B}\ }\textbf {\bibinfo {volume} {774}},\ \bibinfo {pages}
  {351--356} (\bibinfo {year} {2017})},\ \Eprint
  {http://arxiv.org/abs/1701.07145} {arXiv:1701.07145 [nucl-th]} \BibitemShut
  {NoStop}%
\bibitem [{\citenamefont {Romatschke}(2017)}]{Romatschke_2017}%
  \BibitemOpen
  \bibfield  {author} {\bibinfo {author} {\bibfnamefont {P.}~\bibnamefont
  {Romatschke}},\ }\bibfield  {title} {\enquote {\bibinfo {title} {Do nuclear
  collisions create a locally equilibrated quark--gluon plasma?}}\ }\href
  {\doibase 10.1140/epjc/s10052-016-4567-x} {\bibfield  {journal} {\bibinfo
  {journal} {The European Physical Journal C}\ }\textbf {\bibinfo {volume}
  {77}} (\bibinfo {year} {2017}),\ 10.1140/epjc/s10052-016-4567-x}\BibitemShut
  {NoStop}%
\bibitem [{\citenamefont {Romatschke}\ and\ \citenamefont
  {Romatschke}(2017)}]{romatschke2017relativistic}%
  \BibitemOpen
  \bibfield  {author} {\bibinfo {author} {\bibfnamefont {Paul}\ \bibnamefont
  {Romatschke}}\ and\ \bibinfo {author} {\bibfnamefont {Ulrike}\ \bibnamefont
  {Romatschke}},\ }\href@noop {} {\enquote {\bibinfo {title} {Relativistic
  fluid dynamics in and out of equilibrium -- ten years of progress in theory
  and numerical simulations of nuclear collisions},}\ } (\bibinfo {year}
  {2017}),\ \Eprint {http://arxiv.org/abs/1712.05815} {arXiv:1712.05815
  [nucl-th]} \BibitemShut {NoStop}%
\bibitem [{\citenamefont {Florkowski}\ \emph {et~al.}(2018)\citenamefont
  {Florkowski}, \citenamefont {Heller},\ and\ \citenamefont
  {Spalinski}}]{Florkowski:2017olj}%
  \BibitemOpen
  \bibfield  {author} {\bibinfo {author} {\bibfnamefont {Wojciech}\
  \bibnamefont {Florkowski}}, \bibinfo {author} {\bibfnamefont {Michal~P.}\
  \bibnamefont {Heller}}, \ and\ \bibinfo {author} {\bibfnamefont {Michal}\
  \bibnamefont {Spalinski}},\ }\bibfield  {title} {\enquote {\bibinfo {title}
  {{New theories of relativistic hydrodynamics in the LHC era}},}\ }\href
  {\doibase 10.1088/1361-6633/aaa091} {\bibfield  {journal} {\bibinfo
  {journal} {Rept. Prog. Phys.}\ }\textbf {\bibinfo {volume} {81}},\ \bibinfo
  {pages} {046001} (\bibinfo {year} {2018})},\ \Eprint
  {http://arxiv.org/abs/1707.02282} {arXiv:1707.02282 [hep-ph]} \BibitemShut
  {NoStop}%
\bibitem [{\citenamefont {Heller}\ \emph {et~al.}(2012)\citenamefont {Heller},
  \citenamefont {Janik},\ and\ \citenamefont {Witaszczyk}}]{Heller:2011ju}%
  \BibitemOpen
  \bibfield  {author} {\bibinfo {author} {\bibfnamefont {Michal~P.}\
  \bibnamefont {Heller}}, \bibinfo {author} {\bibfnamefont {Romuald~A.}\
  \bibnamefont {Janik}}, \ and\ \bibinfo {author} {\bibfnamefont {Przemyslaw}\
  \bibnamefont {Witaszczyk}},\ }\bibfield  {title} {\enquote {\bibinfo {title}
  {{The characteristics of thermalization of boost-invariant plasma from
  holography}},}\ }\href {\doibase 10.1103/PhysRevLett.108.201602} {\bibfield
  {journal} {\bibinfo  {journal} {Phys. Rev. Lett.}\ }\textbf {\bibinfo
  {volume} {108}},\ \bibinfo {pages} {201602} (\bibinfo {year} {2012})},\
  \Eprint {http://arxiv.org/abs/1103.3452} {arXiv:1103.3452 [hep-th]}
  \BibitemShut {NoStop}%
\bibitem [{\citenamefont {Heller}\ \emph {et~al.}(2013)\citenamefont {Heller},
  \citenamefont {Janik},\ and\ \citenamefont {Witaszczyk}}]{Heller:2013fn}%
  \BibitemOpen
  \bibfield  {author} {\bibinfo {author} {\bibfnamefont {Michal~P.}\
  \bibnamefont {Heller}}, \bibinfo {author} {\bibfnamefont {Romuald~A.}\
  \bibnamefont {Janik}}, \ and\ \bibinfo {author} {\bibfnamefont {Przemyslaw}\
  \bibnamefont {Witaszczyk}},\ }\bibfield  {title} {\enquote {\bibinfo {title}
  {{Hydrodynamic Gradient Expansion in Gauge Theory Plasmas}},}\ }\href
  {\doibase 10.1103/PhysRevLett.110.211602} {\bibfield  {journal} {\bibinfo
  {journal} {Phys. Rev. Lett.}\ }\textbf {\bibinfo {volume} {110}},\ \bibinfo
  {pages} {211602} (\bibinfo {year} {2013})},\ \Eprint
  {http://arxiv.org/abs/1302.0697} {arXiv:1302.0697 [hep-th]} \BibitemShut
  {NoStop}%
\bibitem [{\citenamefont {Heller}\ and\ \citenamefont
  {Spalinski}(2015)}]{Heller:2015dha}%
  \BibitemOpen
  \bibfield  {author} {\bibinfo {author} {\bibfnamefont {Michal~P.}\
  \bibnamefont {Heller}}\ and\ \bibinfo {author} {\bibfnamefont {Michal}\
  \bibnamefont {Spalinski}},\ }\bibfield  {title} {\enquote {\bibinfo {title}
  {{Hydrodynamics Beyond the Gradient Expansion: Resurgence and
  Resummation}},}\ }\href {\doibase 10.1103/PhysRevLett.115.072501} {\bibfield
  {journal} {\bibinfo  {journal} {Phys. Rev. Lett.}\ }\textbf {\bibinfo
  {volume} {115}},\ \bibinfo {pages} {072501} (\bibinfo {year} {2015})},\
  \Eprint {http://arxiv.org/abs/1503.07514} {arXiv:1503.07514 [hep-th]}
  \BibitemShut {NoStop}%
\bibitem [{\citenamefont {Behtash}\ \emph
  {et~al.}(2019{\natexlab{a}})\citenamefont {Behtash}, \citenamefont
  {Cruz-Camacho}, \citenamefont {Kamata},\ and\ \citenamefont
  {Martinez}}]{Behtash_2019a}%
  \BibitemOpen
  \bibfield  {author} {\bibinfo {author} {\bibfnamefont {Alireza}\ \bibnamefont
  {Behtash}}, \bibinfo {author} {\bibfnamefont {C.N.}\ \bibnamefont
  {Cruz-Camacho}}, \bibinfo {author} {\bibfnamefont {Syo}\ \bibnamefont
  {Kamata}}, \ and\ \bibinfo {author} {\bibfnamefont {M.}~\bibnamefont
  {Martinez}},\ }\bibfield  {title} {\enquote {\bibinfo {title}
  {Non-perturbative rheological behavior of a far-from-equilibrium expanding
  plasma},}\ }\href {\doibase 10.1016/j.physletb.2019.134914} {\bibfield
  {journal} {\bibinfo  {journal} {Physics Letters B}\ }\textbf {\bibinfo
  {volume} {797}},\ \bibinfo {pages} {134914} (\bibinfo {year}
  {2019}{\natexlab{a}})}\BibitemShut {NoStop}%
\bibitem [{\citenamefont {Berges}\ \emph {et~al.}(2020)\citenamefont {Berges},
  \citenamefont {Heller}, \citenamefont {Mazeliauskas},\ and\ \citenamefont
  {Venugopalan}}]{berges2020thermalization}%
  \BibitemOpen
  \bibfield  {author} {\bibinfo {author} {\bibfnamefont {J{\"u}rgen}\
  \bibnamefont {Berges}}, \bibinfo {author} {\bibfnamefont {Michal~P.}\
  \bibnamefont {Heller}}, \bibinfo {author} {\bibfnamefont {Aleksas}\
  \bibnamefont {Mazeliauskas}}, \ and\ \bibinfo {author} {\bibfnamefont {Raju}\
  \bibnamefont {Venugopalan}},\ }\href@noop {} {\enquote {\bibinfo {title}
  {Thermalization in qcd: theoretical approaches, phenomenological
  applications, and interdisciplinary connections},}\ } (\bibinfo {year}
  {2020}),\ \Eprint {http://arxiv.org/abs/2005.12299} {arXiv:2005.12299
  [hep-th]} \BibitemShut {NoStop}%
\bibitem [{\citenamefont {Blaizot}\ and\ \citenamefont
  {Yan}(2017)}]{Blaizot:2017lht}%
  \BibitemOpen
  \bibfield  {author} {\bibinfo {author} {\bibfnamefont {Jean-Paul}\
  \bibnamefont {Blaizot}}\ and\ \bibinfo {author} {\bibfnamefont
  {Li}~\bibnamefont {Yan}},\ }\bibfield  {title} {\enquote {\bibinfo {title}
  {{Onset of hydrodynamics for a quark-gluon plasma from the evolution of
  moments of distribution functions}},}\ }\href {\doibase
  10.1007/JHEP11(2017)161} {\bibfield  {journal} {\bibinfo  {journal} {JHEP}\
  }\textbf {\bibinfo {volume} {11}},\ \bibinfo {pages} {161} (\bibinfo {year}
  {2017})},\ \Eprint {http://arxiv.org/abs/1703.10694} {arXiv:1703.10694
  [nucl-th]} \BibitemShut {NoStop}%
\bibitem [{\citenamefont {Blaizot}\ and\ \citenamefont
  {Yan}(2018)}]{Blaizot:2017ucy}%
  \BibitemOpen
  \bibfield  {author} {\bibinfo {author} {\bibfnamefont {Jean-Paul}\
  \bibnamefont {Blaizot}}\ and\ \bibinfo {author} {\bibfnamefont
  {Li}~\bibnamefont {Yan}},\ }\bibfield  {title} {\enquote {\bibinfo {title}
  {{Fluid dynamics of out of equilibrium boost invariant plasmas}},}\ }\href
  {\doibase 10.1016/j.physletb.2018.02.058} {\bibfield  {journal} {\bibinfo
  {journal} {Phys. Lett. B}\ }\textbf {\bibinfo {volume} {780}},\ \bibinfo
  {pages} {283--286} (\bibinfo {year} {2018})},\ \Eprint
  {http://arxiv.org/abs/1712.03856} {arXiv:1712.03856 [nucl-th]} \BibitemShut
  {NoStop}%
\bibitem [{\citenamefont {Blaizot}\ and\ \citenamefont
  {Yan}(2020{\natexlab{a}})}]{Blaizot:2019scw}%
  \BibitemOpen
  \bibfield  {author} {\bibinfo {author} {\bibfnamefont {Jean-Paul}\
  \bibnamefont {Blaizot}}\ and\ \bibinfo {author} {\bibfnamefont
  {Li}~\bibnamefont {Yan}},\ }\bibfield  {title} {\enquote {\bibinfo {title}
  {{Emergence of hydrodynamical behavior in expanding ultra-relativistic
  plasmas}},}\ }\href {\doibase 10.1016/j.aop.2019.167993} {\bibfield
  {journal} {\bibinfo  {journal} {Annals Phys.}\ }\textbf {\bibinfo {volume}
  {412}},\ \bibinfo {pages} {167993} (\bibinfo {year} {2020}{\natexlab{a}})},\
  \Eprint {http://arxiv.org/abs/1904.08677} {arXiv:1904.08677 [nucl-th]}
  \BibitemShut {NoStop}%
\bibitem [{\citenamefont {Kovchegov}\ and\ \citenamefont
  {Taliotis}(2007)}]{Kovchegov_2007}%
  \BibitemOpen
  \bibfield  {author} {\bibinfo {author} {\bibfnamefont {Yuri~V.}\ \bibnamefont
  {Kovchegov}}\ and\ \bibinfo {author} {\bibfnamefont {Anastasios}\
  \bibnamefont {Taliotis}},\ }\bibfield  {title} {\enquote {\bibinfo {title}
  {Early time dynamics in heavy-ion collisions from ads/cft correspondence},}\
  }\href {\doibase 10.1103/physrevc.76.014905} {\bibfield  {journal} {\bibinfo
  {journal} {Physical Review C}\ }\textbf {\bibinfo {volume} {76}} (\bibinfo
  {year} {2007}),\ 10.1103/physrevc.76.014905}\BibitemShut {NoStop}%
\bibitem [{\citenamefont {Kurkela}\ \emph {et~al.}(2020)\citenamefont
  {Kurkela}, \citenamefont {van~der Schee}, \citenamefont {Wiedemann},\ and\
  \citenamefont {Wu}}]{Kurkela:2019set}%
  \BibitemOpen
  \bibfield  {author} {\bibinfo {author} {\bibfnamefont {Aleksi}\ \bibnamefont
  {Kurkela}}, \bibinfo {author} {\bibfnamefont {Wilke}\ \bibnamefont {van~der
  Schee}}, \bibinfo {author} {\bibfnamefont {Urs~Achim}\ \bibnamefont
  {Wiedemann}}, \ and\ \bibinfo {author} {\bibfnamefont {Bin}\ \bibnamefont
  {Wu}},\ }\bibfield  {title} {\enquote {\bibinfo {title} {{Early- and
  Late-Time Behavior of Attractors in Heavy-Ion Collisions}},}\ }\href
  {\doibase 10.1103/PhysRevLett.124.102301} {\bibfield  {journal} {\bibinfo
  {journal} {Phys. Rev. Lett.}\ }\textbf {\bibinfo {volume} {124}},\ \bibinfo
  {pages} {102301} (\bibinfo {year} {2020})},\ \Eprint
  {http://arxiv.org/abs/1907.08101} {arXiv:1907.08101 [hep-ph]} \BibitemShut
  {NoStop}%
\bibitem [{\citenamefont {Romatschke}(2018)}]{Romatschke:2017vte}%
  \BibitemOpen
  \bibfield  {author} {\bibinfo {author} {\bibfnamefont {Paul}\ \bibnamefont
  {Romatschke}},\ }\bibfield  {title} {\enquote {\bibinfo {title}
  {{Relativistic Fluid Dynamics Far From Local Equilibrium}},}\ }\href
  {\doibase 10.1103/PhysRevLett.120.012301} {\bibfield  {journal} {\bibinfo
  {journal} {Phys. Rev. Lett.}\ }\textbf {\bibinfo {volume} {120}},\ \bibinfo
  {pages} {012301} (\bibinfo {year} {2018})},\ \Eprint
  {http://arxiv.org/abs/1704.08699} {arXiv:1704.08699 [hep-th]} \BibitemShut
  {NoStop}%
\bibitem [{\citenamefont {Blaizot}\ and\ \citenamefont
  {Yan}(2020{\natexlab{b}})}]{Blaizot:2020gql}%
  \BibitemOpen
  \bibfield  {author} {\bibinfo {author} {\bibfnamefont {Jean-Paul}\
  \bibnamefont {Blaizot}}\ and\ \bibinfo {author} {\bibfnamefont
  {Li}~\bibnamefont {Yan}},\ }\bibfield  {title} {\enquote {\bibinfo {title}
  {{Analytical attractor for Bjorken expansion}},}\ }\href@noop {} {\
  (\bibinfo {year} {2020}{\natexlab{b}})},\ \Eprint
  {http://arxiv.org/abs/2006.08815} {arXiv:2006.08815 [nucl-th]} \BibitemShut
  {NoStop}%
\bibitem [{\citenamefont {Baier}\ \emph {et~al.}(2001)\citenamefont {Baier},
  \citenamefont {Mueller}, \citenamefont {Schiff},\ and\ \citenamefont
  {Son}}]{Baier:2000sb}%
  \BibitemOpen
  \bibfield  {author} {\bibinfo {author} {\bibfnamefont {R.}~\bibnamefont
  {Baier}}, \bibinfo {author} {\bibfnamefont {Alfred~H.}\ \bibnamefont
  {Mueller}}, \bibinfo {author} {\bibfnamefont {D.}~\bibnamefont {Schiff}}, \
  and\ \bibinfo {author} {\bibfnamefont {D.~T.}\ \bibnamefont {Son}},\
  }\bibfield  {title} {\enquote {\bibinfo {title} {{'Bottom up' thermalization
  in heavy ion collisions}},}\ }\href {\doibase 10.1016/S0370-2693(01)00191-5}
  {\bibfield  {journal} {\bibinfo  {journal} {Phys. Lett. B}\ }\textbf
  {\bibinfo {volume} {502}},\ \bibinfo {pages} {51--58} (\bibinfo {year}
  {2001})},\ \Eprint {http://arxiv.org/abs/hep-ph/0009237}
  {arXiv:hep-ph/0009237} \BibitemShut {NoStop}%
\bibitem [{\citenamefont {Blaizot}\ and\ \citenamefont
  {Tanji}(2019)}]{Blaizot:2019dut}%
  \BibitemOpen
  \bibfield  {author} {\bibinfo {author} {\bibfnamefont {Jean-Paul}\
  \bibnamefont {Blaizot}}\ and\ \bibinfo {author} {\bibfnamefont {Naoto}\
  \bibnamefont {Tanji}},\ }\bibfield  {title} {\enquote {\bibinfo {title}
  {{Angular mode expansion of the Boltzmann equation in the small-angle
  approximation}},}\ }\href {\doibase 10.1016/j.nuclphysa.2019.121618} {\
  (\bibinfo {year} {2019}),\ 10.1016/j.nuclphysa.2019.121618},\ \Eprint
  {http://arxiv.org/abs/1904.08244} {arXiv:1904.08244 [hep-ph]} \BibitemShut
  {NoStop}%
\bibitem [{\citenamefont {Denicol}\ and\ \citenamefont
  {Noronha}(2020)}]{Denicol:2019lio}%
  \BibitemOpen
  \bibfield  {author} {\bibinfo {author} {\bibfnamefont {Gabriel~S.}\
  \bibnamefont {Denicol}}\ and\ \bibinfo {author} {\bibfnamefont {Jorge}\
  \bibnamefont {Noronha}},\ }\bibfield  {title} {\enquote {\bibinfo {title}
  {{Exact hydrodynamic attractor of an ultrarelativistic gas of hard
  spheres}},}\ }\href {\doibase 10.1103/PhysRevLett.124.152301} {\bibfield
  {journal} {\bibinfo  {journal} {Phys. Rev. Lett.}\ }\textbf {\bibinfo
  {volume} {124}},\ \bibinfo {pages} {152301} (\bibinfo {year} {2020})},\
  \Eprint {http://arxiv.org/abs/1908.09957} {arXiv:1908.09957 [nucl-th]}
  \BibitemShut {NoStop}%
\bibitem [{\citenamefont {Chattopadhyay}\ and\ \citenamefont
  {Heinz}(2020)}]{Chattopadhyay:2019jqj}%
  \BibitemOpen
  \bibfield  {author} {\bibinfo {author} {\bibfnamefont {Chandrodoy}\
  \bibnamefont {Chattopadhyay}}\ and\ \bibinfo {author} {\bibfnamefont
  {Ulrich~W.}\ \bibnamefont {Heinz}},\ }\bibfield  {title} {\enquote {\bibinfo
  {title} {{Hydrodynamics from free-streaming to thermalization and back
  again}},}\ }\href {\doibase 10.1016/j.physletb.2019.135158} {\bibfield
  {journal} {\bibinfo  {journal} {Phys. Lett. B}\ }\textbf {\bibinfo {volume}
  {801}},\ \bibinfo {pages} {135158} (\bibinfo {year} {2020})},\ \Eprint
  {http://arxiv.org/abs/1911.07765} {arXiv:1911.07765 [nucl-th]} \BibitemShut
  {NoStop}%
\bibitem [{\citenamefont {Mazeliauskas}\ and\ \citenamefont
  {Berges}(2019)}]{Mazeliauskas_2019}%
  \BibitemOpen
  \bibfield  {author} {\bibinfo {author} {\bibfnamefont {Aleksas}\ \bibnamefont
  {Mazeliauskas}}\ and\ \bibinfo {author} {\bibfnamefont {J{\"u}rgen}\
  \bibnamefont {Berges}},\ }\bibfield  {title} {\enquote {\bibinfo {title}
  {Prescaling and far-from-equilibrium hydrodynamics in the quark-gluon
  plasma},}\ }\href {\doibase 10.1103/physrevlett.122.122301} {\bibfield
  {journal} {\bibinfo  {journal} {Physical Review Letters}\ }\textbf {\bibinfo
  {volume} {122}} (\bibinfo {year} {2019}),\
  10.1103/physrevlett.122.122301}\BibitemShut {NoStop}%
\bibitem [{\citenamefont {Bjorken}(1983)}]{Bjorken:1982qr}%
  \BibitemOpen
  \bibfield  {author} {\bibinfo {author} {\bibfnamefont {J.D.}\ \bibnamefont
  {Bjorken}},\ }\bibfield  {title} {\enquote {\bibinfo {title} {{Highly
  Relativistic Nucleus-Nucleus Collisions: The Central Rapidity Region}},}\
  }\href {\doibase 10.1103/PhysRevD.27.140} {\bibfield  {journal} {\bibinfo
  {journal} {Phys. Rev. D}\ }\textbf {\bibinfo {volume} {27}},\ \bibinfo
  {pages} {140--151} (\bibinfo {year} {1983})}\BibitemShut {NoStop}%
\bibitem [{\citenamefont {Baym}(1984)}]{Baym:1984np}%
  \BibitemOpen
  \bibfield  {author} {\bibinfo {author} {\bibfnamefont {G.}~\bibnamefont
  {Baym}},\ }\bibfield  {title} {\enquote {\bibinfo {title} {{Thermal
  equilibration in ultrarelativistc heavy ion collisions}},}\ }\href {\doibase
  10.1016/0370-2693(84)91863-X} {\bibfield  {journal} {\bibinfo  {journal}
  {Phys. Lett. B}\ }\textbf {\bibinfo {volume} {138}},\ \bibinfo {pages}
  {18--22} (\bibinfo {year} {1984})}\BibitemShut {NoStop}%
\bibitem [{\citenamefont {Arnold}\ \emph {et~al.}(2005)\citenamefont {Arnold},
  \citenamefont {Lenaghan}, \citenamefont {Moore},\ and\ \citenamefont
  {Yaffe}}]{Arnold_2005}%
  \BibitemOpen
  \bibfield  {author} {\bibinfo {author} {\bibfnamefont {Peter}\ \bibnamefont
  {Arnold}}, \bibinfo {author} {\bibfnamefont {Jonathan}\ \bibnamefont
  {Lenaghan}}, \bibinfo {author} {\bibfnamefont {Guy~D.}\ \bibnamefont
  {Moore}}, \ and\ \bibinfo {author} {\bibfnamefont {Laurence~G.}\ \bibnamefont
  {Yaffe}},\ }\bibfield  {title} {\enquote {\bibinfo {title} {Apparent
  thermalization due to plasma instabilities in the quark-gluon plasma},}\
  }\href {\doibase 10.1103/physrevlett.94.072302} {\bibfield  {journal}
  {\bibinfo  {journal} {Physical Review Letters}\ }\textbf {\bibinfo {volume}
  {94}} (\bibinfo {year} {2005}),\ 10.1103/physrevlett.94.072302}\BibitemShut
  {NoStop}%
\bibitem [{\citenamefont {Denicol}\ \emph {et~al.}(2012)\citenamefont
  {Denicol}, \citenamefont {Niemi}, \citenamefont {Molnar},\ and\ \citenamefont
  {Rischke}}]{Denicol:2012cn}%
  \BibitemOpen
  \bibfield  {author} {\bibinfo {author} {\bibfnamefont {G.S.}\ \bibnamefont
  {Denicol}}, \bibinfo {author} {\bibfnamefont {H.}~\bibnamefont {Niemi}},
  \bibinfo {author} {\bibfnamefont {E.}~\bibnamefont {Molnar}}, \ and\ \bibinfo
  {author} {\bibfnamefont {D.H.}\ \bibnamefont {Rischke}},\ }\bibfield  {title}
  {\enquote {\bibinfo {title} {{Derivation of transient relativistic fluid
  dynamics from the Boltzmann equation}},}\ }\href {\doibase
  10.1103/PhysRevD.85.114047} {\bibfield  {journal} {\bibinfo  {journal} {Phys.
  Rev. D}\ }\textbf {\bibinfo {volume} {85}},\ \bibinfo {pages} {114047}
  (\bibinfo {year} {2012})},\ \bibinfo {note} {[Erratum: Phys.Rev.D 91, 039902
  (2015)]},\ \Eprint {http://arxiv.org/abs/1202.4551} {arXiv:1202.4551
  [nucl-th]} \BibitemShut {NoStop}%
\bibitem [{\citenamefont {Behtash}\ \emph {et~al.}(2020)\citenamefont
  {Behtash}, \citenamefont {Kamata}, \citenamefont {Martinez}, \citenamefont
  {Schaefer},\ and\ \citenamefont {Skokov}}]{behtash2020transasymptotics}%
  \BibitemOpen
  \bibfield  {author} {\bibinfo {author} {\bibfnamefont {A.}~\bibnamefont
  {Behtash}}, \bibinfo {author} {\bibfnamefont {S.}~\bibnamefont {Kamata}},
  \bibinfo {author} {\bibfnamefont {M.}~\bibnamefont {Martinez}}, \bibinfo
  {author} {\bibfnamefont {T.}~\bibnamefont {Schaefer}}, \ and\ \bibinfo
  {author} {\bibfnamefont {V.}~\bibnamefont {Skokov}},\ }\href@noop {}
  {\enquote {\bibinfo {title} {Transasymptotics and hydrodynamization of the
  fokker-planck equation for gluons},}\ } (\bibinfo {year} {2020}),\ \Eprint
  {http://arxiv.org/abs/2011.08235} {arXiv:2011.08235 [hep-ph]} \BibitemShut
  {NoStop}%
\bibitem [{\citenamefont {Behtash}\ \emph
  {et~al.}(2019{\natexlab{b}})\citenamefont {Behtash}, \citenamefont {Kamata},
  \citenamefont {Martinez},\ and\ \citenamefont {Shi}}]{Behtash:2019txb}%
  \BibitemOpen
  \bibfield  {author} {\bibinfo {author} {\bibfnamefont {Alireza}\ \bibnamefont
  {Behtash}}, \bibinfo {author} {\bibfnamefont {Syo}\ \bibnamefont {Kamata}},
  \bibinfo {author} {\bibfnamefont {Mauricio}\ \bibnamefont {Martinez}}, \ and\
  \bibinfo {author} {\bibfnamefont {Haosheng}\ \bibnamefont {Shi}},\ }\bibfield
   {title} {\enquote {\bibinfo {title} {{Dynamical systems and nonlinear
  transient rheology of the far-from-equilibrium Bjorken flow}},}\ }\href
  {\doibase 10.1103/PhysRevD.99.116012} {\bibfield  {journal} {\bibinfo
  {journal} {Phys. Rev. D}\ }\textbf {\bibinfo {volume} {99}},\ \bibinfo
  {pages} {116012} (\bibinfo {year} {2019}{\natexlab{b}})},\ \Eprint
  {http://arxiv.org/abs/1901.08632} {arXiv:1901.08632 [hep-th]} \BibitemShut
  {NoStop}%
\bibitem [{\citenamefont {Pines}\ and\ \citenamefont
  {Nozi{\`e}res}(1966)}]{Pines}%
  \BibitemOpen
  \bibfield  {author} {\bibinfo {author} {\bibfnamefont {David}\ \bibnamefont
  {Pines}}\ and\ \bibinfo {author} {\bibfnamefont {Philippe}\ \bibnamefont
  {Nozi{\`e}res}},\ }\href@noop {} {\emph {\bibinfo {title} {The Theory of
  Quantum Liquids}}}\ (\bibinfo  {publisher} {W.A. Benjamin, inc},\ \bibinfo
  {year} {1966})\BibitemShut {NoStop}%
\bibitem [{\citenamefont {Baym}\ and\ \citenamefont
  {Pethick}(1991)}]{baym1991landau}%
  \BibitemOpen
  \bibfield  {author} {\bibinfo {author} {\bibfnamefont {Gordon}\ \bibnamefont
  {Baym}}\ and\ \bibinfo {author} {\bibfnamefont {Christopher}\ \bibnamefont
  {Pethick}},\ }\href@noop {} {\emph {\bibinfo {title} {Landau Fermi-liquid
  theory}}},\ Vol.~\bibinfo {volume} {1}\ (\bibinfo  {publisher} {Wiley Online
  Library},\ \bibinfo {year} {1991})\BibitemShut {NoStop}%
\bibitem [{\citenamefont {Israel}\ and\ \citenamefont
  {Stewart}(1979)}]{Israel:1979wp}%
  \BibitemOpen
  \bibfield  {author} {\bibinfo {author} {\bibfnamefont {W.}~\bibnamefont
  {Israel}}\ and\ \bibinfo {author} {\bibfnamefont {J.M.}\ \bibnamefont
  {Stewart}},\ }\bibfield  {title} {\enquote {\bibinfo {title} {{Transient
  relativistic thermodynamics and kinetic theory}},}\ }\href {\doibase
  10.1016/0003-4916(79)90130-1} {\bibfield  {journal} {\bibinfo  {journal}
  {Annals Phys.}\ }\textbf {\bibinfo {volume} {118}},\ \bibinfo {pages}
  {341--372} (\bibinfo {year} {1979})}\BibitemShut {NoStop}%
\bibitem [{\citenamefont {M{\"u}ller}(1967)}]{muller1967paradoxon}%
  \BibitemOpen
  \bibfield  {author} {\bibinfo {author} {\bibfnamefont {Ingo}\ \bibnamefont
  {M{\"u}ller}},\ }\bibfield  {title} {\enquote {\bibinfo {title} {Zum
  paradoxon der w{\"a}rmeleitungstheorie},}\ }\href@noop {} {\bibfield
  {journal} {\bibinfo  {journal} {Zeitschrift f{\"u}r Physik}\ }\textbf
  {\bibinfo {volume} {198}},\ \bibinfo {pages} {329--344} (\bibinfo {year}
  {1967})}\BibitemShut {NoStop}%
\bibitem [{\citenamefont {Denicol}\ and\ \citenamefont
  {Noronha}(2018)}]{Denicol:2017lxn}%
  \BibitemOpen
  \bibfield  {author} {\bibinfo {author} {\bibfnamefont {Gabriel~S.}\
  \bibnamefont {Denicol}}\ and\ \bibinfo {author} {\bibfnamefont {Jorge}\
  \bibnamefont {Noronha}},\ }\bibfield  {title} {\enquote {\bibinfo {title}
  {{Analytical attractor and the divergence of the slow-roll expansion in
  relativistic hydrodynamics}},}\ }\href {\doibase 10.1103/PhysRevD.97.056021}
  {\bibfield  {journal} {\bibinfo  {journal} {Phys. Rev.}\ }\textbf {\bibinfo
  {volume} {D97}},\ \bibinfo {pages} {056021} (\bibinfo {year} {2018})},\
  \Eprint {http://arxiv.org/abs/1711.01657} {arXiv:1711.01657 [nucl-th]}
  \BibitemShut {NoStop}%
\bibitem [{\citenamefont {Jaiswal}\ \emph {et~al.}(2019)\citenamefont
  {Jaiswal}, \citenamefont {Chattopadhyay}, \citenamefont {Jaiswal},
  \citenamefont {Pal},\ and\ \citenamefont {Heinz}}]{Jaiswal_2019}%
  \BibitemOpen
  \bibfield  {author} {\bibinfo {author} {\bibfnamefont {Sunil}\ \bibnamefont
  {Jaiswal}}, \bibinfo {author} {\bibfnamefont {Chandrodoy}\ \bibnamefont
  {Chattopadhyay}}, \bibinfo {author} {\bibfnamefont {Amaresh}\ \bibnamefont
  {Jaiswal}}, \bibinfo {author} {\bibfnamefont {Subrata}\ \bibnamefont {Pal}},
  \ and\ \bibinfo {author} {\bibfnamefont {Ulrich}\ \bibnamefont {Heinz}},\
  }\bibfield  {title} {\enquote {\bibinfo {title} {Exact solutions and
  attractors of higher-order viscous fluid dynamics for bjorken flow},}\ }\href
  {\doibase 10.1103/physrevc.100.034901} {\bibfield  {journal} {\bibinfo
  {journal} {Physical Review C}\ }\textbf {\bibinfo {volume} {100}} (\bibinfo
  {year} {2019}),\ 10.1103/physrevc.100.034901}\BibitemShut {NoStop}%
\bibitem [{\citenamefont {Baier}\ \emph {et~al.}(2008)\citenamefont {Baier},
  \citenamefont {Romatschke}, \citenamefont {Son}, \citenamefont {Starinets},\
  and\ \citenamefont {Stephanov}}]{Baier:2007ix}%
  \BibitemOpen
  \bibfield  {author} {\bibinfo {author} {\bibfnamefont {Rudolf}\ \bibnamefont
  {Baier}}, \bibinfo {author} {\bibfnamefont {Paul}\ \bibnamefont
  {Romatschke}}, \bibinfo {author} {\bibfnamefont {Dam~Thanh}\ \bibnamefont
  {Son}}, \bibinfo {author} {\bibfnamefont {Andrei~O.}\ \bibnamefont
  {Starinets}}, \ and\ \bibinfo {author} {\bibfnamefont {Mikhail~A.}\
  \bibnamefont {Stephanov}},\ }\bibfield  {title} {\enquote {\bibinfo {title}
  {{Relativistic viscous hydrodynamics, conformal invariance, and
  holography}},}\ }\href {\doibase 10.1088/1126-6708/2008/04/100} {\bibfield
  {journal} {\bibinfo  {journal} {JHEP}\ }\textbf {\bibinfo {volume} {04}},\
  \bibinfo {pages} {100} (\bibinfo {year} {2008})},\ \Eprint
  {http://arxiv.org/abs/0712.2451} {arXiv:0712.2451 [hep-th]} \BibitemShut
  {NoStop}%
\bibitem [{\citenamefont {Heller}\ and\ \citenamefont
  {Janik}(2007)}]{Heller:2007qt}%
  \BibitemOpen
  \bibfield  {author} {\bibinfo {author} {\bibfnamefont {Michal~P.}\
  \bibnamefont {Heller}}\ and\ \bibinfo {author} {\bibfnamefont {Romuald~A.}\
  \bibnamefont {Janik}},\ }\bibfield  {title} {\enquote {\bibinfo {title}
  {{Viscous hydrodynamics relaxation time from AdS/CFT}},}\ }\href {\doibase
  10.1103/PhysRevD.76.025027} {\bibfield  {journal} {\bibinfo  {journal} {Phys.
  Rev. D}\ }\textbf {\bibinfo {volume} {76}},\ \bibinfo {pages} {025027}
  (\bibinfo {year} {2007})},\ \Eprint {http://arxiv.org/abs/hep-th/0703243}
  {arXiv:hep-th/0703243} \BibitemShut {NoStop}%
\bibitem [{\citenamefont {Messiah}(1962)}]{messiah1962quantum}%
  \BibitemOpen
  \bibfield  {author} {\bibinfo {author} {\bibfnamefont {Albert}\ \bibnamefont
  {Messiah}},\ }\href@noop {} {\enquote {\bibinfo {title} {Quantum mechanics,
  vol. ii},}\ } (\bibinfo {year} {1962})\BibitemShut {NoStop}%
\bibitem [{\citenamefont {Brewer}\ \emph {et~al.}(2021)\citenamefont {Brewer},
  \citenamefont {Yan},\ and\ \citenamefont {Yin}}]{Brewer:2019oha}%
  \BibitemOpen
  \bibfield  {author} {\bibinfo {author} {\bibfnamefont {Jasmine}\ \bibnamefont
  {Brewer}}, \bibinfo {author} {\bibfnamefont {Li}~\bibnamefont {Yan}}, \ and\
  \bibinfo {author} {\bibfnamefont {Yi}~\bibnamefont {Yin}},\ }\bibfield
  {title} {\enquote {\bibinfo {title} {{Adiabatic hydrodynamization in
  rapidly-expanding quark\textendash{}gluon plasma}},}\ }\href {\doibase
  10.1016/j.physletb.2021.136189} {\bibfield  {journal} {\bibinfo  {journal}
  {Phys. Lett. B}\ }\textbf {\bibinfo {volume} {816}},\ \bibinfo {pages}
  {136189} (\bibinfo {year} {2021})},\ \Eprint
  {http://arxiv.org/abs/1910.00021} {arXiv:1910.00021 [nucl-th]} \BibitemShut
  {NoStop}%
\bibitem [{\citenamefont {Broniowski}\ \emph {et~al.}(2009)\citenamefont
  {Broniowski}, \citenamefont {Florkowski}, \citenamefont {Chojnacki},\ and\
  \citenamefont {Kisiel}}]{Broniowski_2009}%
  \BibitemOpen
  \bibfield  {author} {\bibinfo {author} {\bibfnamefont {Wojciech}\
  \bibnamefont {Broniowski}}, \bibinfo {author} {\bibfnamefont {Wojciech}\
  \bibnamefont {Florkowski}}, \bibinfo {author} {\bibfnamefont {Mikolaj}\
  \bibnamefont {Chojnacki}}, \ and\ \bibinfo {author} {\bibfnamefont {Adam}\
  \bibnamefont {Kisiel}},\ }\bibfield  {title} {\enquote {\bibinfo {title}
  {Free-streaming approximation in early dynamics of relativistic heavy-ion
  collisions},}\ }\href {\doibase 10.1103/physrevc.80.034902} {\bibfield
  {journal} {\bibinfo  {journal} {Physical Review C}\ }\textbf {\bibinfo
  {volume} {80}} (\bibinfo {year} {2009}),\
  10.1103/physrevc.80.034902}\BibitemShut {NoStop}%
\bibitem [{\citenamefont {Basar}\ and\ \citenamefont
  {Dunne}(2015)}]{Basar:2015ava}%
  \BibitemOpen
  \bibfield  {author} {\bibinfo {author} {\bibfnamefont {Gokce}\ \bibnamefont
  {Basar}}\ and\ \bibinfo {author} {\bibfnamefont {Gerald~V.}\ \bibnamefont
  {Dunne}},\ }\bibfield  {title} {\enquote {\bibinfo {title} {{Hydrodynamics,
  resurgence, and transasymptotics}},}\ }\href {\doibase
  10.1103/PhysRevD.92.125011} {\bibfield  {journal} {\bibinfo  {journal} {Phys.
  Rev. D}\ }\textbf {\bibinfo {volume} {92}},\ \bibinfo {pages} {125011}
  (\bibinfo {year} {2015})},\ \Eprint {http://arxiv.org/abs/1509.05046}
  {arXiv:1509.05046 [hep-th]} \BibitemShut {NoStop}%
\bibitem [{\citenamefont {Lublinsky}\ and\ \citenamefont
  {Shuryak}(2007)}]{Lublinsky:2007mm}%
  \BibitemOpen
  \bibfield  {author} {\bibinfo {author} {\bibfnamefont {Michael}\ \bibnamefont
  {Lublinsky}}\ and\ \bibinfo {author} {\bibfnamefont {Edward}\ \bibnamefont
  {Shuryak}},\ }\bibfield  {title} {\enquote {\bibinfo {title} {{How much
  entropy is produced in strongly coupled Quark-Gluon Plasma (sQGP) by
  dissipative effects?}}}\ }\href {\doibase 10.1103/PhysRevC.76.021901}
  {\bibfield  {journal} {\bibinfo  {journal} {Phys. Rev. C}\ }\textbf {\bibinfo
  {volume} {76}},\ \bibinfo {pages} {021901} (\bibinfo {year} {2007})},\
  \Eprint {http://arxiv.org/abs/0704.1647} {arXiv:0704.1647 [hep-ph]}
  \BibitemShut {NoStop}%
\bibitem [{\citenamefont {Dash}\ and\ \citenamefont
  {Roy}(2020)}]{Dash:2020zqx}%
  \BibitemOpen
  \bibfield  {author} {\bibinfo {author} {\bibfnamefont {Ashutosh}\
  \bibnamefont {Dash}}\ and\ \bibinfo {author} {\bibfnamefont {Victor}\
  \bibnamefont {Roy}},\ }\bibfield  {title} {\enquote {\bibinfo {title}
  {{Hydrodynamic attractors for Gubser flow}},}\ }\href {\doibase
  10.1016/j.physletb.2020.135481} {\bibfield  {journal} {\bibinfo  {journal}
  {Phys. Lett. B}\ }\textbf {\bibinfo {volume} {806}},\ \bibinfo {pages}
  {135481} (\bibinfo {year} {2020})},\ \Eprint
  {http://arxiv.org/abs/2001.10756} {arXiv:2001.10756 [nucl-th]} \BibitemShut
  {NoStop}%
\bibitem [{\citenamefont {Gubser}(2010)}]{Gubser:2010ze}%
  \BibitemOpen
  \bibfield  {author} {\bibinfo {author} {\bibfnamefont {Steven~S.}\
  \bibnamefont {Gubser}},\ }\bibfield  {title} {\enquote {\bibinfo {title}
  {{Symmetry constraints on generalizations of Bjorken flow}},}\ }\href
  {\doibase 10.1103/PhysRevD.82.085027} {\bibfield  {journal} {\bibinfo
  {journal} {Phys. Rev. D}\ }\textbf {\bibinfo {volume} {82}},\ \bibinfo
  {pages} {085027} (\bibinfo {year} {2010})},\ \Eprint
  {http://arxiv.org/abs/1006.0006} {arXiv:1006.0006 [hep-th]} \BibitemShut
  {NoStop}%
\bibitem [{\citenamefont {Bhalerao}\ \emph {et~al.}(2005)\citenamefont
  {Bhalerao}, \citenamefont {Blaizot}, \citenamefont {Borghini},\ and\
  \citenamefont {Ollitrault}}]{Bhalerao:2005mm}%
  \BibitemOpen
  \bibfield  {author} {\bibinfo {author} {\bibfnamefont {R.~S.}\ \bibnamefont
  {Bhalerao}}, \bibinfo {author} {\bibfnamefont {Jean-Paul}\ \bibnamefont
  {Blaizot}}, \bibinfo {author} {\bibfnamefont {Nicolas}\ \bibnamefont
  {Borghini}}, \ and\ \bibinfo {author} {\bibfnamefont {Jean-Yves}\
  \bibnamefont {Ollitrault}},\ }\bibfield  {title} {\enquote {\bibinfo {title}
  {{Elliptic flow and incomplete equilibration at RHIC}},}\ }\href {\doibase
  10.1016/j.physletb.2005.08.131} {\bibfield  {journal} {\bibinfo  {journal}
  {Phys. Lett. B}\ }\textbf {\bibinfo {volume} {627}},\ \bibinfo {pages}
  {49--54} (\bibinfo {year} {2005})},\ \Eprint
  {http://arxiv.org/abs/nucl-th/0508009} {arXiv:nucl-th/0508009} \BibitemShut
  {NoStop}%
\bibitem [{\citenamefont {York}\ and\ \citenamefont
  {Moore}(2009)}]{York:2008rr}%
  \BibitemOpen
  \bibfield  {author} {\bibinfo {author} {\bibfnamefont {Mark~Abraao}\
  \bibnamefont {York}}\ and\ \bibinfo {author} {\bibfnamefont {Guy~D.}\
  \bibnamefont {Moore}},\ }\bibfield  {title} {\enquote {\bibinfo {title}
  {{Second order hydrodynamic coefficients from kinetic theory}},}\ }\href
  {\doibase 10.1103/PhysRevD.79.054011} {\bibfield  {journal} {\bibinfo
  {journal} {Phys. Rev. D}\ }\textbf {\bibinfo {volume} {79}},\ \bibinfo
  {pages} {054011} (\bibinfo {year} {2009})},\ \Eprint
  {http://arxiv.org/abs/0811.0729} {arXiv:0811.0729 [hep-ph]} \BibitemShut
  {NoStop}%
\bibitem [{\citenamefont {Teaney}\ and\ \citenamefont
  {Yan}(2014)}]{Teaney:2013gca}%
  \BibitemOpen
  \bibfield  {author} {\bibinfo {author} {\bibfnamefont {Derek}\ \bibnamefont
  {Teaney}}\ and\ \bibinfo {author} {\bibfnamefont {Li}~\bibnamefont {Yan}},\
  }\bibfield  {title} {\enquote {\bibinfo {title} {{Second order viscous
  corrections to the harmonic spectrum in heavy ion collisions}},}\ }\href
  {\doibase 10.1103/PhysRevC.89.014901} {\bibfield  {journal} {\bibinfo
  {journal} {Phys. Rev. C}\ }\textbf {\bibinfo {volume} {89}},\ \bibinfo
  {pages} {014901} (\bibinfo {year} {2014})},\ \Eprint
  {http://arxiv.org/abs/1304.3753} {arXiv:1304.3753 [nucl-th]} \BibitemShut
  {NoStop}%
\bibitem [{\citenamefont {Giacalone}\ \emph {et~al.}(2019)\citenamefont
  {Giacalone}, \citenamefont {Mazeliauskas},\ and\ \citenamefont
  {Schlichting}}]{Giacalone:2019ldn}%
  \BibitemOpen
  \bibfield  {author} {\bibinfo {author} {\bibfnamefont {Giuliano}\
  \bibnamefont {Giacalone}}, \bibinfo {author} {\bibfnamefont {Aleksas}\
  \bibnamefont {Mazeliauskas}}, \ and\ \bibinfo {author} {\bibfnamefont
  {S\"oren}\ \bibnamefont {Schlichting}},\ }\bibfield  {title} {\enquote
  {\bibinfo {title} {{Hydrodynamic attractors, initial state energy and
  particle production in relativistic nuclear collisions}},}\ }\href {\doibase
  10.1103/PhysRevLett.123.262301} {\bibfield  {journal} {\bibinfo  {journal}
  {Phys. Rev. Lett.}\ }\textbf {\bibinfo {volume} {123}},\ \bibinfo {pages}
  {262301} (\bibinfo {year} {2019})},\ \Eprint
  {http://arxiv.org/abs/1908.02866} {arXiv:1908.02866 [hep-ph]} \BibitemShut
  {NoStop}%
\bibitem [{\citenamefont {Abramowitz}(1974)}]{10.5555/1098650}%
  \BibitemOpen
  \bibfield  {author} {\bibinfo {author} {\bibfnamefont {Milton}\ \bibnamefont
  {Abramowitz}},\ }\href@noop {} {\emph {\bibinfo {title} {Handbook of
  Mathematical Functions, With Formulas, Graphs, and Mathematical Tables,}}}\
  (\bibinfo  {publisher} {Dover Publications, Inc.},\ \bibinfo {address}
  {USA},\ \bibinfo {year} {1974})\BibitemShut {NoStop}%
\bibitem [{DLM(2020)}]{DLMF}%
  \BibitemOpen
  \href {http://dlmf.nist.gov/} {\enquote {\bibinfo {title} {Nist digital
  library of mathematical functions},}\ } (\bibinfo {year} {2020})\BibitemShut
  {NoStop}%
\bibitem [{\citenamefont {Silverstone}\ \emph {et~al.}(1985)\citenamefont
  {Silverstone}, \citenamefont {Nakai},\ and\ \citenamefont
  {Harris}}]{PhysRevA.32.1341}%
  \BibitemOpen
  \bibfield  {author} {\bibinfo {author} {\bibfnamefont {Harris~J.}\
  \bibnamefont {Silverstone}}, \bibinfo {author} {\bibfnamefont {Sachiko}\
  \bibnamefont {Nakai}}, \ and\ \bibinfo {author} {\bibfnamefont {Jonathan~G.}\
  \bibnamefont {Harris}},\ }\bibfield  {title} {\enquote {\bibinfo {title}
  {Observations on the summability of confluent hypergeometric functions and on
  semiclassical quantum mechanics},}\ }\href {\doibase
  10.1103/PhysRevA.32.1341} {\bibfield  {journal} {\bibinfo  {journal} {Phys.
  Rev. A}\ }\textbf {\bibinfo {volume} {32}},\ \bibinfo {pages} {1341--1345}
  (\bibinfo {year} {1985})}\BibitemShut {NoStop}%
\end{thebibliography}%

\end{document}